\documentclass[letterpaper,accepted=2024-01-12]{quantumarticle}
\pdfoutput=1

\usepackage[utf8]{inputenc}
\usepackage[english]{babel} %
\usepackage[T1]{fontenc}

\usepackage{amsmath,amsthm,amssymb,amsfonts,mathtools}
\usepackage{graphicx}
\usepackage[breaklinks=true,colorlinks=true,linkcolor=blue,urlcolor=blue,citecolor=blue%
]{hyperref}
\usepackage{color}
\usepackage{multirow,hhline}
\usepackage[table]{xcolor}
\usepackage{float}
\usepackage{algpseudocode}

\usepackage[numbers]{natbib}

\DeclareUnicodeCharacter{2009}{\,} 

\newcommand{\mat}[1]{\mathbf{#1}}
\newcommand{\tra}{\intercal} %
\newcommand{\Tr}{\operatorname{Tr}} %
\renewcommand{\arctan}{\mathrm{atan}}
\renewcommand{\vec}[1]{\mathbf{#1}}

\newcommand{\ie}{{\it i.e.},\ }

\newcommand{\eg}{{\it e.g.},\ }

\newcommand{\Hf}{\hat H_\mathrm{f}}
\newcommand{\Hi}{\hat H_\mathrm{i}}

\newcommand{\alad}{\hat a_\mathrm{A}} %
\newcommand{\pil}{ {\hat\pi_+} } %
\newcommand{\pir}{ {\hat\pi_-} } %
\newcommand{\nn}{\nonumber\\} %

\newcommand{\hint}{\hat H_\mathrm{i}}
\newcommand{\hd}{\hat H_\mathrm{d}}
\newcommand{\hf}{\hat H_\mathrm{f}}

\newcommand{\ket}[1]{\left| {#1} \right\rangle}
\newcommand{\bra}[1]{\left\langle {#1} \right|}
\newcommand{\braket}[2]{\left\langle \vphantom{#2} {#1}\left|\vphantom{#1}{#2}\right.\right\rangle}
\newcommand{\ketbra}[2]{\left| {#1}\vphantom{#2} \right\rangle\!\left\langle {#2}\vphantom{#1} \right|}

\newcommand{\exptval}[1]{\left<#1\right>}
\newcommand{\normord}[1]{\,:#1 :\,}
\newcommand{\ii}{\mathrm{i}}
\newcommand{\ee}[1]{\mathrm{e}^{#1}} %

\newcommand{\hc}{\mathrm{h.c.}}

\newcommand{\id}{\mathbb{I}} %

\newcommand{\comm}[2]{\left[{#1},{#2}\right]}

\newcommand{\integral}[3]{\int_{#2}^{#3} \!\! \mathrm{d} #1 \,}
\newcommand{\difffrac}[2]{\frac{\mathrm{d} #1}{\mathrm{d} #2}}

\newcommand{\partialfrac}[2]{\frac{\partial #1}{\partial #2}}
\newcommand{\diff}{\text{d}}

\newcommand{\sgn}{\operatorname{sgn}}

\newcommand{\be}{\hat b^{(\mathrm{e})}_k}
\newcommand{\dk}{\hat d_k}
\newcommand{\dkd}{\hat d_k^\dagger}

\newcommand{\dmkd}{\hat d_{-k}^\dagger}

\newcommand{\rhj}[1]{{\color{magenta}#1}}

\begin{document}

\title{Chain-mapping methods for relativistic light-matter interactions}

\author{Robert H. Jonsson}
\email{robert.jonsson@su.se}
\affiliation{Max-Planck-Institut für Quantenoptik, Hans-Kopfermann-Str.~1, 85748 Garching, Germany}
\affiliation{Nordita, Stockholm University and KTH Royal Institute of Technology, Hannes Alfv\'ens v\"ag 12, SE-106 91 Stockholm, Sweden}

\author{Johannes Knörzer}
\email{jknoerzer@ethz.ch}
\affiliation{Institute for Theoretical Studies, ETH Zurich, 8092 Zurich, Switzerland}

\begin{abstract}
The interaction between localized emitters and quantum fields, both in relativistic settings and in the case of ultra-strong couplings, requires non-perturbative methods beyond the rotating-wave approximation.
In this work we employ chain-mapping methods to achieve a numerically exact treatment of the interaction between a localized emitter and a scalar quantum field.
We extend the application range of these methods beyond emitter observables and apply them to study field observables.
We first provide an overview of chain-mapping methods and their physical interpretation, and discuss the thermal double construction for systems coupled to thermal field states.
Modelling the emitter as an Unruh-DeWitt particle detector, we then calculate the energy density emitted by a detector coupling strongly to the field.
As a stimulating demonstration of the approach's potential, we calculate the radiation emitted from an accelerated detector in the Unruh effect, which is closely related to the thermal double construction as we discuss.
We comment on prospects and challenges of the method.
\end{abstract}

\maketitle

\section{Introduction}

\begin{figure*}[t!]
\centering
\includegraphics[width=15.5cm]{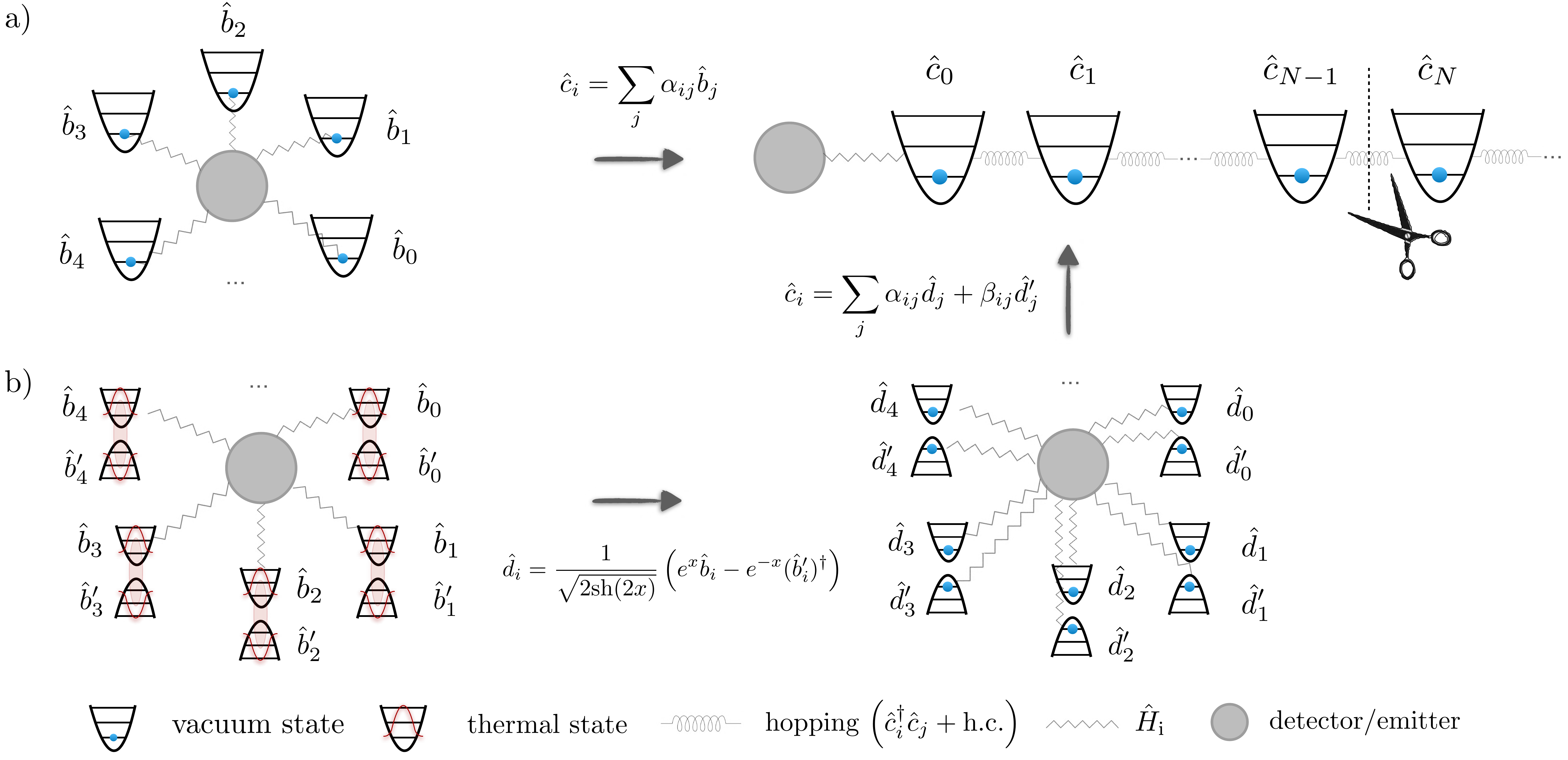}
\caption{Schematic depiction of chain transformation for field in a) vacuum or b) thermal state, respectively.
\textit{Upper panel:} field mode operators $\hat b_i$ and chain mode operators $\hat c_i$ are related by \eqref{eq:definition-chain-operators} and \eqref{eq:be_from_chain}, see Sec.~\ref{ssec:chain-mapping}.
\textit{Lower panel:} in the thermal double construction, each $\hat b_i$ mode is assigned a partner mode $\hat b_i^\prime$, see Sec.~\ref{ssec:thermal-double}.
New modes $\hat d_i$ are introduced by means of a two-mode squeezing transformation with respect to the $\hat b_i$ and $\hat b_i^\prime$ modes, cf.~Eq.~\eqref{eq:dmodes_thermal}.
The chain modes $\hat c_i$ for the thermal case are then expressed in terms of the $\hat d_i$ modes, cf.~Eq.~\eqref{eq:thermal_chain_modes_from_poly}.
The coupling $\hat H_\mathrm{i}$ is equivalently described by Eqs.~\eqref{eq:interaction-general}, \eqref{eq:interaction-star}, \eqref{eq:interaction-chain} and \eqref{eq:hint-dmodes} in the main text.
}
\label{fig:schematic_transformations}
\end{figure*}

Interacting quantum systems are ubiquitous in nature.
Yet their dynamics is challenging to predict beyond simplifying approximations.
A versatile set of computational tools is offered by the theory of open quantum systems, in which a physical system of interest is described as being coupled to its environment \cite{breuer_theory_2002,breuer_colloquium_2016,weimer_simulation_2021}.
Common approaches in the study of open systems rely on an effective description of the system which may be obtained by tracing out the environmental degrees of freedom yielding a quantum master equation.
Its validity is usually restricted to weak system-bath couplings and short-lived bath correlations within the Born-Markov approximation.
Physically it describes scenarios of low entanglement between system and environment and, being an effective description of the reduced state of the system, yields access only to system and not to bath observables.

Physical systems do not necessarily satisfy the underlying assumptions of weak coupling and Markovianity as encountered in, \eg quantum optics \cite{gustafsson_propagating_2014,andersson_non-exponential_2019,gonzalez-tudela_engineering_2019}, condensed-matter physics \cite{de_vega_matter-wave_2008,groblacher_observation_2015,del_pino_tensor_2018}, quantum chemistry and biology \cite{huelga_vibrations_2013,chen_using_2015,pollock_non-markovian_2018}, or acceleration-induced quantum effects \cite{lopp_quantum_2021,soda_acceleration-induced_2022}.
Under such conditions, predicting the time evolution is challenging.
While the Nakajima-Zwanzig generalized master equation \cite{nakajima_quantum_1958,zwanzig_ensemble_1960} provides an exact framework for the simulation of quantum dynamics, it is usually hard to derive.
For special cases, \eg if the environment may be described by independent quantum harmonic oscillators, there exist numerically convergent methods to calculate the dynamics within the non-Markovian and strong-coupling regimes \cite{tanimura_time_1989,tanimura_numerically_2020,prior_efficient_2010,chin_exact_2010,feynman_theory_1963}.

While most approaches aim at solving the dynamics of the reduced system only, some physical phenomena require a detailed analysis of bath observables.
Apart from exact diagonalization, which becomes intractable for moderate system sizes, the total system dynamics may be obtained by unitarily mapping the underlying model onto a one-dimensional chain Hamiltonian and performing time evolution with respect to this tight-binding chain.
Tracing back to the numerical renormalization group \cite{wilson_renormalization_1975,vojta_quantum_2005,bulla_numerical_2005,bulla_numerical_2008}, so-called star-to-chain transformations may even be performed analytically and without previous discretization of the environment \cite{prior_efficient_2010,chin_exact_2010},  and they are closely related to \emph{reaction coordinate mappings} (see, for example,~\cite{nazir_reaction_2018, puebla_spin-boson_2019, strasberg_nonequilibrium_2016}).
The resulting semi-infinite chain may be truncated and the thereby obtained model can be evolved efficiently by matrix-product state (MPS) simulations \cite{vidal_efficient_2004,cirac_matrix_2021}.
For a large class of Gaussian bosonic environments, the validity of this truncation can be certified by appropriate error bounds \cite{woods_simulating_2015,woods_dynamical_2016,mascherpa_open_2017,de_vega_how_2015,trivedi_convergence_2021}.

Chain-mapping approaches have recently been utilized to investigate a variety of different problems, \eg for non-perturbative studies of light-matter interaction at strong couplings \cite{sanchez_munoz_resolution_2018,lambert_modelling_2019,noachtar_nonperturbative_2022} and quantum impurity problems in structured environments \cite{busser_lanczos_2013,allerdt_kondo_2015,allerdt_numerically_2019}.
Other works have focused on the extension of this approach to thermal baths via a thermofield transformation \cite{bargmann_hilbert_1961,araki_representations_1963,takahashi_thermo_1996}, which may be used to map an initially thermal chain to two empty chains \cite{de_vega_thermofield-based_2015,tamascelli_efficient_2019,landi_nonequilibrium_2022}.
The latter case of two initially empty chains provides a useful starting point for MPS-based numerical simulations in the presence of thermal baths as demonstrated, \eg in Refs.~\cite{guo_stable-unstable_2018,schwarz_nonequilibrium_2018,chen_steady-state_2020}.
Yet most of the previous works have focused on the reduced system's dynamics or explored only coarse-grained bath observables, while the approach's broad access to bath observables remains to be fully leveraged.
In this direction, some more recent works started recognizing the insights into emergent dissipative phenomena and non-perturbative effects that can be obtained by combining thermofield-based chain mappings with MPS simulations for treatments of bath dynamics in various settings ranging from chemistry to condensed matter \cite{dunnett_matrix_2021,lacroix_unveiling_2021,riva_thermal_2023}.

In this work, we demonstrate the potential of chain mappings for detailed, non-perturbative studies of field observables, such as the energy density radiated from an emitter interacting strongly with a quantum field.
As a physically interesting and stimulating first example, we consider the question what kind of radiation is emitted from a uniformly accelerated emitter in the context of the Unruh effect.
To model the emitter we use the Unruh-DeWitt (UDW) particle detector model~\cite{unruh_notes_1976,hawking_quantum_1979,hu_relativistic_2012}.
The UDW detector model is a central tool in relativistic quantum information, used to address a wide range of phenomena including the Unruh effect \cite{crispino_unruh_2008}, but also Hawking radiation, vacuum entanglement, relativistic communication, particle and radiation creation  or even superposition of trajectories and temporal orders \cite{mann_relativistic_2012}.
It consists of a single emitter, modelled as a two-level system (TLS) or a harmonic oscillator (HO), which couples via its monopole operator to a scalar quantum field.
After being first posed, the question whether a accelerated detectors emit radiation inspired various works whose development is summarized, for example, in~\cite{crispino_unruh_2008} and~\cite{lin_accelerated_2006}.
The general conclusion was that uniformly accelerated detectors do not emit radiation when they are in an equilibrium state.
However, radiation emission is expected in transients, for example, immediately after the detector-field interaction is switched on.
With respect to the present work, it is interesting to note that these conclusions were based on works considering HO detectors for which the model can be solved exactly~\cite{raine_does_1991, hinterleitner_inertial_1993,massar_problem_1993, massar_vacuum_1996,audretsch_radiation_1994,  kim_radiation_1997,kim_quantum_1999, lin_accelerated_2006}, whereas no exact, non-perturbative solution is known for TLS detectors.

Many questions in relativistic quantum information, in particular questions concerning the extraction and transmission of entanglement, require the non-perturbative treatment of the detector-field interaction.
For TLS detectors, whose Hamiltonian is a type of spin-boson model, non-perturbative solutions are challenging and only a limited number of solutions are known, as recently summarized in~\cite{tjoa_non-perturbative_2023}.
Here we treat them using MPS-based approaches.
For HO detectors, Gaussian state methods can be employed for non-perturbative treatments~\cite{brown_detectors_2013,bruschi_time_2013} which we also build upon in our work.

Here we show that, employing star-to-chain transformations, it is possible to calculate non-perturbatively the time evolution of the joint detector-field state.
Most interestingly, the approach introduces no approximations to the model but allows for (\textit{i}) a treatment which is numerically exact up to a time scale determined by the numerical resources available and (\textit{ii}) a precise control over the simulation error.
Since the UDW model is prototypical for many models in quantum optics, these results lead the way to future applications, for example, in the treatment of ultra-strong matter-light couplings.
Specifically in the following we calculate the time evolution of the detector state and the energy density emitted by both resting and accelerated detectors into the Minkowski vacuum state of a massless scalar field in 1+1 dimensions.
We consider TLS and HO detectors, initialized in their ground or (first) excited states, and verify that the applied coupling strength is significantly beyond the regime of leading-order time-dependent perturbation theory.

This work is organized as follows:
Sec.~\ref{sec:theory} summarizes the employed chain mapping combined with the thermofield approach.
Sec.~\ref{sec:truncation_error} discusses errors arising in the approach and how they restrict the maximal simulation times, for both a free field and an emitter coupled to the vacuum field.
Subsequently, thermal field states are considered and the relation to the Unruh effect is made explicit in Sec.~\ref{sec:unruh}.
The results are summarized in Sec.~\ref{sec:outlook}, where we also provide future perspectives.

\section{Theoretical framework \label{sec:theory}}

In this section, we introduce our theoretical approach and numerical methods.
In Sec.~\ref{ssec:chain-mapping} we briefly review chain transformations and employ them to cast the UDW detector model into a form that can be studied efficiently using our numerical methods.
To account for the coupling to thermal field states we summarize the thermal double construction and subsequent chain transformation in Sec.~\ref{ssec:thermal-double}.
This is followed by a brief discussion of the numerical methods we utilize in Sec.~\ref{ssec:num_methods}, and a proof-of-principle demonstration in Sec.~\ref{ssec:numerical-example}, in which we calculate the energy density of a detector coupled to the vacuum.

\subsection{Chain mapping\label{ssec:chain-mapping}}
\textit{UDW model}.\textemdash
Chain mappings can be applied to systems coupled bilinearly to a harmonic bath.
Here we apply them to the UDW detector model, which phenomenologically describes a monopole detector coupled to a massless scalar field.
We start by considering a generic Hamiltonian
\begin{equation}\label{eq:general-model}
    \hat H = \hf + \hd + \hint,
\end{equation}
which contains the free field described by $\hf$, a detector modeled by $\hd$, and an interaction Hamiltonian $\hint$.

We here consider a massless scalar field in 1+1D for which the field Hamiltonian reads, in the Schrödinger picture,
\begin{equation}\label{eq:field-hamiltonian-1}
    \hf = \integral{k}{-\infty}\infty |k|\,  \hat b^\dagger_k \hat b_k,
\end{equation}
and is described by bosonic annihilation (creation) operators $\hat b_k^{(\dagger)}$.
(When applying this method to massive fields or higher spacetime dimensions, the general dispersion relation $\omega_k$ appears in the field Hamiltonian instead of $|k|$, see~\cite{chin_exact_2010}.)

The field is coupled to a detector, or emitter, with which we mean either a two-level system (TLS) or a harmonic oscillator (HO) that can emit or absorb energy by interacting with the field.
For these two cases, we consider the detector models (in the following, $\hbar = c = 1$):
\begin{equation}\label{eq:detector-models}
    \hd^{(\mathrm{TLS})} = \frac{\Omega_{\mathrm{d}}}{2} \hat \sigma_z, \quad \hd^{(\mathrm{HO})} = \Omega_{\mathrm{d}} \left ( \hat a^\dagger \hat a + \frac{1}{2} \right ).
\end{equation}
Here, $\Omega_{\mathrm{d}}$ is the level spacing, $\hat a^{(\dagger)}$ are ladder operators of the oscillator, $\hat \sigma$ denotes the Pauli spin operator and $\hat \sigma_z$ its $z$ component.
Finally, the bilinear interaction between field and detector is modeled by
\begin{equation}\label{eq:interaction-general}
    \hint = \lambda \ \hat X \otimes \int \mathrm{d}x f(x) \hat \pi(x),
\end{equation}
with the dimensionless coupling constant $\lambda$, field momentum $\hat \pi$ and the \textit{smearing function} $f(x)$ that describes the shape of the coupling in real space.
The field couples to the detector via the system operators $\hat X = \hat \sigma^+ + \hat \sigma^-$ (TLS) and $\hat X = \hat a^\dagger + \hat a$ (HO), respectively.
Note that this interaction Hamiltonian is not number-conserving. %
Generally speaking, at strong couplings this constitutes a formidable challenge for numerical treatments.

\textit{Lorentzian coupling profile}.\textemdash
In the following, we choose to model the interaction by a Lorentzian smearing function, \ie
\begin{equation}\label{eq:lorentzian-smearing}
    f(x) = \frac{L}{\pi(L^2 + x^2)},
\end{equation}
with length scale $L$, see also Table~\ref{tab:overview_constants}.
For the purpose of this section, which concerns the coupling to the vacuum state of the field, also other choices of smearing functions, \eg a Gaussian function, yield analytical expressions in the following. 
Our choice of a Lorentzian smearing is motivated by the fact that it yields certain closed-form solutions also for thermal states (see Eq.~\eqref{eq:thermal_weight_moment_integral} below).

Because this smearing function is even, \ie $f(x)=f(-x)$, 
the detector only couples to the even sector of the field.
With $\hat b^{(e/o)}_k =\left(\hat b_k\pm\hat b_{-k}\right)/\sqrt2$, which yields $\hf=\integral{k}0\infty k \left(\hat b^{(e)\dagger}_k\hat b^{(e)}_k+\hat b^{(o)\dagger}_k\hat b^{(o)}_k\right)=:\hf^{(e)} +\hf^{(o)} $, the interaction Hamiltonian only couples to even modes,
\begin{equation}\label{eq:interaction-star}
    \hint = \lambda \ \hat X \otimes \left (\sqrt2 \integral{k}0\infty f_k \hat b^{(e)}_k + f_k^* \hat b^{(e)\dagger}_k \right ),
\end{equation}
and we can discard the (dynamics of) the odd sector of the field henceforth.
Here, the coupling coefficients $f_k$ are
\begin{equation}\label{eq:def-fk}
    f_k = \frac{-\ii\sqrt{k}}{\sqrt{4\pi}}\left( \integral{x}{}{}\ee{\ii kx}f(x)\right)= - \mathrm{i} \sqrt{\frac{k}{4\pi}} \mathrm{e}^{-Lk}.
\end{equation}

\textit{Chain modes}.\textemdash
The model captured by Eqs.~(\ref{eq:general-model},\ref{eq:field-hamiltonian-1},\ref{eq:detector-models},\ref{eq:interaction-star}) describes a (harmonic or two-level) detector coupled to independent harmonic oscillators, as schematically depicted in Fig.~\ref{fig:schematic_transformations}(a).
This Hamiltonian may be transformed such that the new model takes the form of a semi-infinite chain with only nearest-neighbor interactions.
To this end, we introduce the chain mode operators as
\begin{equation}\label{eq:definition-chain-operators}
\hat c_i=\sqrt2 \integral{k}0{\infty}  f_k p_i(k) \hat b^{(e)}_k.
\end{equation}
As originally presented and detailed in Ref.~\cite{chin_exact_2010}, the functions $p_i(k)$ ($i = 0, 1, 2, ...$) form a family of orthogonal polynomials,
\begin{equation}\label{eq:orthonormality-polys}
2 \integral{k}0{\infty} |f_k|^2 \, p_i(k)\, p_j(k)\,=\delta_{ij},
\end{equation}
which for the Lorentzian detector profile~\eqref{eq:lorentzian-smearing} is given by rescaled and normalized Laguerre polynomials (see~\eqref{eq:pn_fron_laguerre_long}):
\begin{equation}\label{eq:pn_from_laguerre}
    p_n(k)=\frac{L\sqrt{8\pi}}{ \sqrt{n+1}}L^1_n(2L k).
\end{equation}

\textit{Chain Hamiltonian}.\textemdash
The chain form of the field Hamiltonian is obtained by plugging the inverse Bogoliubov transformation,
\begin{equation}\label{eq:be_from_chain}
    \hat b_k^{(e)} =\sqrt2 \sum_i f_k^* p_i(k) \hat c_i,
\end{equation}
into Eq.~\eqref{eq:field-hamiltonian-1}, and using both Eq.~\eqref{eq:orthonormality-polys} and the recurrence relations \cite{chin_exact_2010}
\begin{equation}
    k p_n(k) = \gamma_n p_{n+1}(k) +\nu_n p_n(k) +\gamma_{n-1} p_{n-1}(k),
\end{equation}
where, by convention, $\gamma_{-1}=0$.
The Hamiltonian then takes the form
\begin{equation}\label{eq:hfe_chain_form}
    \hf^{(e)} = \sum_{i=0,1,\dots} \nu_i \hat c_i^\dagger \hat c_i + \gamma_i \left ( \hat c_i^\dagger \hat c_{i+1} + \hat c_{i+1}^\dagger \hat c_i \right ),
\end{equation}
which describes the anticipated chain with nearest-neighbor interactions.
For the Lorentzian detector, the Laguerre polynomials' recurrence relations yield
\begin{equation}
\gamma_n =-   \frac{\sqrt{(n+2)(n+1)}}{2L},\quad \nu_n= \frac{n+1}L.
\end{equation}
Correspondingly, the interaction Hamiltonian now takes the form of a detector coupled only to the first chain mode,
\begin{equation}\label{eq:interaction-chain}
    \hint = \lambda\, \kappa \ \hat X \otimes \left ( \hat c_0 + \hat c_0^\dagger \right ),
\end{equation}
with the normalization constant $\kappa = 1/(L\sqrt{8\pi})$ for the Lorentzian coupling profile.
With this we arrive at the chain-mode representation of our general model \eqref{eq:general-model}, which is $\hat H_\mathrm{chain} = \hat H_\mathrm{f}^{(\mathrm{e})} + \hat H_\mathrm{d}^{(\mathrm{TLS/HO})} + \hint$, combining Eqs.~\eqref{eq:detector-models}, \eqref{eq:hfe_chain_form} and \eqref{eq:interaction-chain}.

\subsection{Thermal double construction \label{ssec:thermal-double}}

In this work, we investigate the coupling of a detector to thermal field states.
Specifically, in Sec.~\ref{sec:unruh}, we consider a detector which due to its uniform acceleration is coupled to the Rindler modes of the field which are in a thermal state.

The chain transformation as introduced in the previous section for the vacuum state of the field, however, is inapt for a direct treatment of thermal field states, since (\textit{i}) their representation in terms of chain modes may be non-trivial or inefficient, and (\textit{ii}) since they can contain a large number of excitations, whereas our numerical MPS simulations are restricted to a small number of field excitations.
This problem can be circumvented by resorting to a thermal double construction \cite{de_vega_thermofield-based_2015}, in which the original environment is viewed as a subsystem of an enlarged environment in its vacuum state.
This enlarged environment can again be treated efficiently using chain transformations and numerical simulations.
This subsection reviews how to apply the thermal double construction to our model for a thermal field state with inverse temperature $\beta$.
App.~\ref{app:thermal_edensity} discusses how the energy density emitted from a detector at rest, which couples to a thermal state of the field, can be evaluated numerically and derives the necessary expressions.

\textit{Double construction}.\textemdash
The enlargement of the environment is given by a doubling of the field modes:
For each field mode ($\hat b_k$) we introduce a partner mode ($\hat b_k^\prime$) with opposite excitation energy.
As indicated in Fig.~\ref{fig:schematic_transformations}(b), each pair of partner modes is in a two-mode squeezed state such that the individual modes' partial state is a thermal state.
The overall state of the doubled field, however, is pure and corresponds to the vacuum state of the 'unsqueezed' $\hat d_k$ and $\hat d'_k$ modes.

Before going through the individual steps of these transformations, as above, we make use of the fact that the detector only couples to even field modes. 
Thus we can discard the odd sector of the field and apply the double construction to the even sector only.
The even-sector, doubled-field Hamiltonian reads
\begin{equation}\label{eq:hf_doubled_even}
    \Hf'^{(e)} =\integral{k}0\infty k\, \left ( \left(\hat{b}_k^{(e)}{}\right)^\dagger  \hat{b}_k^{(e)} -\left(\hat{b}_k^{(e)'}{}\right)^\dagger \hat{b}_k^{(e)'} \right ).
\end{equation}
By acting with two-mode squeezing transformations on each pair of partner modes, we obtain a new basis of  canonically commuting operators
\begin{align}\label{eq:dmodes_thermal}
    \hat d_k&=\frac{ \ee{\frac{\beta k}{4}}\hat{b}_k^{(e)} -\ee{-\frac{\beta k}{4}}\hat{b}_k^{(e)'}{}^\dagger }{\sqrt{2\sinh(\beta k/2)}}
    ,\quad
    \hat {d'}_k =\frac{\ee{\frac{\beta k}{4}} \hat{b}_k^{(e)'}  -\ee{-\frac{\beta k}{4}} \hat{b}_k^{(e)}{}^\dagger }{\sqrt{2\sinh(\beta k/2)}},
\end{align}
under which the field Hamiltonian remains invariant, %
\begin{align}\label{eq:hf-dmodes-1}
\Hf'^{(e)} &=\integral{k}0\infty k \, \left ( \hat{d}_k^\dagger  \hat{d}_k -\hat{d'}_k^\dagger  \hat{d'}_k \right ).
\end{align}
The squeezing parameter $(\beta k)/4$ is chosen such that the vacuum $\ket{0_{\mathrm{D}}}$ of these new modes (\ie the state $\ket{0_\mathrm{D}}$ for which $\hat d_k^{(\prime)} \ket{0_\mathrm{D}} = 0$) is the thermal state of $\hf$ with inverse temperature $\beta$ on the original field modes $\hat b^{(e)}_k$, %
\begin{equation}\label{eq:nk_thermal_exptval}
    \bra{0_{\mathrm{D}}} \hat b_k^{(e)\dagger} \hat b_{k'}^{(e)} \ket{0_{\mathrm{D}}} = \delta(k-k^\prime) \left ( e^{\beta k} - 1 \right )^{-1}.
\end{equation}

The interaction Hamiltonian $\Hi$ remains unchanged in the thermal double construction.
To express $\Hi$ in terms of the new modes, we invert~\eqref{eq:dmodes_thermal} and obtain $\be = (e^{\beta k/4}\dk + e^{-\beta k/4} \dmkd)/\sqrt{2\sinh(\beta k/2)}$,
which we insert in~\eqref{eq:interaction-star},
\begin{align}\label{eq:hint-dmodes}
    &\Hi =  \lambda \hat X \otimes \integral{k}{-\infty}\infty \frac{\sgn(k) f_{|k|} \ee{\frac{\beta k}{4}}   }{\sqrt{|\sinh(\beta k/2)|}} \hat d_k + \hc\,,
\end{align}
where we used, from \eqref{eq:def-fk}, that $f_k^*=-f_k$ is purely imaginary, and place the primed $\hat d_k^\prime$ operators on the negative-$k$ axis via the identification $k<0: \,\hat d_k = \hat d^\prime_{-k}$.
With this identification, the doubled field Hamiltonian~\eqref{eq:hf-dmodes-1} takes the form $\hf'^{(e)} = \integral{k}{-\infty}\infty k\, \dkd \dk$.

By doubling the number of field modes the thermal double construction has enlarged the system we need to simulate from the total Hamiltonian $\hat H=\hf^{(e)}+\hd+\Hi$ to the total Hamiltonian $\hat H'=\hf'^{(e)}+\hd+\Hi$.
However, the $\hat d_k$-modes are eigenmodes of $\hf'^{(e)}$, and the initial state of the field is their vacuum state.
Hence, as depicted in Fig.~\ref{fig:schematic_transformations}, the enlarged system can again be treated efficiently with chain transformations, where the chain modes $\hat c_i$ are constructed from the $\hat d_k$-modes.

\textit{Chain modes}.\textemdash
The chain modes are obtained by the same procedure as outlined above, in Sec.~\ref{ssec:chain-mapping}.
However, instead of the Laguerre polynomials from \eqref{eq:pn_from_laguerre}, in the thermal case the polynomials $q_i(k)$ are defined on the entire real line and need to obey
\begin{equation}\label{eq:thermal_inner_product}
    \integral{k}{-\infty}\infty w(k) q_n(k) q_m(k) =\delta_{nm},
    \quad w(k)= \frac{k \ee{-2L|k|}\ee{\beta k/2} }{4\pi\sinh(\beta k/2) }.
\end{equation}
In contrast to the vacuum case, the weight function $w(k)$ does, to our knowledge, not correspond to one of the well known and studied families of orthogonal polynomials.
Hence, the polynomial coefficients
\begin{equation}\label{eq:defn_thermal_polys}
    q_n(k)=\sum_{i=0}^n P_{n,i} k^i
\end{equation}
have to be determined numerically.
For this it is useful that the   moments of the weight function for the Lorentzian detector profile have a closed-form solution in terms of Polygamma functions:
\begin{equation}\label{eq:thermal_weight_moment_integral}
\begin{split}
    &{(-1)\pi2^{n+3} L^{n+2}} \integral{k}{-\infty}\infty w(k) k^n 
    \\ &=   {(-1)^{n}(n+1)!-\left(1+(-1)^n\right) \left(\tfrac{2L}{\beta}\right)^{n+2}\,\psi^{(n+1)}\left(\tfrac{2L}{\beta}\right) }.
    \end{split}
\end{equation}
Based on numerical evaluations of these, the coefficients $P_{n,i}$ can be obtained from a Cholesky decomposition of the moment matrix, as detailed in App.~\ref{app:choleskydecomp}.
This step requires large numerical precision because the size of the weight moments~\eqref{eq:thermal_weight_moment_integral} spans a large range of orders of magnitude, as do the resulting coefficients $P_{n,i}$.
In fact, this is to be expected from the analytical solution of the vacuum field state in the previous Sec.~\ref{ssec:chain-mapping}:
The size of the coefficients $\left|P_{249,i}\right|$ ranges from $\log_{10}\left|P_{249,20}\right|\approx 18$ to $\log_{10}\left|P_{249,249}\right|\approx -414$.

To numerically calculate the coefficients $P_{n,i}$ in the thermal double field construction we use Mathematica~\cite{Mathematica} to obtain several hundreds of digits of precision.
This high precision in the beginning of the calculations, which may appear as an overhead at this point, is later consumed, for example, in the evaluation of the energy density emitted from the detector.
When evaluating expressions for the field energy density, such as~\eqref{eq:app_mk_edensity_from_chain} or~\eqref{eq:emergy_density_from_thermal_chain} which we derive below, the coefficients $P_{n,i}$ get multiplied by coefficients $I_i$ spanning a similar range of orders of magnitude.
In this step, many digits of precision are lost, hence a high precision in the initial calculation of $P_{n,i}$ (and $I_i$) is required in order to still be able to extract the energy densities from the numerically calculated covariance matrices of the state with good precision.

With the polynomials at hand, the chain modes for the thermal case are given by
\begin{equation}\label{eq:thermal_chain_modes_from_poly}
    \hat c_i = \integral{k}{-\infty}{\infty} \frac{\sgn(k) f_{|k|} \ee{\frac{\beta k}{4}}   }{\sqrt{|\sinh(\beta k/2)|}} p_i(k) \dk.
\end{equation}
In terms of these mode operators, the interaction Hamiltonian $\hint$ in~\eqref{eq:hint-dmodes} takes the same form as~\eqref{eq:interaction-chain}, and the double field Hamiltonian $\hf'^{(e)}$ takes the same form as in~\eqref{eq:hfe_chain_form}, where the normalization and coupling constants
\begin{align}
\kappa=\frac1{P_{0,0}},\quad \!\! \gamma_n= \frac{P_{n,n}}{P_{n+1,n+1}},\quad \!\! \nu_n=\frac{P_{n,n-1}}{P_{n,n}}-\frac{P_{n+1,n}}{P_{n+1,n+1}},
\end{align}
now follow from the polynomial recurrence relations~\cite{chin_exact_2010}.

\subsection{Numerical methods for time evolution\label{ssec:num_methods}}

We consider two different detectors, \ie a two-level system (TLS) and a harmonic oscillator (HO), as introduced in \eqref{eq:detector-models}.
To treat the composite detector-field system numerically, we utilize two different approaches: matrix product states methods for the TLS, and Gaussian state methods for the HO.
In the following we will briefly highlight these methods, and their different sources of numerical errors.

In addition to the method-specific errors, all numerical methods share a common error which arises because only chains of finite length can be treated numerically. Sec.~\ref{sec:truncation_error} discusses this general truncation error separately and in detail.

\textit{Two-level detector}.\textemdash
The two-level system is described by $\hat H_\mathrm{d}^{(\mathrm{TLS})}$ given in~\eqref{eq:detector-models}.
To compute the time evolution $\ket{\psi_{t+dt}} = \hat U(dt) \ket{\psi_t} = \ee{-i\hat H_\mathrm{chain} dt} \ket{\psi_t}$, we time-evolve the MPS $\ket{\psi_t}$ at time $t$ using the Trotter method, \ie a second-order Trotter-Suzuki decomposition of the time-evolution operator $\hat U(dt)$.
It is well-known that this method is prone to two main sources of error \cite{paeckel_time-evolution_2019}:
(\textit{i}) a total time-step error of order $O(dt^2)$ per unit time, and
(\textit{ii}) a truncation error of the time-evolved state to a manageable bond dimension.
In order to reduce the first type of error, choosing a small time step $dt$ is desirable.
However this increases the required number of time steps to evolve.
Moreover, if $dt$ is chosen too small, the state truncated to a given bond dimension does not properly time-evolve since the truncation error becomes too large in comparison.
We discuss this effect in more detail in App.~\ref{app:mps-dt}, where we also comment on the choice of $dt$, with which we obtain the results in this work.

\textit{Harmonic detector}.\textemdash
The harmonic detector is modeled by $\hat H_\mathrm{d}^{(\mathrm{HO})}$ in \eqref{eq:detector-models}.
Since this Hamiltonian is quadratic and since the initial states we consider are Gaussian states, the state remains Gaussian throughout its time evolution and is fully characterized by its covariance matrix.
This allows us to calculate the time evolution highly efficiently using Gaussian-state methods by a direct numerical exponentiation of the Hamiltonian generator (see, \eg~\cite{hackl_bosonic_2021}).
Using Mathematica~\cite{Mathematica} for these calculations allows us to obtain the time-evolved covariance matrix of the state with very high (hundreds of digits) precision.
For this reason, we expect the presented numerical results for HO detectors to be essentially unaffected by numerical errors, but to be only subject to the truncation error discussed in Sec.~\ref{sec:truncation_error}.
This difference between HO and TLS data is, \eg noticeable in the energy densities discussed in the subsequent subsection.

\subsection{Case study: energy density\label{ssec:numerical-example}}
As a first demonstration of our approach, in this section we consider the energy densities emitted from  detectors at rest of the different models in~\eqref{eq:detector-models}. 
This study provides a good basis for the detailed discussion of the truncation error in the following section, before we consider the radiation emitted by accelerated detectors in Sec.~\ref{sec:unruh}.

\begin{table}[t]
    \centering
    \begin{tabular}{|l|c|c|}
    \toprule
         \textbf{Quantity} & \textbf{ %
         Default
         Value
         } & \textbf{Eq. %
         }    \\ \hline \hline
           Detector width & $L$ & \eqref{eq:lorentzian-smearing} \\ %
        Detector energy gap& $\Omega_{\mathrm{d}}=2 \pi/(5L)$ &\eqref{eq:detector-models} \\
         Coupling constant & $\lambda=2$& \eqref{eq:interaction-general} 
    \\    \botrule
    \end{tabular}
    \caption{Overview of model constants for detector model together with their default numerical values which are used throughout this work, unless stated otherwise.
    The detector width $L$ is used as unit for other numerical values and results. %
    }
    \label{tab:overview_constants}
\end{table}

\textit{Energy density of massless field}.\textemdash
The energy density of the massless Klein-Gordon field~\cite{birrell_quantum_1982},
\begin{equation}\label{eq:Mk_Edensity}
    \hat T_{00}(x)= \frac12\left( \hat\pi(x) ^2+\left(\partial_x\hat\phi(x)\right)^2\right)
     =  \pir(x)^2+\pil(x)^2,
\end{equation}
decouples into the right-moving energy density $\pir^2$ and the left-moving energy density $\pil^2$, 
which are the squares of the right- and left-moving sectors of the field momentum,
\begin{equation}
    \hat\pi_\mp= -\ii\integral{k}0\infty \sqrt{\frac{k}{4\pi}} \left( \ee{\ii \pm{k} x} \hat b_{\pm{k}}-  \ee{\mp\ii {k} x} \hat b_{\pm{k}}^\dagger\right) .
\end{equation}
We consider the normal-ordered energy density, 
\begin{equation}
    \begin{split}\label{eq:norm_ord_pi_density_squared}
    \normord{\hat\pi_\mp^2(x)} &=\integral{k}0\infty\integral{k'}0\infty \frac{\sqrt{kk'}}{4\pi} \left(2 \ee{\mp \ii (k-k') x} \hat b^\dagger_{\pm k}\hat b_{\pm k'} 
    \right.\\&\quad \left. 
    - \ee{\pm\ii (k+k')x} \hat b_{\pm k} \hat b_{\pm k'} - \ee{\mp \ii (k+ k')x} \hat b_{\pm k}^\dagger \hat b_{\pm k'}^\dagger \right),
    \end{split}
\end{equation}
which when integrated up over all space  $\Hf=\integral{x}{}{}\normord{\hat T_{00}(x)}$ yields the field Hamiltonian \eqref{eq:field-hamiltonian-1}. 
To evaluate $\exptval{\normord{\hat\pi_\mp^2(x)}}$ from the numerical data, we rewrite the operator in terms of the chain mode operators $\hat c_i$, as detailed in App.~\ref{app:energy_density_coefffs_atrest}.
Because the coupling between detector and field is even, the expectation value of the left-moving energy density $\exptval{\normord{\hat\pi^2_+(x,t)}}=-\exptval{\normord{\hat\pi^2_+(-x,t)}}$ is simply the mirror image of the right-moving energy density, and we need only consider the latter.

\begin{figure}
\centering
\includegraphics[width=\columnwidth]{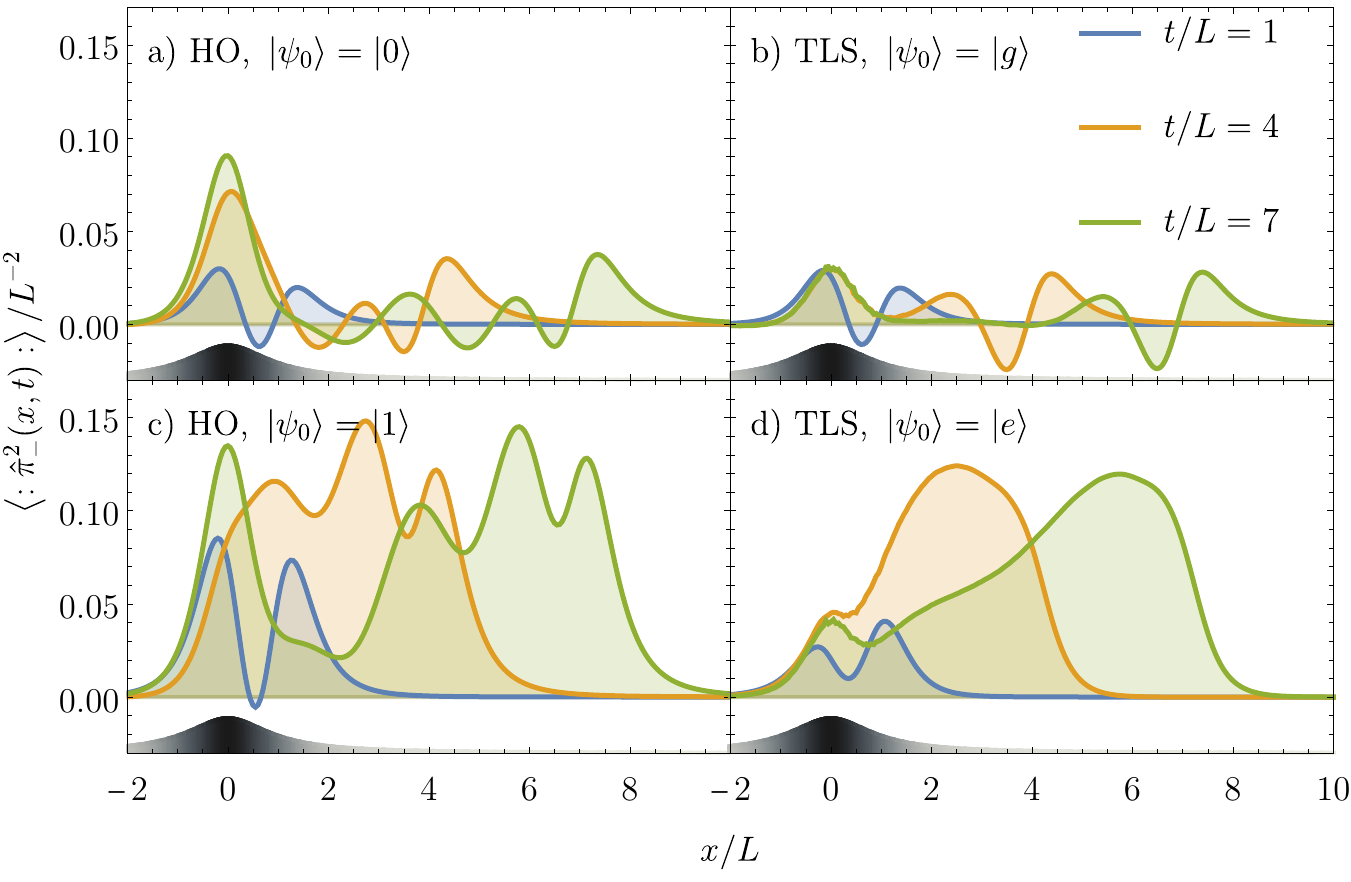}
\caption{Energy densities in vacuum. For MPS calculations $dt=10^{-3}L$ was used. The values of $t$ plotted in subplots b) and d) for the TLS are $t=1+dt,4+dt,7+dt$.
Physical parameters: $\lambda=2, \Omega_{\mathrm{d}}=2\pi/(5L)$. The chain was truncated to $250$ modes.}
\label{fig:edensities_vacuum}
\end{figure}

\textit{Numerical results}.\textemdash
Fig.~\ref{fig:edensities_vacuum} shows the right-moving energy density for the case of a detector coupled to the vacuum with a Lorentzian profile, for a harmonic detector (panels a and c) and a two-level detector (panels b and d) that initially is either in its ground state (panels a and b) or first-excited state (panels c and d).
The spatial profile of the detector, which determines the region with significant coupling between field and detector, is depicted in the lower panel of each subfigure. 
Each panel shows the energy density for an early time ($t/L=1$), an intermediate time ($t/L=4$) and a late  time ($t/L=7$) after the interaction begins at $t/L=0$.

In this illustrative example, we can make several observations.
First, as can be seen by comparing the upper with the lower rows, the initially excited emitter naturally radiates much more energy into the field.
In contrast, when the emitter is initialized in its ground state and early excitations stem from counter-rotating terms in \eqref{eq:interaction-chain}, little energy is being emitted overall.
Moreover, in the case of ground states, we find negative densities propagating to the right, that are more pronounced for the two-level emitter than for the harmonic oscillator.

Once the right-moving density has left the region in which the coupling to the detector is significant, it maintains its shape and simply propagates to the right.
This behavior can be seen for all depicted cases and propagation times.
On the one hand, this reassures that the number of 250 chain modes used in the numerical simulation (which for consistency we use throughout the paper) is sufficient to reliable represent the full time evolution within the selected times.
On the other hand, it also allows us to extrapolate that radiation once it has propagated past $x/L\gtrsim L$ will maintain its shape as it propagates further. 
Hence, the early-time radiation will maintain the profile observed in Fig.~\ref{fig:edensities_vacuum} for $t/L=7$, and only to assess the radiation emanating from the detector at this late time numerical calculations employing a larger number of modes in the chain would be necessary.

\begin{figure}
\centering
\includegraphics[width=\columnwidth]{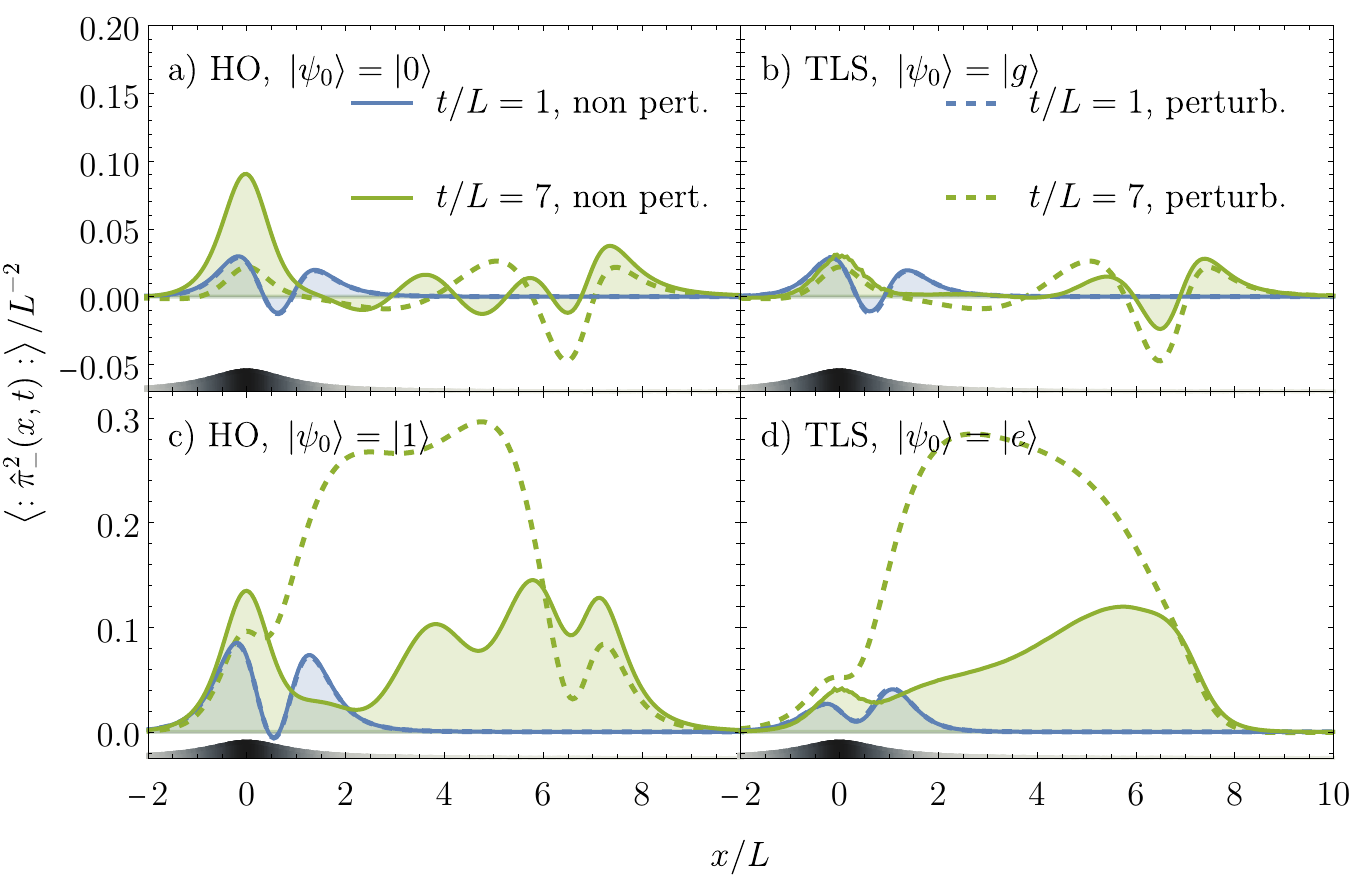}
\caption{Comparison of energy densities resulting from leading order perturbation theory and the non-perturbative numerical results from Fig.~\ref{fig:edensities_vacuum}.
Same parameters as in Fig.~\ref{fig:edensities_vacuum}.
}
\label{fig:edensities_vacuum_compare_perturbative}
\end{figure}

To conclude this section, we illustrate the need for advanced non-perturbative numerical methods in strong coupling regimes in Fig.~\ref{fig:edensities_vacuum_compare_perturbative} which contrasts the results of Fig.~\ref{fig:edensities_vacuum} with the values obtained from leading-order time-dependent perturbation theory (see App.~\ref{app:pert_edensity}).
Whereas at early times leading order perturbation theory still captures the energy density accurately, due to the strong coupling the regime of validity ends soon thereafter.
For late times, such as $t=7L$ in the figure, only the tails of the emitted energy density are captured accurately.
However, for the bulk part of the radiation the leading-order result is misleading.
Here, if these features are captured by perturbation theory at all, then higher-order perturbative calculations would be required.

\section{The truncation error \label{sec:truncation_error}}

The star-to-chain transformation as introduced above (see Sec.~\ref{ssec:chain-mapping} and Sec.~\ref{ssec:thermal-double}) is exact and introduces no approximations or simplifications to the original Hamiltonian.
Up to numerical errors, which in the MPS simulations are introduced by finite Trotter steps, the approach thus yields a faithful and numerically exact representation of the time evolution.
However, in numerical studies the derived infinite chains have to be truncated as also indicated in Fig.~\ref{fig:schematic_transformations}, because only a finite number of modes can be represented on a computer.
This necessarily degrades the accuracy of numerical calculations at sufficiently long simulation times.
This section is devoted to the consequencues of the \emph{truncation error} that is thereby introduced.

\begin{figure*}
\centering
\includegraphics[width=16cm]{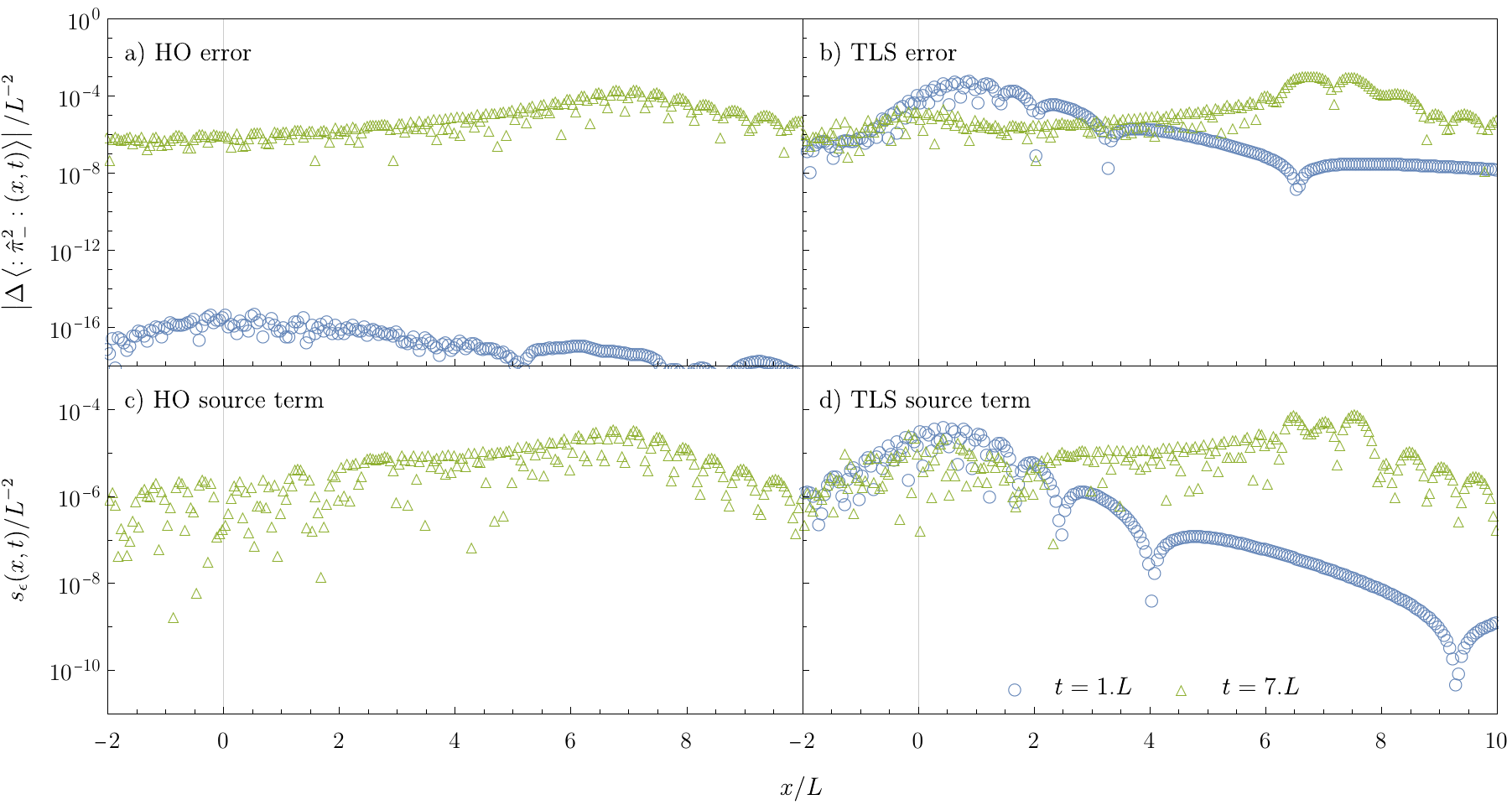}
\caption{Comparison of error in free field energy density and source term~\eqref{eq:discrete_source_term}, with $\epsilon=L/20$. For MPS calculations $dt=10^{-3}L$ was used. }
\label{fig:error_and_source}
\end{figure*}

\subsection{Heuristic of truncation error}\label{sec:truncation_heuristic}
In the study of dynamics, the truncation error stems from the difference between time evolution according to the infinite system and the truncated Hamiltonian.
This difference is rooted in neglecting the hopping term between the last considered mode ($\hat c_{N-1}$ in Fig.~\ref{fig:schematic_transformations}) and the first truncated mode ($\hat c_N$ in Fig.~\ref{fig:schematic_transformations}).
Therefore, the truncation error can be understood intuitively and treated analytically to some extent.

The intuitive picture of the truncation picture is as follows: Initially, the time evolution of the truncated system agrees well with the time evolution of the exact, infinite system.
Since the chain starts out in the vacuum state, this holds up to the time which it takes excitations, created by the interaction with the emitter, to propagate from the front to the chain to the truncated end.
After this time, the excitations in the truncated, numerically implementated model are reflected back to the front of the chain, whereas they had propagated further down the original infinite chain, thus causing the truncation error.

\textit{Free field without detector}.\textemdash
To make this picture more exact it is helpful to consider the excitation dynamics of the chain.
This approach was pursued in~\cite{tamascelli_excitation_2020} to deepen the understanding of chain-mapping methods.
For our purpose it suffices to consider the chain for the free field to which no emitter is coupled, \ie we put $\lambda=0$ above, and to consider the evolution of the  state $\hat c_0^\dagger\ket0$ which has one excitation in the first mode of the chain at $t=0$.
To this end, it is convenient to work in the Heisenberg picture and express $\hat c_0(t)=\sum_{j=0}^\infty \rho_j(t)\hat c_j(0)$.
From~\eqref{eq:definition-chain-operators} it follows that
$\hat c_0(t)=\sqrt2\integral{\omega}0\infty f_\omega p_0(\omega) \hat b^{(e)}_\omega \ee{-\ii \omega t}$, and we obtain
\begin{equation}
    \rho_j(t)=\comm{\hat c_0(t)}{\hat c_j^\dagger}%
=\frac{4 \sqrt{j+1} (\ii t/L)^j}{\left(2+\ii t/L\right)^{2+j}} \,.
\end{equation}
The absolute value squared of these coefficients $\left|\rho_j(t)\right|^2=\bra0 \hat c_0(-t)^\dagger \hat c_j^\dagger\hat c_j\hat c_0(-t)\ket0$ yields the number expectation value of the $j$th chain mode at time $t$. This distribution spreads and flattens out quickly over the chain, as can be characterised by the center of mass of the distribution 
$
\sum_j j |\rho_j(t)|^2 = t^2/(2 L^2)
$
growing quadratically in time. Also, the peak of the distribution $|\rho_{J}(t)|^2:=\sup_j |\rho_j(t)|^2$ has a position which asymptotically behaves as $J\sim t^2/(4L^2)$ for large times and takes the value $|\rho_{J}(t)|^2\sim 4L^2/(\ee{} t^2)$ as $t\to\infty$.
These observations indicate that in order to avoid the truncation error in numerical calculations, the number of required chain modes may scale quadratically in the duration of the time evolution.

In the following Sec.~\ref{sec:truncation_error_bound} we discuss how the error in the state arising due to the truncation may be bounded from above.
This rather straightforward bound, however, (\textit{i}) is only useful for bounded observables, and (\textit{ii}) does not take into account that chain modes near the front of the chain are affected by the truncation much later than modes near the chain end.
Both these points render the state error bound not useful for the energy density of the field which we are interested in here.
Therefore, in Sec.~\ref{sec:source_term_for_error}, we discuss how the truncation error arising in the energy density of the field can be assessed heuristically by a wave-equation source term.

\subsection{Bounding the state truncation error}\label{sec:truncation_error_bound}

The truncation of the chain after $N$ chain modes corresponds to subtracting 
\begin{equation}\label{eq:DeltaH}
\Delta\hat H= \gamma_{N-1} \left(\hat c_{N-1}^\dagger \hat c_{N}+h.c.\right)
\end{equation}
from the full Hamiltonian $\hat H$.
The system thus evolves from its initial state $\ket{\psi_0}$ at $t=0$ into a defective state 
\begin{equation}
    \ket{\psi^\epsilon}=\exp\left(-\ii t (\hat H-\Delta \hat H)\right) \ket{\psi_0}    
\end{equation}
instead of the correct state $\ket{\psi}=\exp\left(-\ii t \hat H\right)\ket{\psi_0}$. The error $\ket\epsilon=\ket\psi-\ket{\psi^\epsilon}$ evolves as
\begin{equation}
    \difffrac{}t\ket\epsilon=\difffrac{}t\left(\ket\psi-\ket{\psi^\epsilon}\right)
    =-\ii H\ket{\epsilon} -\ii \Delta H \ket{\psi^\epsilon}.
\end{equation}
As detailed in App.~\ref{app:state_error_bound}, the norm of the state error evolves as
\begin{equation}
    \difffrac{}t\left\|\ket\epsilon\right\|\leq \sqrt{\braket{\Delta \hat H\psi^\epsilon}{\Delta \hat  H\psi^\epsilon}},
\end{equation}
and its norm at time $t$ is lower or equal to the integral
\begin{equation}\label{eq:errorbound}
    \left\|\ket\epsilon\right\|\leq 
    \epsilon_t:= %
        \left|\gamma_{N-1}\right| \integral{t'}0t \sqrt{\bra{\psi^\epsilon}\hat c_{N-1}^\dagger\hat c_{N-1}\ket{\psi^\epsilon}}.
\end{equation}

\textit{Advantages of error bound}.\textemdash
The expression \eqref{eq:errorbound} achieves something practically useful, since numerically we have access to the expectation value $\langle \hat c_{N-1}^\dagger \hat c_{N-1}\rangle$ with respect to the state we propagate, $\ket{\psi^\epsilon}$, at each available time step.
From this the integrated error bound can be obtained straightforwardly.

Moreover, if the emitter is  itself a harmonic oscillator, then a bound on the error in the (Frobenius) norm $\|\mat G\|$ of the covariance matrix
$\mat G_{ij}%
=\exptval{\hat\xi^i\hat\xi^j+\hat\xi^j\hat\xi^i}$
of the total system state can be derived.\footnote{Here $\hat\xi^\tra=(\hat q_1,\hat p_1,...)$  represents a basis of quadrature operators.}
As detailed in App.~\ref{app:quadratic_error_bound}, this uses that the system remains Gaussian both under the true and the truncated time evolution.
The bound on the error in $\|\mat G\|$ translates into a bound on the error in the expectation value of quadratic observables $\hat O=\frac12\sum_{i,j} \mat O_{ij}\hat\xi^i\hat\xi^j$, provided that the norm of $\mat O$ is bounded.
For example, this allows to bound the error in the expectation values of number operators of chain mode ladder operators, or of other collective mode operators $\hat B=\integral\omega{}{} g(\omega)\hat b_\omega$ (with $\integral\omega{}{}\left|g(\omega)\right|^2=1$) which can be one way to characterize emitted radiation.

\textit{Drawbacks of error bound}.\textemdash
Since the above error bound concerns the norm of the state, it only allows us to bound the error in the expectation values of observables with finite operator norm.
This excludes many operators of interest such as the number and quadrature operators of individual field modes, as well as the energy density of the field, all of which are quadratic in the mode ladder operators.

Moreover, whereas this bound may be interesting and practically useful for identifying the regimes of validity of simulations, it appears to be too rigorous for many applications.
This is because it does not take into account the decomposition of an observable in terms of the chain mode operators.
However, operators acting on modes at the front of the chain are affected by the truncation error much later than the modes at the truncated end of the chain.
An important and interesting subject for future research would therefore be to derive error bounds which take into account the decomposition and support of observables with respect to the chain mode operators.
A natural first step in this direction may well be to investigate a generalization of results from the literature regarding observables acting only on the emitter~\cite{woods_simulating_2015,woods_dynamical_2016,mascherpa_open_2017,trivedi_convergence_2021}.

\subsection{Truncation error in the energy density}\label{sec:source_term_for_error}

\begin{figure*}
\centering
\includegraphics[width=16cm]{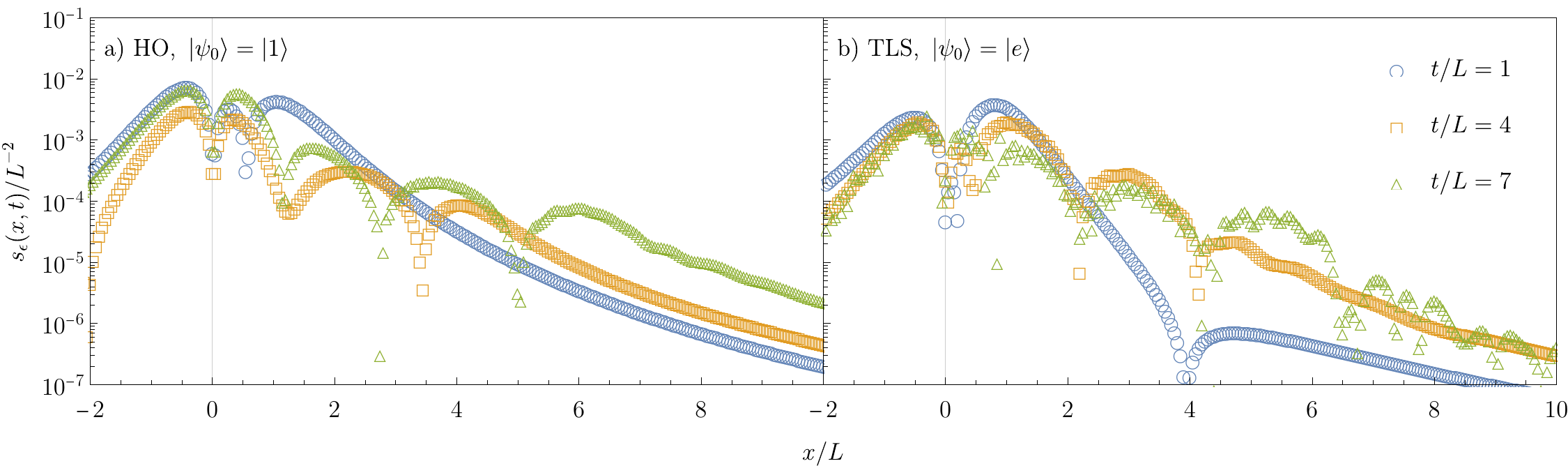}
\caption{Source term~\eqref{eq:discrete_source_term} for scenario of Fig.~\ref{fig:edensities_vacuum} c) and d) with $\epsilon=L/20$. 
}
\label{fig:source_vacuum}
\end{figure*}

Since the coefficients of the field energy density with respect to the chain modes are not bounded, cf.~Eq.~\eqref{eq:right_edensity_mkvacuum_chainladder}, the error bounds from above do not apply to the energy density.
To understand how it is impacted by the truncation error, we first consider the free field with initial state $\hat c_0^\dagger\ket0$.
For this case, we know the exact solution from Sec.~\ref{sec:truncation_heuristic}, which allows to precisely quantify the errors arising in the numerical simulations of the truncated chain.
We find that the errors constitute themselves in the shape of oscillatory features which have a short wave length, tend to arise away from the location of the detector and can be recognized as contributions to the source term of the wave equation.

If the free field ($\lambda=0$ in \eqref{eq:interaction-general}, \ie no emitter-field coupling) is prepared in the initial state $\hat c_0^\dagger\ket0$ at time $t=0$, then the exact expectation value of the right-moving field energy density is
\begin{equation}
\begin{split}
        \exptval{\normord{\pir^2(x,t)}}_{\mathrm{ex}}&=\bra{0}\hat c_0(-t) \normord{\pir^2(x)}\hat c_0^\dagger(-t)\ket{0}
    \\&
    =\frac{L^2}{\pi (L^2+(x-t)^2)^2}.
\end{split}
\end{equation}
The error 
$\Delta \exptval{\normord{\pir^2(x,t)}}= \exptval{\normord{\pir^2(x,t)}}_{\mathrm{ex}}- \exptval{\normord{\pir^2(x,t)}}_{\mathrm{num}}$ 
which arises in the numerical calculations for the truncated chain is shown in Fig.~\ref{fig:error_and_source}(a), for the Gaussian methods applied for HO emitters, and in Fig.~\ref{fig:error_and_source}(b), for the MPS methods applied for TLS emitters.
The figures compare the error $\Delta \exptval{\normord{\pir^2(x,t)}}$, in their upper panel, to the difference between the energy density at a given point and the density's value traced back a small distance ($\epsilon=L/20$) along a light ray:
\begin{equation}\label{eq:discrete_source_term}
    s_{\epsilon}(x,t) %
    =\exptval{\normord{\pir^2(x,t)}}_{\mathrm{num}}-\exptval{\normord{\pir^2(x-\epsilon,t-\epsilon)}}_{\mathrm{num}}.
\end{equation}
In the (exact solution of) the free field, this term always vanishes since the right-moving energy density is simply translated along light rays in time.
However, for the truncated chain, in Fig.~\ref{fig:error_and_source}(c) and Fig.~\ref{fig:error_and_source}(d) we see that a non-zero value of this difference builds up as the simulation time increases.
In particular, the behavior of the source term is highly parallel to the behavior of the absolute error. 
Both signal the effects of the truncation error by the appearance of highly oscillatory features away from $x=0$ where the emitter is centered.

This observation motivates our use of the source term as a heuristic measure for the error arising in numerical simulations in scenarios where the emitter is coupled to the field and no analytical solution is available.
When the emitter is coupled to the field, the term~\eqref{eq:discrete_source_term} serves as an approximation to the expectation value of the source term of the wave equation,
\begin{equation}\label{eq:source_term}
    \begin{split}
        \left(\partial_t+\partial_x\right)\exptval{\normord{\pir^2(x)}} &= \ii\exptval{\comm{\Hi}{\pir^2(x)}}\\
    &= -\lambda \difffrac{f}x \exptval{\hat X\otimes \pir(x)}.
    \end{split}
\end{equation}
In the exact solution of the model, the source term is restricted to the support of the (derivative of the) smearing function $f(x)$. Thus, a non-zero source term away from the support of the emitter signals the appearance of numerical errors.

Fig.~\ref{fig:source_vacuum} shows the numerical source term~\eqref{eq:discrete_source_term} for the data in Fig.~\ref{fig:edensities_vacuum} which showed the energy density emitted by an HO and a TLS emitter at rest into the vacuum of the field. Based on the rise of oscillating features in the source term well away from the emitter's support around a total simulation duration of up to $t=7L$, we decide to only consider results up to this simulation time, here and in the following.
Also below, for detectors coupled to thermal field states, we checked the source term and energy densities for highly oscillatory features to ensure that the truncation error has no significant impact within this simulation time.

\section{Detector radiation in the Unruh effect\label{sec:unruh}}

The previous sections discussed basic properties of chain transformations applied to relativistic fields, and applied them to non-perturbatively calculate the energy density emitted from a particle detector at rest. In this section we use chain transformations to address the Unruh effect as a paradigmatic phenomenon of relativistic quantum fields, and calculate the radiation emitted from a uniformly accelerated detector.

While the Unruh effect itself happens in flat spacetime, it captures a central lesson of quantum field theory in curved spacetimes which is that particles are an observer-dependent concept.
At its core the Unruh effect is the observation that what an inertial observer (which we refer to as Minkowski observer) describes as the vacuum state of the field, a uniformly accelerated observer (Rindler observer) describes as a thermal state of the field.
Famously, the associated Unruh temperature $T_U=a/(2\pi)$ is proportional to the proper acceleration $a$ of the observer (see, \eg~\cite{crispino_unruh_2008}).
In fact, the Unruh effect exhibits intriguing parallels to the thermal double construction of Sec.~\ref{ssec:thermal-double}. 
For a self-contained and detailed review of the Unruh effect and this perspective we refer to App.~\ref{app:unruh_review}. 
In the following,  we summarize it in a high-level overview to introduce and motivate our modeling of the radiation emitted from a uniformly accelerated detector.

\subsection{Modeling the coupling of an accelerated detector}
The Unruh effect takes place in ordinary, flat Minkowski spacetime. 
We restrict ourselves to the (1+1)-dimensional case and use $(t,x)$ as the standard coordinates  for the Minkowski observer.
The quantum field is in the vacuum state $\ket{0_{\mathrm{M}}}$ with respect to the Minkowski observer. That means that the mode operators $\hat a_k$, that the Minkowski observer uses to expand the field in, annihilate the vacuum state: $\hat a_k\ket{0_{\mathrm M}}=0$. 
As will be clear shortly, the Minkowski modes have no equivalent in the framework as discussed so far and depicted in Fig.~\ref{fig:schematic_transformations}, which is why we intentionally denote them as  $\hat a_k$.

\begin{figure}[t]
\centering
\includegraphics[width=.9\columnwidth]{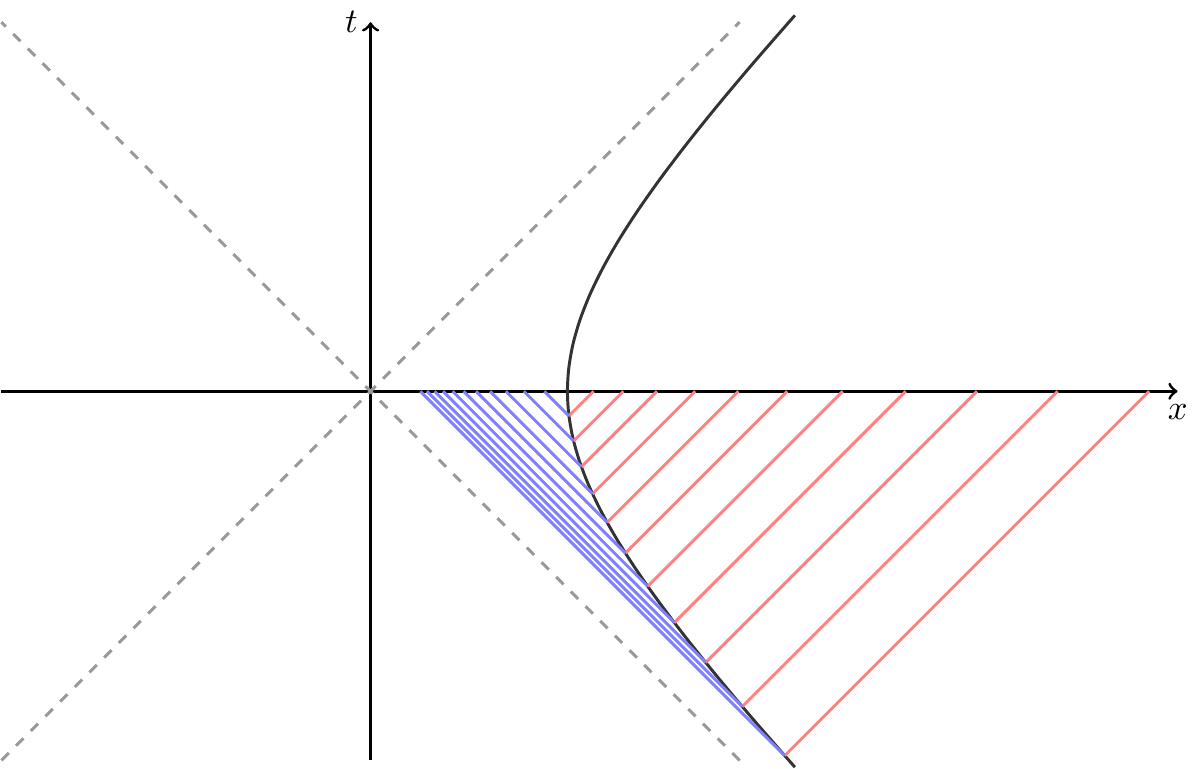}
\caption{Spacetime diagram of the worldline~\eqref{eq:unif_a_observer_wordline} of a uniformly accelerated detector. The detector is coupled to the field for a proper time interval $-T\leq\tau\leq0$, and we evaluate the energy density an inertial observer measures on the hyperplane $t=0$. The lines indicate the emitted radiation. The left-moving energy density is boosted to higher values due to the Doppler shift, whereas the right-moving density is lowered. }
\label{fig:spacetimediagram}
\end{figure}

A wordline of a uniformly accelerated observer (see Fig.~\ref{fig:spacetimediagram}), \ie an observer undergoing constant proper acceleration $a$, is 
\begin{equation}\label{eq:unif_a_observer_wordline}
t=\frac1a\sinh(a\tau),\quad x=\frac1a\cosh(a\tau),
\end{equation}
where $\tau$ is the proper time of the accelerated observer. 
The so-called Rindler coordinates 
$(\tau,\xi)$ in \eqref{eq:Mk_to_rindler_coordinates} (for details see App.~\ref{app:unruh_review}) are the natural choice of coordinates for a uniformly accelerated observer, rather than the Minkowski coordinates.
Similarily, such an observer will use so-called Rindler modes $\hat b^R_\Omega$ to expand the field, rather than  the Minkowski $\hat a_k$-modes.
Again we choose this notation intentionally because the Rindler modes  play exactly the role of the modes labelled as $\hat b_k$ earlier, in the thermal double construction and in Fig.~\ref{fig:schematic_transformations}(b):
Because the Rindler annihilation operators are linear combinations both of  Minkowski annihilation and of Minkowski creation operators (see~\eqref{eq:bogos_rindler_minkowski}), they  do not share the vacuum state with the Minkowski modes.
Instead the Minkowski vacuum is a thermal state with respect to the Rindler modes, whose temperature is the Unruh temperature  $T_U=a/(2\pi)$, as seen from the expectation value (see~\eqref{eq:Rindler_num_exptval})
\begin{equation}
\bra{0_{\mathrm{M}}} \hat b_\Omega^{R\dagger}\hat b_\Omega^R\ket{0_{\mathrm{M}}}  =  \frac{\delta(\Omega-\Omega') }{\ee{\frac{2\pi\Omega}a} -1}\,,
\end{equation}
where $\Omega$ is the Rindler mode frequency.

The thermal $\hat b_k$-modes in the thermal double construction are purified by their partner $\hat b'_k$-modes. 
Where are then  the partner modes of the Rindler modes $\hat b^R_\Omega$ found?
The uniformly accelerated observer above is restricted to the right Rindler wedge, \ie the spacetime region of $|t|<x$, and the $\hat b^R_\Omega$-modes completely capture the field in this region.
Their purifying partner modes $\hat b^L_{-\Omega}$ pertain analogously to the left Rindler wedge, \ie the region $|t|<-x$, to which the mirror image (along the origin $x=0$) of our uniformly accelerated observer~\eqref{eq:unif_a_observer_wordline} is restricted.
As indicated by the notation, these modes have negative Rindler frequency and play exactly the role of the $\hat b'_k$-modes in our discussion of the thermal double construction above.

Exactly as the $\hat d$-modes are constructed in the thermal double construction, also the Rindler partner mode pairs can be transformed into pairs of so-called Unruh modes $\hat d_\Omega$ (see~\eqref{eq:wop}).
For these modes the  field state is the vacuum state, \ie $\hat d_\Omega\ket{0_\mathrm{M}}=0$, and the chain modes  for the numerical simulation of the system are constructed as linear combinations of Unruh modes.

Building on the relations summarized above, our approach to modeling the interaction of a uniformly accelerated detector with the quantum field in its Minkowski vacuum state is to numerically simulate it as the interaction of a detector at rest with field modes in a thermal state. That is, we use the total model Hamiltonian $\hat H = \hf + \hd + \hint$ with its three parts exactly in the same form as introduced in Sec.~\ref{ssec:chain-mapping} and Sec.~\ref{ssec:thermal-double}, respectively.
However, the role of the Minkowski coordinates $(t,x)$ is now played  by the Rindler coordinates $(\tau,\xi)$, and the role of the eigenmodes of the field operator is played by the Rindler modes, which are in a thermal state.

As discussed in detail  at the end of App.~\ref{app:unruh_review},  the interaction Hamiltonian $\hint$ takes the form (see~\eqref{eq:hint_in_rindler})
\begin{equation}
    \hat H_i=\lambda \hat X\otimes\integral\xi{}{} f(\xi) \partial_\tau \hat\phi(\xi)\,,
\end{equation}
where the detector smearing is performed with respect to Rindler coordinates.
The worldline of constant Rindler coordinate $\xi=0$ exactly is the detector worldline~\eqref{eq:unif_a_observer_wordline}. 
Note that worldlines of constant Rindler coordinate $\xi_0$ correspond to a constant proper acceleration of $a\ee{-a \xi_0}$.
Hence, for our ansatz to model a detector experiencing a single constant proper acceleration, the width of the detector profile needs to be small, \ie we require $aL \ll1$.

A consequence of our approach is also, that our calculations now yield time evolution with respect to Rindler time $\tau$ as opposed to Minkowski coordinate time $t$.
Concerning detector observables,  the action of the time evolution operator $\exp(-\ii T \hat H)$ is to simply evolve the detector state forward with respect to detector proper time by an amount $T$, since along $\xi=0$  the Rindler time coordinate $\tau$ equals the detector's proper time.
Concerning field observables, because the Rindler field Hamiltonian $\hf'=\integral{\Omega}{-\infty}\infty \Omega \left(\hat b_\Omega^{R\dagger}\hat b_\Omega^R-\hat b_\Omega^{L\dagger}\hat b_\Omega^L\right)$ generates Lorentz boosts in Minkowski spacetime, the action of $\exp(-\ii T \hat H)$ is to, for example, transform a state defined on the hyperplane $t=\tau=0$ to the hyperplane $\tau=T$, which in Minkowski coordinates is the hyperplane $t=\tanh(aT) x$.

In App.~\ref{app:mk_observables_vs_rindler_evol} we discuss in detail how observables like the energy density of the field with respect to an inertial Minkowski observer are affected.
The easy way in which we handle this issue here is, figuratively speaking, to move the start of the interaction back in time: We move the onset of the interaction back to proper time $\tau=-T$ of the detector, at which point we assume the detector and field to be in a product initial state $\ket{\psi_0}\otimes\ket{0_\mathrm{M}}$, and then numerically calculate the action of $\exp(-\ii T\hat H)$ on this state, which results in a state defined on the hyperplane $t=0$.

\begin{figure}
\centering
\includegraphics[width=\columnwidth]{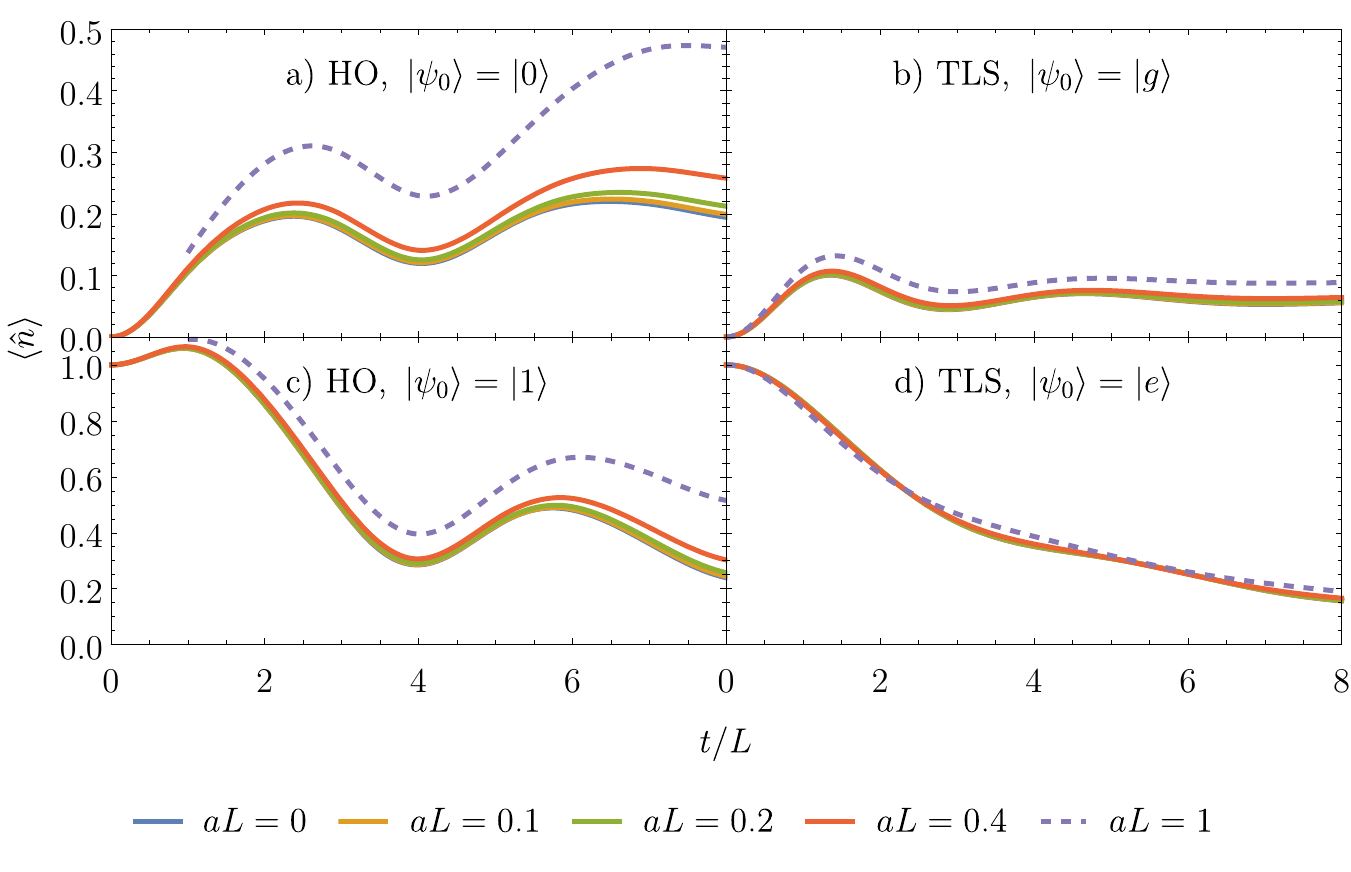}
\caption{The emitter occupation expectation value $\exptval{\hat n}$, as defined in~\eqref{eq:defn_det_occupation}, shows the deviation of the detector response in thermal field states from the field vacuum.
The figure shows data for HO and TLS detectors in response to thermal states with  inverse temperatures $\beta=\infty,\, 20\pi L,\,10\pi L,\,5\pi L,\,2\pi L$, which in the Unruh effect corresponds to acceleration values $aL=0,\,0.1,\,0.2,\,0.4,\,1$.
(For MPS calculations $dt=10^{-3}L$ was used.)
}
\label{fig:emitter_occupation}
\end{figure}

\subsection{Results}

In this section, we discuss the numerical results we obtained for three different acceleration values, $aL=0.1,\,0.2,\,0.4$.
The largest of these values is interesting to understand the numerical performance of our method, even if it may well be viewed as being in conflict with our modeling requirement that $aL\ll1$, as discussed above.

Nevertheless, by considering the dynamics of the occupation number expectation value of the detector $\exptval{\hat n}$, which is 
\begin{equation}\label{eq:defn_det_occupation}
    \hat n^{\mathrm{HO}} =\hat a^\dagger\hat a,\quad \hat n^{\mathrm{TLS}}=\tfrac12\left(\hat \sigma_z+\id\right)
\end{equation}
 for the HO detector  the TLS detector respectively,
we see that the thermal response of the detector due to the Unruh effect is not too pronounced at these accelerations, for the numerical detector parameters that we consider.
Fig.~\ref{fig:emitter_occupation} shows the expectation value for initial states with zero and with one excitation for both detector types for a detector with the same coupling parameteres ($\Omega_{\mathrm{d}}=2\pi/5,\,\lambda=2$) as we considered in Sec.~\ref{ssec:numerical-example} for a detector at rest.
For the TLS the detector, the response of the detector occupation for the three acceleration values $aL=0.1,\,0.2,\,0.4$ is hardly distinguishable from a resting detector ($aL=0$). And even in the case of $aL=1$ which we present there for reference, and which corresponds to an inverse Unruh temperature of $\beta=2\pi L$ the difference is relatively small still.
For the HO detector the differences are somewhat more pronounced and already the case of $aL=0.4$, corresponding to an inverse Unruh temperature of $\beta=5 \pi L $ are noticeable.

\begin{figure*}
\centering
\includegraphics[width=.9\textwidth]{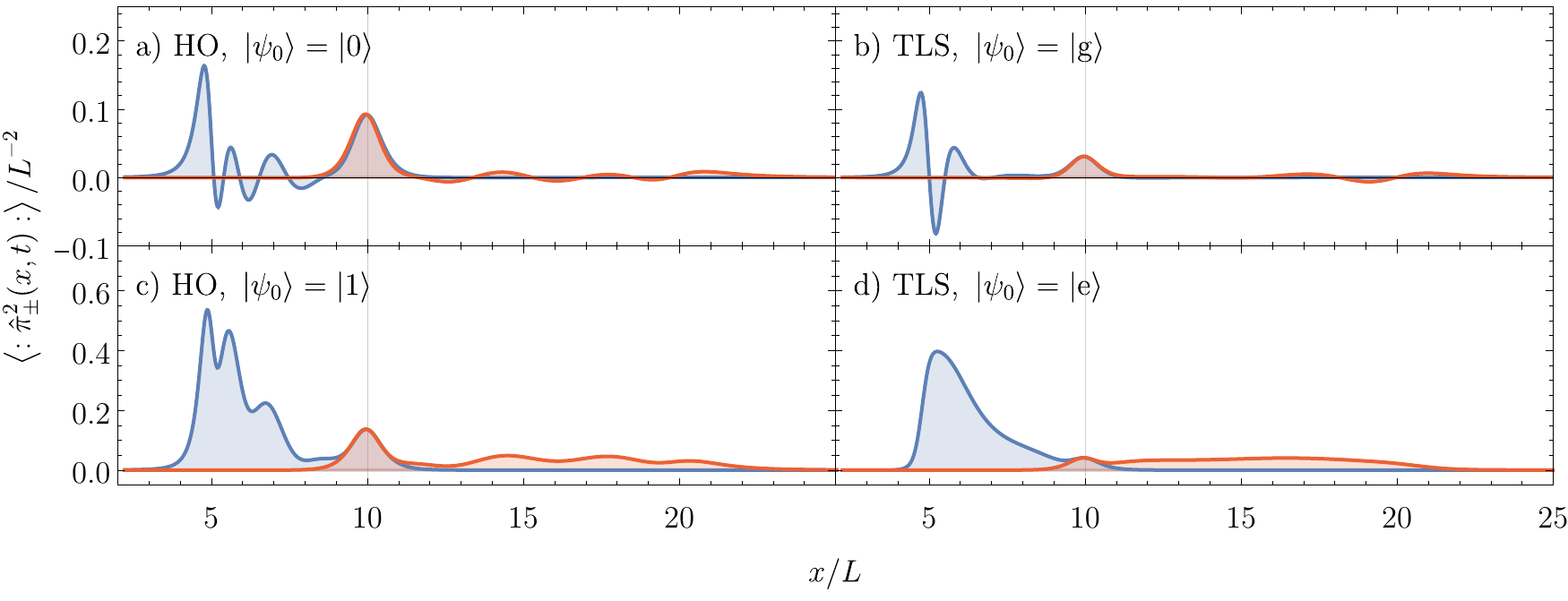}
\caption{Comparison of energy densities emitted from uniformly accelerated emitter with $aL=1/10$, switched on at $\tau=-7L$ and switched off at $\tau=0$. For the TLS system $dt=0.005$ is used in the MPS calculations.}
\label{fig:unruh_compare}
\end{figure*}

Based on this observation, we would expect the radiation from our accelerated detectors to correspond  to the profiles observed in Fig.~\ref{fig:edensities_vacuum} for resting detectors, after undergoing a Lorentz boost (or Doppler shift) which along each light ray in the emitted radiation depends on the detector's velocity at the intersection between the detector's worldline with the light ray, \ie the point in time at which the light ray would have been emitted from the detector.
In fact, the energy densities in Fig.~\ref{fig:unruh_compare}, that shows the results for all four combinations of detector types and initial states for an acceleration of $aL=0.1$, shows the expected similarities.
And Fig.~\ref{fig:synth_vac} and Fig.~\ref{fig:synth_exc} in App.~\ref{app:mk_observables_vs_rindler_evol} confirm that to a very high degree this expectation agrees with our numerical results for the energy density emitted from a uniformly accelerated detector.

\begin{figure*}
\centering
\includegraphics[width=.9\textwidth]{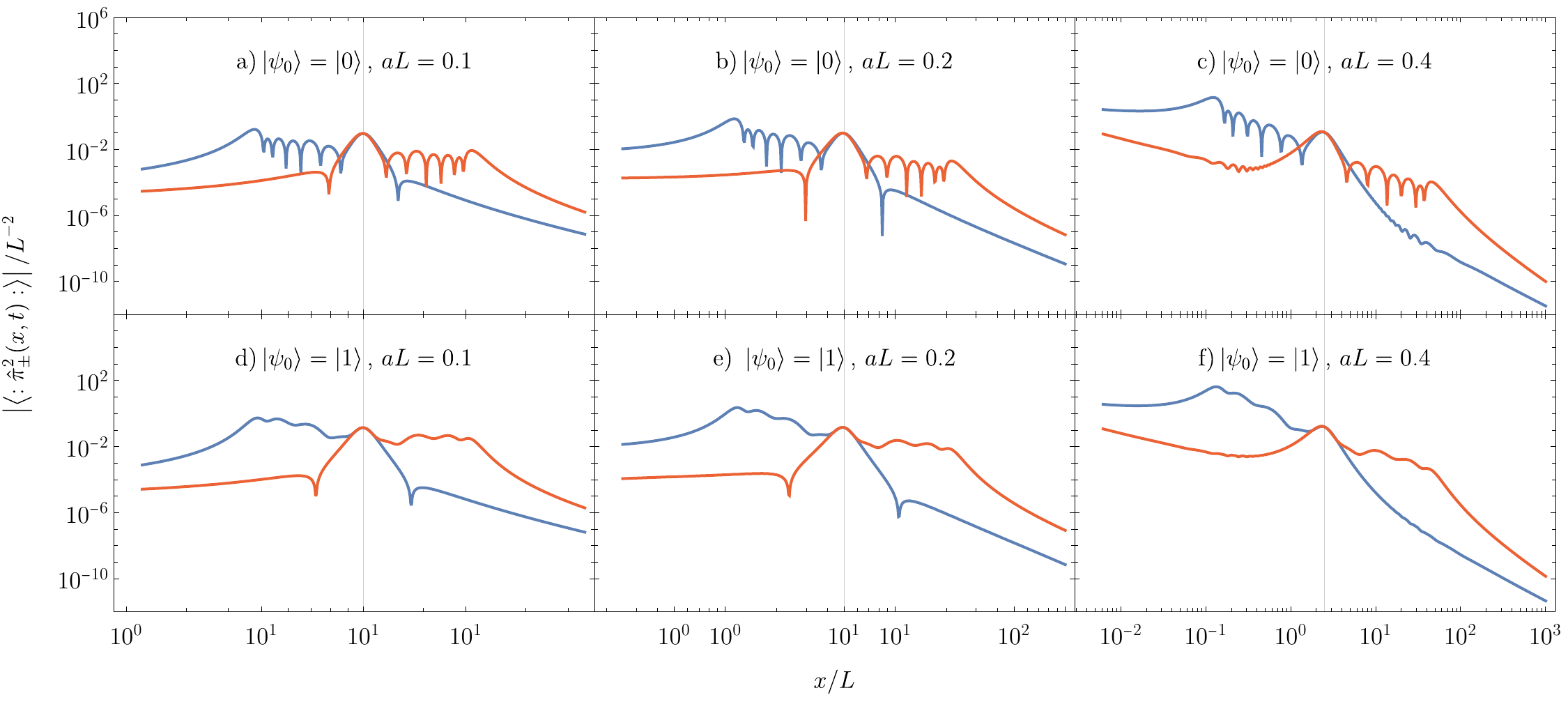}
\caption{Double logarithmic plot of the left-moving (in blue) and right-moving (in red) energy density emitted from a uniformly accelerated HO detector, for different acceleration values (in the different columns) and different initial states (upper row $\ket{\psi_0}\ket0$ and lower row $\ket{\psi_0}\ket1$).
The symmetry between left-moving and right-moving densities follows from~\eqref{eq:symmetry_unruh_densities}.
Note that a) and d) here, are equal to Fig.~\ref{fig:unruh_compare}(a) and~(c).
}
\label{fig:unruh_ho_loglog}
\end{figure*}

Fig.~\ref{fig:unruh_ho_loglog} shows how the emitted energy density profile changes as the acceleration increases.
Furthermore, its double logarithmic plots exhibit some characteristic features more clearly, which we observe for both HO and TLS detectors. 
First, as a consequence of the accelerated detector coupling evenly to the Rindler modes, the observed (Minkowski) energy density exhibits the following symmetry between left-moving and right-moving energy densities:
\begin{equation}\label{eq:symmetry_unruh_densities}
    \exptval{\normord{\hat\pi^2_+\left(x=\tfrac1a\ee{-a\xi},t\right)}}
    =\ee{2a\xi}\exptval{\normord{\hat\pi^2_-\left(x=\tfrac1a\ee{a\xi},t\right)}},
\end{equation}
which can be read of directly from expressions~\eqref{eq:I_terms_unruh_edensity} and~\eqref{eq:app_mk_edensity_from_chain}.

Second, for $aL=0.4$, we see that both left-moving and right-moving energy densities appear to diverge as $x\to0$.
This behaviour is in fact to be expected for all accelerations towards the coordinate origin $x\to0^+$, if one takes into account that the end of the time evolution on the hyperplane $t=0$ is equivalent to a sudden switch-off of the interaction between detector at field:
Since we applied the detector smearing function~\eqref{eq:lorentzian-smearing} with respect to Rindler coordinates, in terms of Minkowski coordinates it reads
\begin{equation}
    f'(x)=f(\xi)=\frac{L}{\pi (L^2+\xi^2)}=\frac{L}{\pi\left(L^2+\tfrac1{4a^2}\ln(a^2 x^2)^2\right)}.
\end{equation}
The derivative of this function $\lim_{x\to0^+}\difffrac{f'}x=\infty$ diverges towards the coordinate origin. However, infinitely steep smearing functions  lead to diverging energy densities for instantaneous interaction switch-offs for the detector model we employ here.

Furthermore, we highlight the oscillatory features appearing in the data for $aL=0.4$ at $x\approx0.2L$ in the right-moving energy density and $x\approx30L$ in the left-moving energy density. 
These features grow more dominant when the simulation is continued further and they appear at earlier simulation times for higher accelerations (respectively later for lower accelerations).
Based on our investigation of the truncation error above, we interpret them as indicating the onset of the truncation error effects at simulation times beyond $t=7L$ for the chosen coupling parameters of our model and chosen chain length for our numerical simulations.

As seen in the detector responses in Fig.~\ref{fig:emitter_occupation} and in the emitted energy density profiles, the coupled detector-field system has not reached a stationary state at this time yet.
To reliably extend the simulation time \rhj{into} %
this regime one would therefore have to use more chain modes in the numerical calculations.
This could also be  of interest, for example, for further investigations dedicated to the radiation arising in scenarios in which the Unruh temperatures are larger relative to the detector energy gap $\Omega_{\mathrm{d}}$, because a longer chain would allow for longer simulation times which, in turn, would allow to cover an equal number of detector periods $2\pi/\Omega_{\mathrm{d}}$ for detectors with lower $\Omega_{\mathrm{d}}$.

In summary, our numerical results for the enery density emitted from a uniformly accelerated detector which couples to the field for a finite amount of time, agree with physical expectations and earlier results in the literature.
On physical grounds, as discussed above, we expect the density in our scenario to correspond to the Lorentz boosted density emitted from a detector at rest. The high level of agreement with which this expectation is confirmed (Figs.~\ref{fig:synth_vac} and~\ref{fig:synth_exc}) in data originating from entirely independent numerical calculations demonstrates the reliability of the employed method.
In the literature exact solutions for HO detectors coupled to the field amplitude (as opposed to the field momentum coupling we employ) found that radiation is emitted from the detector only in transient stages but not in equilibrium states, and that a polarization cloud of radiation forms around a coupled detector.
This agrees well with the radiation we obtain in our results (Figs.~\ref{fig:unruh_compare} and~\ref{fig:unruh_ho_loglog}) where the radiation burst from the detector originates from the interaction switch-on, and which shows the radiation cloud around the detector at its final position $x=10L$.
Due to the limited interaction time, however, the detector does not reach an equilibrium state within our calculations as discussed above.
An adaptation of exact results in 1+1 dimensions~\cite{hinterleitner_inertial_1993, massar_vacuum_1996,kim_radiation_1997,kim_quantum_1999}, or~\cite{lin_accelerated_2006} in higher dimensions, may further provide a valuable check and benchmark for our method.

\section{Conclusions \& outlook \label{sec:outlook}}
We have utilized chain-mapping methods to numerically study the interaction between a scalar quantum field, and localized quantum emitters both at rest and undergoing uniform acceleration.
The numerically exact treatment of the entire system, including the field, allows efficient access to a large variety of system and field observables.
In addition, while our main focus rests on the emission and absorption of excitations from an emitter, which we monitor by calculating and time-evolving the field energy density, the method is not restricted to these observables.
While we focus on a two-level or harmonic emitter, respectively, coupled to its bath via a Lorentzian coupling profile, for which we find convenient expressions within the chain-mapping approach, other emitters may be considered as well. 
Similarly, whereas we here considered a massless field in 1+1 spacetime dimensions, the method can be extended to massive fields and higher dimensions.
Future works may use our approach to study, \eg bath or system-bath correlation functions, or to calculate the entanglement dynamics of multiple emitters coupled to a thermal bath.
In this context, an interesting question is whether the chain mapping can be efficiently implemented for two emitters coupled to the same continuum of bath modes.
This would pave the way for a new non-perturbative approach to many questions regarding communication, correlation or entanglement transfer between localized emitters and the quantum field.
In particular in the context of relativistic scenarios the present approach has the advantage to introduce no further causality violating approximations to the model such as a UV cutoff~\cite{jonsson_quantum_2014,martin-martinez_causality_2015}.
Instead, the model is treated exactly within the maximal achievable simulation time determined by the number of chain modes used in the numerical simulation.

In Sec.~\ref{sec:truncation_error} we discussed an error bound which can be practically evaluated along with the numerical simulations and which rigorously controls the total error introduced to the time-evolved state due to the truncation of the chain.
Since this error bound appears to be too rigorous for many applications of interest, it would be useful to derive error bounds that are tailored towards specific observables by taking into account their decomposition in terms of the chain modes.
This may be achieved building on existing error bounds~\cite{woods_simulating_2015,woods_dynamical_2016,mascherpa_open_2017,trivedi_convergence_2021}.

\section*{Acknowledgments}
R.H.J. gratefully acknowledges support by the Wenner-Gren Foundations and, in part, by the Wallenberg Initiative on Networks and Quantum Information (WINQ).
Nordita is supported in part by NordForsk.
J.K. gratefully acknowledges support from Dr. Max Rössler, the Walter Haefner Foundation and the ETH Zürich Foundation.
We thank Mari-Carmen Bañuls for fruitful discussions.

\appendix

\section{Polynomial coefficients from numerical weight moment matrix via Cholesky decomposition}\label{app:choleskydecomp}

First we calculate a vector $\vec{w}$ with $2N+1$ entries containing the weights as given in~\eqref{eq:thermal_weight_moment_integral},
\begin{equation}
    \vec{w}=\left[w_k\right]_{k=0,\dots,2N},\quad w_k=\integral{k}{-\infty}\infty w(k).
\end{equation}
Then we arrange these into an $(N+1)\times(N+1)$-matrix $\vec{M}$ which represents the scalar product defined by~\eqref{eq:thermal_inner_product} with respect to the polynomials $k^n$,
\begin{equation}
    \vec{M}_{ij}=\left[w_{(i+j)}\right]_{i,j=0,\dots,N}.
\end{equation}
We need the Cholesky decomposition of this matrix $\vec{M}=\vec{L}\vec{L}^\tra$. $\vec{L}$ is a lower triangular matrix and it corresponds to a basis change matrix from the basis given by the polynomials $1,k,k^2,\dots,k^{2N}$ to the polynomials $p_0(k),\dots,p_{2N}(k)$ which are orthonormal with respect to the inner product~\eqref{eq:thermal_inner_product}.
In particular,
\begin{equation}
    p_i(k)=\sum_{n=0}^{N}  \left(\vec{L}\right)^{-1}_{i,n} k^n,\quad \Rightarrow \left(\vec{L}\right)^{-1}_{i,n}=P_{i,n},
\end{equation}
\ie the rows of $\vec{L}^{-1}$ contain the coefficients of the orthonormal polynomials $P_{i,n}$ as defined in~\eqref{eq:defn_thermal_polys}.

In practice, we obtained the best performance, in terms of speed and precision, by directly implementing the standard algorithm for the Cholesky transform and its inverse.
For the matrix $\vec L$ that is
\begin{algorithmic}
    \For{$0\leq i\leq N$} 
        
        \For{$0\leq j\leq i$}
            \If{$i=j$}
                $\vec{L}_{ii}=\sqrt{\vec{M}_{ii}-\sum_{k=0}^{j-1} \left(\vec{L}_{jk}\right)^2}$
            \Else 
                {$\,\vec{L}_{ij}= \left(\vec{M}_{ij}-\sum_{k=0}^{j-1} \vec{L}_{ik}\vec{L}_{jk}\right)/\vec{L}_{jj}$}
            \EndIf
        \EndFor
    \EndFor,
\end{algorithmic}
and its inverse can then be constructed as
\begin{algorithmic}
    \For{$0\leq i\leq N$}
         $\,\vec{L}^{-1}_{ii}=1/\left(\vec{L}_{ii}\right)$
        \For{$0\leq j<i$}
         $ \vec{L}^{-1}_{ij}= (-1)\vec{L}^{-1}_{ii}\cdot \sum_{k=j}^{i-1} \vec{L}_{ik}\vec{L}^{-1}_{kj}$
        \EndFor
    \EndFor.
\end{algorithmic}

\section{MPS simulations and choice of time step \label{app:mps-dt}}

\begin{figure*}
\centering
\includegraphics[width=.99\textwidth]{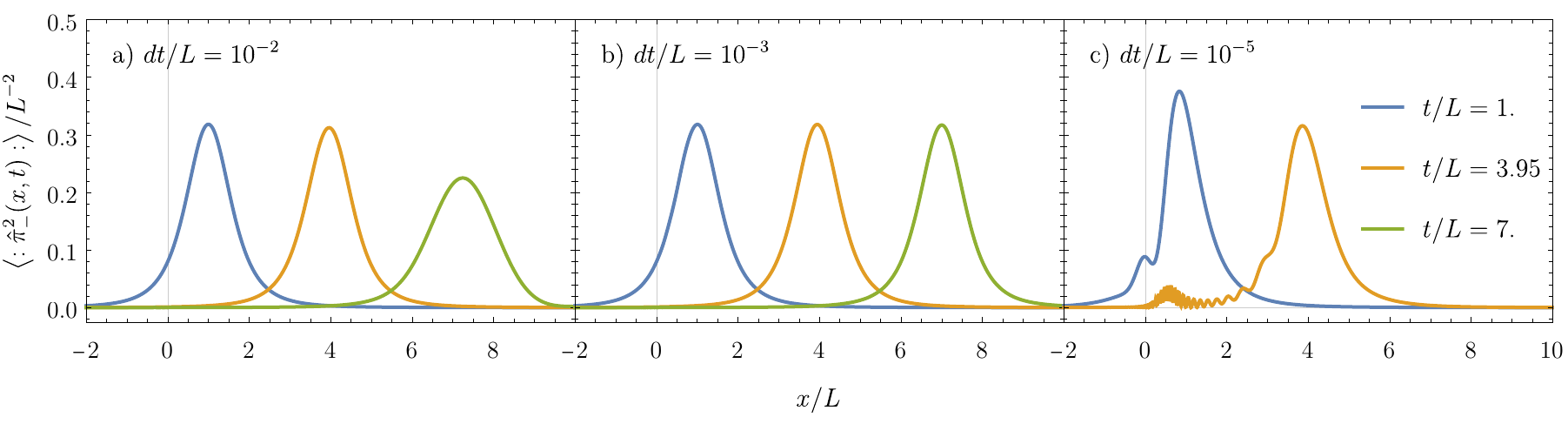}
\caption{Right-moving energy density for a chain without emitter, in which an initial excitation is placed into the first mode and spreads through the system during real-time evolution.
Results shown at times $t/L = 1$, $t/L = 3.95$ and $t/L=7$ for a time step of
(\textit{a}) $dt/L = 10^{-2}$,
(\textit{b}) $dt/L = 10^{-3}$,
(\textit{c}) $dt/L = 10^{-5}$.
Other parameters are the same as in the main text.
}
\label{fig:mps_dt_densities}
\end{figure*}

\begin{figure*}
\centering
\includegraphics[width=.99\textwidth]{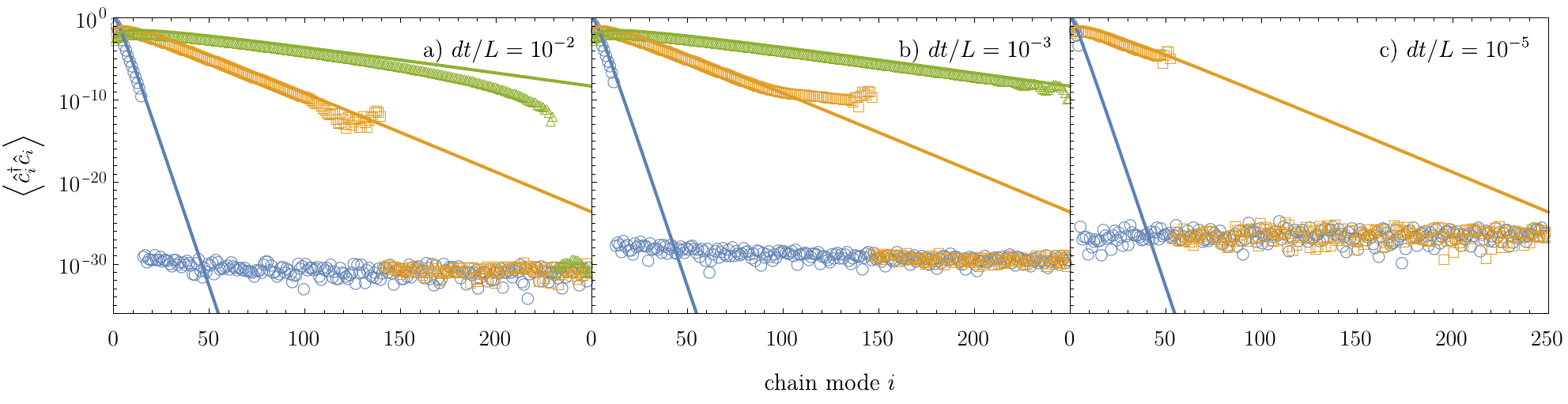}
\caption{Chain occupation $\langle \hat n_i \rangle = \langle \hat c_i^\dagger \hat c_i \rangle$ of all $250$ modes used in the simulation, for a chain without emitter as in Fig.~\ref{fig:mps_dt_densities}.
Results shown at times $t/L = 1$, $t/L = 3.95$ and $t/L=7$ for a time step of
(\textit{a}) $dt/L = 10^{-2}$,
(\textit{b}) $dt/L = 10^{-3}$,
(\textit{c}) $dt/L = 10^{-5}$.
Other parameters are the same as in the main text.
The solid lines represent the exact solution for the infinite chain (without truncation).
}
\label{fig:mps_dt_occupation2}
\end{figure*}

Real-time evolution of matrix product states has been reviewed in Ref.~\cite{paeckel_time-evolution_2019}, where the Trotter or \textit{time-evolving block decimation} (TEBD) method is discussed with its strengths and weaknesses.
One of the critical numerical parameters within TEBD is the time step $dt$, for which the usual trade-off consists in keeping the introduced errors per time step small, while maintaining a reasonable and manageable number of time-evolution steps for the total simulation period of interest.
Here we comment on our choice of suitable time steps $dt$, for time evolving the state $\ket{\psi_{t+dt}} = \hat U(dt) \ket{\psi_t}$, which we use in the simulations with which we obtain the results in the main text.

\textit{Too large time steps.}\textemdash
When decomposing the time-evolution operator using the Trotter method, ideally the time steps should be \textit{sufficiently small}.
When comparing panels (\textit{a}) and (\textit{b}) in Fig.~\ref{fig:mps_dt_densities}, we indeed find that a shorter time step ($dt = 0.001$) reproduces the profile of a simply right-moving energy density more faithfully than a larger time step ($dt = 0.01$), up to a final propagation time $t/L = 7$.
For our simulations, this contains a first lesson:
(\textit{i}) The choice of a suitable time step is always tied to the total propagation time, as time-step errors accumulate during time evolution.
Since we focus on simulation times of up to $t/L = 7$ throughout most of this work, the time step $dt = 0.001$ seems preferable (over $dt = 0.01$ and any larger time steps) based on numerical examples like this.

\textit{Too small time steps.}\textemdash
On the other hand, and based on the same example of Fig.~\ref{fig:mps_dt_densities}, we find that too small time steps lead to inaccurate predictions of the energy density.
When comparing panels (b) and (c) in Fig.~\ref{fig:mps_dt_densities}, we see that the energy density deviates from its expected behavior already for relatively short times, $t/L = 1$.
In order to understand this, in Fig.~\ref{fig:mps_dt_occupation2} we show the occupations $\langle \hat n_i \rangle = \langle \hat c_i^\dagger \hat c_i \rangle$ of all $250$ chain modes for the same three propagation times as in Fig.~\ref{fig:mps_dt_densities}.
When comparing Fig.~\ref{fig:mps_dt_occupation2}(\textit{c}) with the two remaining panels, we find that the excitation, which is at the first chain mode at $t=0$, does not propagate through the chain for the smallest time step ($dt = 10^{-5}$).
We interpret this time step to be too small given the maximum bond dimension of $\chi = 300$, used to obtain the two figures \ref{fig:mps_dt_densities} and \ref{fig:mps_dt_occupation2}.
When the truncation error associated with a given bond dimension is larger than the error induced by the time evolution, the Trotter method fails to meaningfully evolve the MPS.
As a result, in the above example the initial excitation almost does not propagate through the chain.
This provides us with a second useful lesson:
(\textit{ii}) The choice of $dt$ must take into account the truncation error due to restricting the MPS to a realistic bond dimension.
If the time step is too small, the latter dominates and further decreasing the step size is counterproductive.
Here we showed the result for a very small time step, $dt=10^{-5}$.
Based on further numerical experiments that we do not show, we finally choose a time step between $10^{-3} \leq dt \leq 5 \cdot 10^{-3}$ which we use throughout the main text.

\section{Minkowski, Rindler and Unruh modes in the Unruh effect}\label{app:unruh_review}

The purpose of this appendix is to give a brief, but self-contained review of the different basis sets of modes relevant to the Unruh effect, \ie Minkowski, Rindler and Unruh modes, and their relation to the chain modes and thermal double construction of the previous sections.
Table~\ref{tab:modes} gives a compact overview of the modes appearing, and the relations between them. 
Finally, the appendix arrives at the Bogoliubov transformation expressing the Minkowski mode operators in terms of the chain mode operators used in the numerical calculations. 

\textit{Bogolubov transformations}.\textemdash
It is central to the Unruh effect, as it is to many phenomena in quantum field theory in curved spacetime, that different observers may choose different sets of modes to expand the field observables, and to interpret the quantum state of the field \cite{birrell_quantum_1982,wald_quantum_1994_manual}.
In 1+1-dimensional Minkowski spacetime, the general expansion of the amplitude of the massless scalar Klein-Gordon field that we consider here, in the Heisenberg picture, takes the form
\begin{equation}
\hat\phi (t,x)=\integral{k}{}{} u_k(t,x)\hat a_k+u_k^*(t,x)\hat a_k^\dagger,
\end{equation}
where the $u_k$ and their complex conjugates form  a complete basis of complex solutions to the Klein-Gordon field equation, and $\hat a_k$ are the associated mode operators. That is, the mode operators fullfill the canoncial commutation relations $\comm{\hat a_k}{\hat a_{k'}^\dagger}=\delta(k-k')$, and the set of solutions are orthonormal with respect to the Klein-Gordon inner product
\begin{equation}
\left(u_k,u_{k'}\right) = -\ii\integral{x}{}{} \left( u_k\partial_tu_{k'}^* -\left(\partial_t u_k\right)u_{k'}^*\right)=\delta(k-k')\,,
\end{equation}
where the integral  is evaluated on a hyperplane of constant Minkowski coordinate time $t$.
(The inner product can be evaluated on any other Cauchy surface of the spacetime, and the result is independent of this choice \cite{birrell_quantum_1982,wald_quantum_1994_manual}.)
Given a second complete basis of solutions, say $v_l(t,x)$, with associated mode operators $\hat a'_l$, expressions can be transformed from one basis to the other by the Bogoliubov transformations \cite{birrell_quantum_1982}
\begin{equation}
v_l=\integral{k}{}{} \alpha_{lk}u_k+\beta_{lk}u_k^* 
,\quad
\hat a'_l= \integral{k}{}{} \alpha_{lk}^*\hat a_k-\beta^*_{lk}\hat a_k^\dagger
\end{equation}
where the Bogoliubov coefficients are given by
\begin{equation}
\alpha_{lk} = \left( v_l,u_k\right),\quad \beta_{lk}=-\left(v_l,u^*_k\right).
\end{equation}
The inverse transformations read
\begin{equation}
u_k=\integral{l}{}{} \alpha_{lk}^* v_l-\beta_{lk} v^*_l,
\quad \hat a_k=\integral{l}{}{} \alpha_{lk} \hat a'_l+ \beta^*_{lk}\hat a'_l{}^\dagger .
\end{equation}

\textit{Minkowski modes}.\textemdash
With respect to the standard coordinates $(t,x)$, the Minkowski metric reads $\diff s^2=\diff t^2-\diff x^2$ and the massless Klein-Gordon wave equation reads
\begin{equation}
\left(\partial_t^2-\partial_x^2\right)\phi(t,x)=0.
\end{equation}
The plane wave solutions
$
u_k(t,x)= \ee{-\ii|k|t+\ii k x} /\sqrt{4\pi|k|} %
$
yield the orthonormal complete set of solutions which is the canonical choice of basis for inertial observers.
Since $\ii\partial_t u_k = |k|u_k$ they are eigenmodes of positive frequency with respect to the generator of translations along coordinate time $t$, \ie they are eigenmodes of the Hamiltonian which generates time evolution with respect to the proper time of observers at rest relative to the $(t,x)$ coordinates.
These modes separate into left-moving modes $u_\omega^+$ and right-moving modes $u_\omega^-$, with positive frequency $\omega>0$,
\begin{equation}\label{eq:upm}
u_\omega^\pm(t,x)=u_{\mp \omega}(t,x)= \frac1{\sqrt{4\pi\omega}} \ee{-\ii \omega (t\pm x)}.
\end{equation}
To these mode functions we associate the mode operators $\hat a^\pm_\omega$, resulting in the mode operator expansion for the amplitude operator of the quantum field,
\begin{equation}
\hat \phi(t,x)= \integral{\omega}0\infty\sum_{\pm=+,-} u_\omega^\pm(t,x)\,\hat a^\pm_\omega+u_\omega^\pm (t,x)^*\, \hat a^\pm_\omega{}^\dagger\,.
\end{equation}

Here, in the context of the Unruh effect, we denote the Minkowski mode operators by the letter $\hat a$ rather than the letter $\hat b$, because 
the Minkowski modes  neither generically appear in the interaction part of the Hamiltonian, nor in the part that generates the relevant time evolution. Instead, this role is played by the Rindler modes, which are the generic choice of field modes for uniformly accelerated observer, and arise as plane wave solutions with respect to the Rindler coordinates. Hence, consistent with our notation throughout the article, we denote the Rindler modes by  $\hat b$.

\begin{table*}%
    \centering
    \renewcommand{\arraystretch}{1.5}
    \begin{tabular}{ |p{.14\columnwidth} p{.13\columnwidth}| >{\centering}p{.28\columnwidth} | p{.28\columnwidth}<{\centering} |  p{.1\columnwidth}<{\centering} |} 
        \toprule
       \multicolumn{2}{|l|}{ } & \multicolumn{2}{c|}{\bf\underline{Wave function / Operators}} & {\bf\underline{Eqn.}} \\
       &&left-moving $+$ &right-moving $-$ & \\
        \hline %
         \multicolumn{2}{|l|}{ \bf Minkowski}   & $u_\omega^+\,/ \hat a_\omega^+$  & $u_\omega^-\,/\hat a_\omega^-$ & \eqref{eq:upm}\\
        \hline %
        \multirow{4}{*}{\bf Rindler} & \multirow{2}{*}{$|t|\!<\!x$} & $v_\Omega^{R+}\,/ \hat b_\Omega^{R+}$  & $v_\Omega^-\,/ \hat b_\Omega^{R-}$ & \eqref{eq:vor} \\
        \hhline{|~~|---|}
        & & \multicolumn{2}{c|}{ \cellcolor{gray!10} $\hat b^{R\text{\oe{}}}_\Omega= \left(\hat b_\Omega^{R+}\mp\hat b_\Omega^{R-}\right)/\sqrt2$}& \\
        \hhline{|~-|---|}
          &  \multirow{2}{*}{$|t|\!<\!-x$} & $v_\Omega^{L+}\,/\hat b_\Omega^{L+}$  & $v_\Omega^{L-}\,/\hat b_\Omega^{L-}$ &\eqref{eq:vol}\\
          \hhline{|~~|---|}
          & &  \multicolumn{2}{c|}{  \cellcolor{gray!10} $\hat b^{L\text{\oe{}}}_\Omega= \left(\hat b_\Omega^{L+}\mp\hat b_\Omega^{L-}\right)/\sqrt2$} &\\
        \hline %
        \multirow{4}{*}{\bf Unruh} & \multirow{2}{*}{$\Omega>0$} & $w_\Omega^+\,/\hat d_\Omega^{+}$  & $w_\Omega^-\,/\hat d_\Omega^{-}$ & \eqref{eq:wop} \\
        \hhline{|~~|---|}
        & & \multicolumn{2}{c|}{ \cellcolor{gray!10} $\hat d_\Omega^{\text{\oe{}}}=\cosh(r) \hat b^{R\text{\oe{}}}_\Omega - \sinh(r)\hat b^{L\text{\oe{}}\,\dagger}_\Omega  $}&\eqref{eq:dep}\\
        \hhline{|~-|---|}
          &  \multirow{2}{*}{$\Omega<0$} & $w_{-|\Omega|}^+\,/\hat {d}_{-|\Omega|}^{+}$  & $w_{-|\Omega|}^-\,/\hat{d}_{-|\Omega|}^{-}$ & \eqref{eq:wom}\\
          \hhline{|~~|---|}
          & & \multicolumn{2}{c|}{ \cellcolor{gray!10} $\hat{d}_{(-\Omega)}^{\text{\oe{}} }=\cosh(r) \hat b^{L\text{\oe{}}}_\Omega - \sinh(r)\hat b^{R\text{\oe{}}\,\dagger}_\Omega   $}&\eqref{eq:dem}\\
        \hline %
        \multicolumn{2}{|l|}{\bf Chain}  & \multicolumn{2}{c|}{\cellcolor{gray!10} $\hat c_i=\sqrt2 \integral\Omega{-\infty}\infty l_\Omega p_i(\Omega)\hat d^{(e)}_\Omega $} & \\
        \hline
    \end{tabular}
    \caption{Overview of different modes in the Unruh effect}
    \label{tab:modes}
\end{table*}

\textit{Rindler modes}.\textemdash
The worldline $t( \tau)=\sinh(a\tau)/a,$ $x(\tau)=\cosh(a\tau)/a$ describes the wordline of an observer moving through Minkowski spacetime with constant proper acceleration $a$. This wordline is entire located within the region $|t|<x<\infty$, the so called right Rindler wedge, and is, in fact, causally separated from the left Rindler wedge, where $-\infty<x<-|t|$.

The right Rindler wedge is covered by the Rindler coordinates
\begin{equation}\label{eq:Mk_to_rindler_coordinates}
\begin{split}
    &t=\frac{\ee{a\xi}}a\sinh(a\tau),\quad x=\frac{\ee{a\xi}}a\cosh(a\tau),\\
&\Leftrightarrow \tau=\frac1{2a}\ln\frac{x+t}{x-t},\quad \xi=\frac1{2a}\ln\left(a^2(x^2-t^2)\right).
\end{split}
\end{equation}
with the timelike coordinate $\tau$ and the spacelike coordinate $\xi$. 
These coordinates are the canonical choice of coordinates for uniformly accelerated observers because worldlines of constant Rindler coordinats are worldlines of constant proper acceleration. In fact, the proper time $\sigma$ of an observer moving along the worldline of constant $\xi(\tau)=\xi_0$ is $\sigma(\tau)= \ee{a\xi_0}\tau$ and its proper acceleration is $a_0=a\ee{-a\xi_0}$.
In particular, at $\xi=0$, we recover the worldline with proper acceleration $a$.

With respect to the Rindler coordinates the Minkowski metric takes the form $\diff s^2=\diff t^2-\diff x^2=\ee{2a\xi}\left(\diff \tau^2-\diff \xi^2\right)$. Accordingly, the massless Klein-Gordon equation takes the same form as with respect to Minkowski coordinates, namely,\footnote{For a treatment of the detector response in the Unruh effect for massive fields and in higher dimensions see, \eg~\cite{takagi_response_1984} among others.}
\begin{equation}
\left(\partial_\tau^2-\partial_\xi^2\right)\phi=0.
\end{equation}
This form of the wave equation suggests to consider the left-moving and right-moving  Rindler  plane wave modes 
\begin{equation}
    v^{R \pm}_\Omega(\tau,\xi)=\frac{1}{\sqrt{4\pi\Omega}} \ee{-\ii\Omega (\tau\pm\xi)},
\end{equation}
for the expansion of field operators.
From
$
    \tau\pm\xi=\pm\ln(a(x\pm t))/a,
$
one sees that these wavefunctions, with respect to Minkowski coordinates, extend to left-moving and right-moving solutions with support to the right of the null lines $t=-x$ and  $t=x$, respectively:
\begin{equation}\label{eq:vor}
    v^{R \pm}_\Omega(t,x)=\frac{1}{\sqrt{4\pi\Omega}} \ee{\mp \ii\Omega/a \ln(a(x\pm t))}, \quad \text{if } (x\pm t) >0.
\end{equation}
Together, with their mirrored versions which have support on the left side of these null lines
\begin{equation}\label{eq:vol}
    v^{L \pm}_\Omega(t,x)=\frac{1}{\sqrt{4\pi\Omega}} \ee{\pm \ii\Omega/a \ln|a(x\pm t)|},\quad\text{if } x\pm t<0,
\end{equation}
these modes form a complete orthonormal set of modes, \ie
$ %
    \left( v_\Omega^{S\pm}, v_{\Omega'}^{S'\pm'}\right) = \delta_{\pm,\pm'} \delta_{S,S'} \delta(\Omega-\Omega'),
$ %
which can be used to expand the field operator as~\cite{birrell_quantum_1982}
\begin{equation}
\hat \phi(t,x)=\integral{\Omega}0\infty \sum_{\substack{\pm=+,-\\S=L,R}}  v_\Omega^{S\pm}(t,x)\,\hat b^{S\pm}_\Omega+v_\Omega^{S\pm} (t,x)^*\, \hat b^{S\pm}_\Omega{}^\dagger.
\end{equation}
The Bogoliubov coefficients for the transformation from Minkowski to Rindler modes
\begin{equation}
    v_\Omega^{R\pm}=\integral{\omega}0\infty \alpha_{\Omega\omega}^{\pm R} u^{\pm}_\omega +\beta_{\Omega\omega}^{\pm R} u^{\pm}_\omega{}^*
\end{equation}
are given in App.~\ref{app:bogo_rindler}.
For the right Rindler wedge they are
\begin{equation}\label{eq:bogos_rindler_minkowski}
\begin{split}
    & \alpha^{R\pm}_{\Omega\omega}=  \frac{\sqrt{\Omega}\ee{ \pi \Omega/(2a)}}{2\pi a\sqrt \omega}   \left(\frac{\omega}a\right)^{\pm\ii \Omega/a } \Gamma(\mp\ii \Omega/a),
\\
    &\beta^{R\pm}_{\Omega \omega}= -\frac{\sqrt{\Omega} \ee{ -\pi \Omega/(2a)}}{2\pi a\sqrt \omega}  \left(\frac{\omega}a\right)^{\pm\ii \Omega/a } \Gamma(\mp\ii \Omega/a),
\end{split}
\end{equation}
and for the left Rindler wedge modes they are related by complex conjugation $\alpha_{\Omega\omega}^{L\pm}= \left(\alpha^{R\pm}_{\Omega\omega}{}\right)^*$ and $\beta_{\Omega\omega}^{L\pm}= \left(\beta^{R\pm}_{\Omega\omega}{}\right)^*$.

\textit{Lorentz boost}.\textemdash
As discussed above, the Rindler coordinates are closely related to uniformly accelerated observers: 
Worldlines of fixed Rindler spatial coordinate correspond to uniformly accelerated worldllines. Moreover, these worldlines correspond to orbits of the Lorentz boost operator and the Rindler modes are, in fact, eigenmodes of the Lorentz boost operator. We use this relation for our numerical calculation and use the Lorentz boost operator as the Hamiltonian which generates time evolution  along the accelerated detector's worldline.

To illustrate this relation, consider the Lorentz boost, parametrized by a real parameter $T$, acting on points in Minkowski spactime as
\begin{equation}\label{eq:Lorentz_boost_in_Mk_coord}
    \begin{pmatrix} t\\ x\end{pmatrix}\mapsto  \begin{pmatrix} t'\\ x'\end{pmatrix} =
    \begin{pmatrix} \cosh(a T)&\sinh(a T)\\ \sinh(a T) & \cosh(a T)   \end{pmatrix} \begin{pmatrix} t\\ x\end{pmatrix} 
    .
\end{equation}
Inside the right Rindler wedge it transforms points exactly such as to add $T$ to their Rindler time coordinate
\begin{equation}
   \frac{\ee{a\xi}}a \begin{pmatrix}  \sinh(a \tau) \\    \cosh(a \tau) \end{pmatrix}\mapsto  \frac{\ee{a\xi}}a \begin{pmatrix} \sinh(a (T+\tau) ) \\   \cosh(a (T+\tau)) \end{pmatrix}.
\end{equation}
Accordingly, under this Lorentz boost the Rindler modes in the right Rindler wedge acquire only a complex phase
\begin{equation}
    v_{\Omega}^{R\pm}(t',x')= \ee{-\ii\Omega T} v_{\Omega}^{R\pm}(t,x).
\end{equation}
Hence, they are positive frequency modes with respect to Lorentz boosts and we refer to $\Omega$ as their Rindler frequency, which here in the right Rindler wedge is positive.

In the left Rindler wedge, however, the Rindler modes $v^{L\pm}_\Omega$ have a negative Rindler frequency. To see this we cover the left Rindler wedge by the coordinates $(\tilde\tau,\tilde\xi)$ with
\begin{equation}
t=-\frac{\ee{a\tilde\xi}}a\sinh(a\tilde\tau),\quad x=-\frac{\ee{a\tilde\xi}}a\cosh(a\tilde\tau)
.
\end{equation}
The Lorentz boost~\eqref{eq:Lorentz_boost_in_Mk_coord} still increases the parameter $\tilde\tau$,
\begin{equation}
   \frac{\ee{a\tilde\xi}}a \begin{pmatrix}  \sinh(a \tilde\tau) \\    \cosh(a \tilde\tau) \end{pmatrix}\mapsto  \frac{\ee{a\tilde\xi}}a \begin{pmatrix} \sinh(a (T+\tilde\tau) ) \\   \cosh(a (T+\tilde\tau)) \end{pmatrix},
\end{equation}
however, since the left Rindler modes with respect to these coordinates read
\begin{equation}
    v_\Omega^{L\pm}(\tilde\tau,\tilde\xi) %
    =\frac1{\sqrt{4\pi\Omega}} \ee{\ii\Omega (\tilde\tau\pm\tilde\xi)}
\end{equation}
they have a negative Rindler frequency, acquiring the phase
\begin{equation}
    v_{\Omega}^{L\pm}(t',x')= \ee{\ii\Omega T} v_{\Omega}^{L\pm}(t,x)
\end{equation}
under the Lorentz boost,
which moves points in the left Rindler wedge into their causal past.

With these relations at hand we can express the Lorentz boost Hamiltonian in terms of the mode operators associated with the left and right Rindler modes:
\begin{equation}
    \hat H_L=\sum_{\pm=+,-}\integral\Omega0\infty  \Omega \left( \hat b^{R\pm}_\Omega{}^\dagger \hat b^{R\pm}_\Omega-\hat b^{L\pm}_\Omega{}^\dagger \hat b^{L\pm}_\Omega\right).
\end{equation}
In the same way, as the Minkowski field Hamiltonian $\hat H_f=\integral\omega0\infty \sum_\pm \omega \hat a^\pm_\omega{}^\dagger a^\pm_\omega$ generates translations along the Minkowski time coordinate $t$, $\hat H_L$ generates translations along the Rindler time coordinates $\tau$ and $\tilde \tau$.
In particular, this Hamiltonian generates time evolution with respect to the proper time of the uniformly accelerated observer moving along the wordlline at $\xi=0$  with proper acceleration $a$ in the right Rindler wedge.

\textit{Unruh temperature and Unruh modes}.\textemdash
At its core, the Unruh effect is the observation that when the field is in the vacuum state with respect to Minkowski modes, then the Rindler modes of one wedge are in a thermal state with respect to the Lorentz boost operator.
In fact, %
we find that in the Minkowski vacuum (see~\eqref{eq:app_beta_thermal_expt})
\begin{equation}\label{eq:Rindler_num_exptval}
    \bra{0_M} \hat b^{R\pm}_{\Omega'}{}^\dagger \hat b^{R\pm}_\Omega\ket{0_M} = \integral{\omega}0\infty \beta^{R\pm}_{\Omega \omega} \beta^{R\pm}_{\Omega' \omega}{}^*
    =  \frac{\delta(\Omega-\Omega') }{\ee{\frac{2\pi\Omega}a} -1}\,,
\end{equation}
the number expectation value of the Rindler modes equals  thermal expectation value with the celebrated Unruh temperature given by $T_U=a/(2\pi)$.
The analogous relation also holds for the left Rindler wedge modes.
Note here also, that the Rindler modes not having any cross-correlations, identifies them  as the natural basis of normal modes to use in the Rindler wedges.

The Hamiltonian $\hat H_L$ thus takes the same role as the doubled Hamiltonian~\eqref{eq:hf_doubled_even} in the scenario of an inertial detector coupling to a thermal field state.
In the case of a thermal field, we use pairwise squeezing to transform a pair $\hat b_i$ and $\hat b_i'$ of thermal eigenmodes of the Hamiltonian into a pair of eigenmodes $\hat d_i$ and $\hat d_i'$ which are in their vacuum state.
In the same way, in the context of the Unruh effect, we can use pairwise squeezing of two Rinder modes $\hat b_\Omega^{R\pm}$ and $\hat b_\Omega^{L\pm}$ to transform them into a pair of Unruh modes $\hat d_\Omega^\pm$ and $\hat d_{-\Omega}^\pm$ which share their vacuum state with the Minkowski modes, \ie $\hat d^\pm_\Omega\ket{0_M}=0$.

Following the same steps as in the thermal case, the mode functions associated to $\hat d^\pm_\Omega$ which have   positive Rindler frequency $\Omega>0$ are
\begin{equation}\label{eq:wop}
\begin{split}
    w_\Omega^{\pm} &=\frac{ \ee{\Omega\pi/(2a)} v_\Omega^{R\pm} +\ee{-\Omega\pi/(2a)}   \left(v_\Omega^{L\pm} \right)^*}{\sqrt{2\sinh(\Omega\pi/a)} }
    \nn& =\cosh(r) v_\Omega^{R\pm} +\sinh(r)  \left(v_\Omega^{L\pm} \right)^*
    \end{split}\,,
\end{equation}
where we introduced $r$ such that $\cosh(2r)=2/\tanh(\Omega \pi/a)$ for compact notation. %
The Unruh modes associated to $\hat d^\pm_{-\Omega}$ with negative Rindler frequency, $\Omega<0$, are accordingly
\begin{equation}\label{eq:wom}
    w_{\Omega}^{\pm}  =\cosh(r) v_{(-\Omega)}^{L\pm} +\sinh(r)  \left(v_{(-\Omega)}^{R\pm} \right)^*.
\end{equation}
By construction, the Unruh modes are thus linear combinations of positive frequency Minkowski modes only,
\begin{equation}\label{eq:bogo_u_to_w}
    w_\Omega^{\pm} =\integral\omega0\infty \gamma_{\Omega\omega}^\pm u^\pm_\omega\,,
\end{equation}
with $\gamma_{\Omega\omega}^\pm$ as in~\eqref{eq:gamma_bogo}, for both the positive and the negative Rindler frequency modes.
This implies that, most importantly, $\hat d_\Omega^\pm\ket{0_M}$, \ie the vacuum state of the Unruh modes coincides with the Minkowski vacuum.
And the the Lorentz boost Hamiltonian, in terms of the Unruh mode operators, reads
\begin{equation}
    \hat H_L =\sum_{\pm=+,-} \integral{\Omega}{-\infty}\infty \Omega \hat d_\Omega^\pm{}^\dagger\hat d_\Omega^\pm{} \,.
\end{equation}

\textit{Coupling to even Rindler modes}.\textemdash
To couple the detector to the field along a uniformly accelerated oberserver's worldline, we now replace the Minkowksi coordinates, used for the detector at rest in~\eqref{eq:interaction-general}
by Rindler coordinates, and obtain
\begin{equation}\label{eq:hint_in_rindler}
    \hat H_i=\lambda \hat X\otimes\integral\xi{}{} f(\xi) \partial_\tau \hat\phi(\xi)\,.
\end{equation}
Since the detector's smearing function $f$ extends over a length scale $L$, the different points of the detector actually experience different proper accelerations. Therefore, we assume that $aL\ll 1$ so that such finite-size effects can be neglected.

Also note, that at this point we switch to the Schrödinger picture which is also employed in the numerical calculations. We take the observables in the Schrödinger picture to be equal to the observables in the Heisenberg picture one the spacelike hypersurface $t=\tau=0$.

The field observable $\partial_\tau \hat\phi(\xi)=\ii\comm{\hat H_L}{\hat\phi(\xi)}$ to which the detector couples can be expanded as
\begin{equation}
\begin{split}
    \partial_\tau \hat\phi(\xi) &= \sum_\pm\integral\Omega0\infty (-\ii\Omega) v_\Omega^{R\pm} (\xi)\hat b^{R\pm}_\Omega+ \mathrm{H.c.}
    \\&
    =  \sum_\pm\integral\Omega0\infty \frac{-\ii\sqrt{\Omega}\ee{\mp\ii\Omega \xi} }{\sqrt{4\pi}} \hat b^{R\pm}_\Omega+ \mathrm{H.c.}\,.
    \end{split}
\end{equation}
from which it is clear that the time evolution of this model is equal to the time evolution of a detector at rest coupling to a field in a thermal state.

As before, in the numerical calculations we use that the atom couples symmetrically to left- and right-moving modes, \ie it couples only to the even sector of the field, spanned by $\hat b_\Omega^{R e}$ but not the odd sector, spanned by $\hat b_\Omega^{R o}$, where
\begin{equation}
    \hat b_\Omega^{R e}=\frac1{\sqrt2}\left(\hat b_\Omega^{R+}+\hat b_\Omega^{R-}\right),\quad
    \hat b_\Omega^{R o}=\frac1{\sqrt2}\left(\hat b_\Omega^{R+}-\hat b_\Omega^{R-}\right).
\end{equation}
Accordingly, for the chain transformation and the numerical calculations, we use the corresponding even Unruh modes
\begin{align}
    &\hat d_\Omega^{e}%
    =\cosh(r) \hat b_\Omega^{Re} -\sinh(r) \hat b_\Omega^{Le}{}^\dagger
    =\frac{ \hat d^+_\Omega+\hat d_\Omega^- }{\sqrt2}, 
    \label{eq:dep}\\
    &\hat d_{-|\Omega|}^{e}
    =\cosh(r) \hat b_\Omega^{Le} -\sinh(r) \hat b_\Omega^{Re}{}^\dagger
    =\frac{ \hat d^+_{-|\Omega|}+\hat d_{-|\Omega|}^-}{\sqrt2}, \label{eq:dem}
\end{align}
whose relation to the other modes is summarized also in Tab.~\ref{tab:modes}.
The chain modes for the Unruh effect are thus the same chain modes as for a detector at rest coupling to a thermal field, when setting $\beta=2\pi/a$. %
They are composed from the even Unruh modes only,
\begin{align} 
    \hat c_i &= \integral{\Omega}{-\infty}{\infty} \frac{\sgn(\Omega) f_{|\Omega|} \ee{\frac{\Omega\pi}{2a}}   }{\sqrt{|\sinh(\Omega\pi/a)|}} p_i(\Omega)\hat d^{e}_\Omega.
\end{align}
The odd Unruh modes $\hat d^o_\Omega=\left(\hat d_\Omega^+-\hat d_\Omega^-\right)/\sqrt2$, however, do not couple to the atom, and remain in their vacuum state.

App.~\ref{app:mk_density_from_unruh_chain}, based on the Bogolubov transformations reviewed in App.~\ref{app:bogo_rindler}, derives closed form expressions for the energy density $\exptval{\normord{\hat\pi^\mp(x)}}$ as measured by an observer at rest with respect to the Minkowski coordinates.

\section{Lorentz boost of Minkowski observables under Rindler time evolution}\label{app:mk_observables_vs_rindler_evol}

\begin{figure*}
\centering
\includegraphics[width=.99\textwidth]{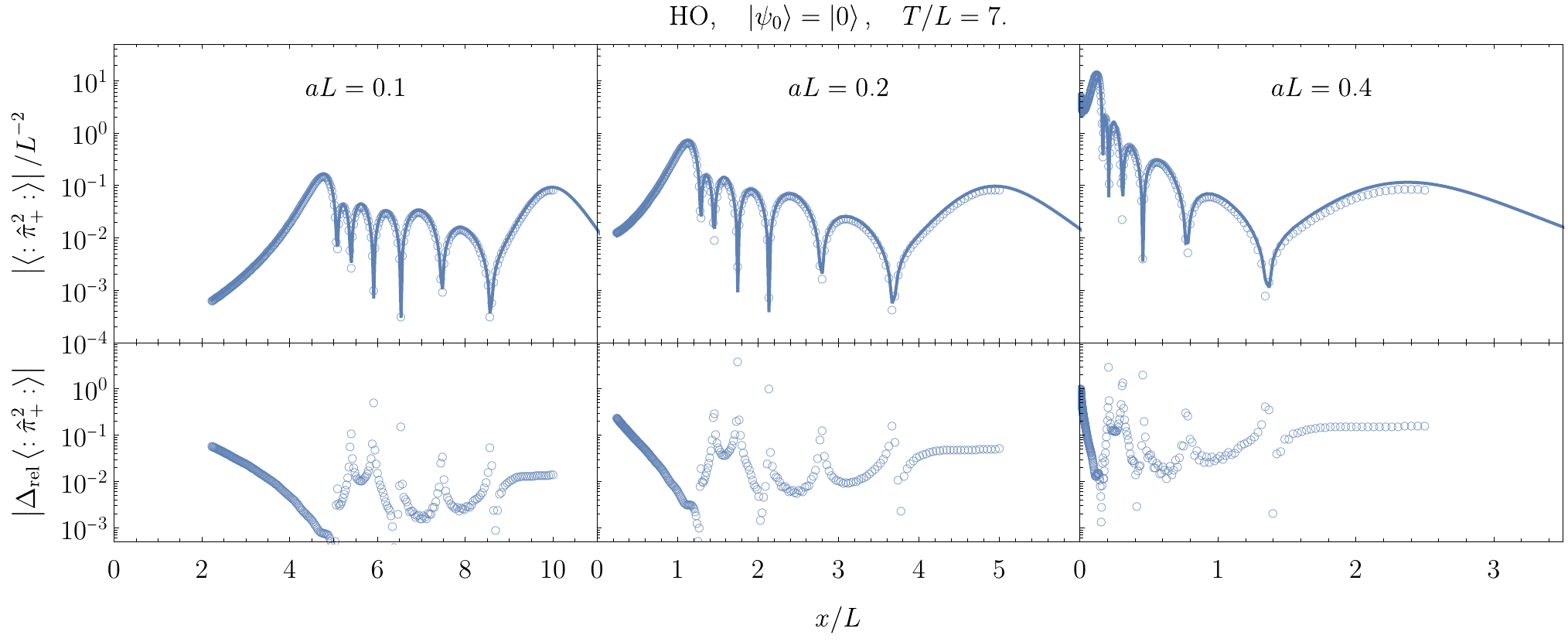}
\caption{Left-moving radiation $\exptval{\normord{\hat\pi^2_+}}$ from a uniformly accelerated harmonic oscillator detector compared to 
the radiation obtained by boosting  the radiation emitted by a resting detector  as described in App.~\ref{app:mk_observables_vs_rindler_evol}.
The detector is initialized in its ground state. The upper row of panels shows the expectation values for the energy density. The line plot shows $\exptval{\normord{\hat\pi^2_+}}_{\mathrm{num}}$ which are the results from Sec.~\ref{sec:unruh}, and the circle points show $\exptval{\normord{\hat\pi^2_+}}_{\mathrm{bst}}$ the 'boosted' data.
The lower panels show the relative difference $\Delta_{\mathrm{rel}}\left|\exptval{\normord{\hat\pi^2_+}}\right|=1-\exptval{\normord{\hat\pi^2_+}}_{\mathrm{bst}}/\exptval{\normord{\hat\pi^2_+}}_{\mathrm{num}}$.
}
\label{fig:synth_vac}
\end{figure*}

\begin{figure*}
\centering
\includegraphics[width=.99\textwidth]{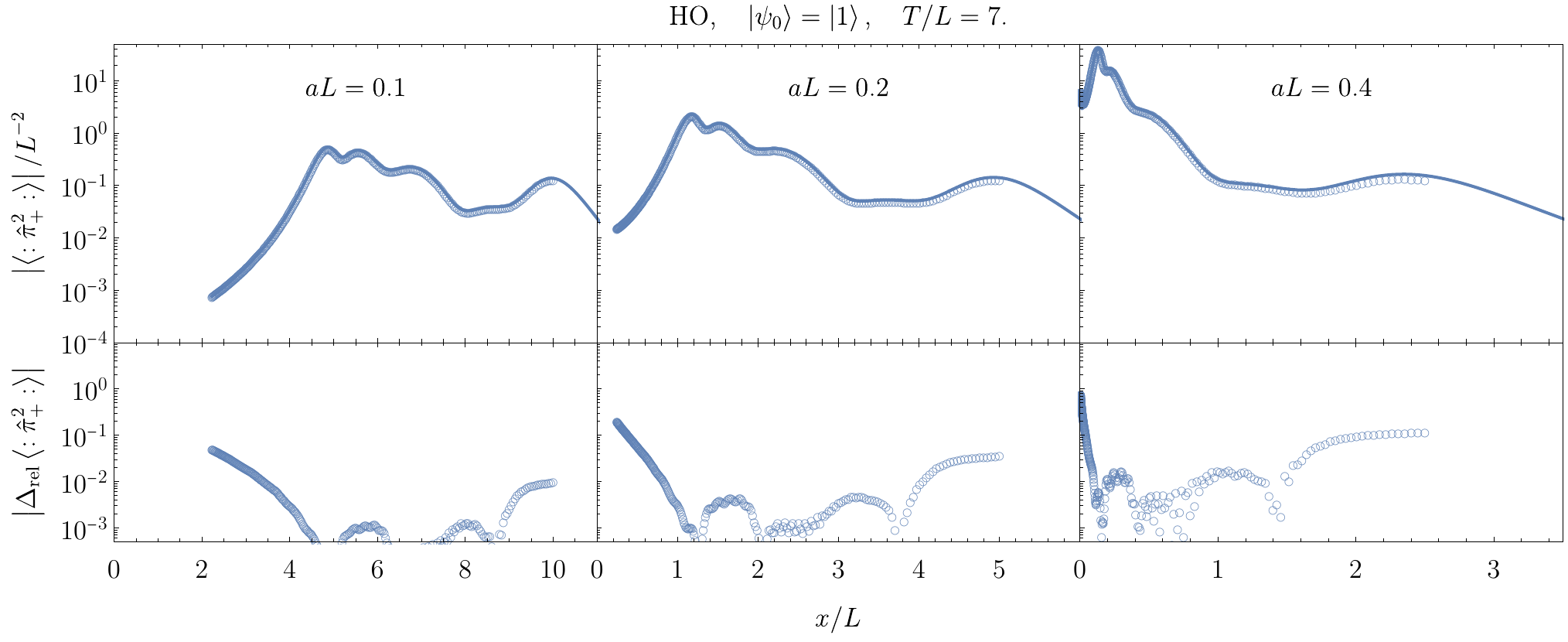}
\caption{Comparison of radiation as in Fig.~\ref{fig:synth_vac}, for initial state $\ket{\psi_0}=\ket1$ of the harmonic detector.}
\label{fig:synth_exc}
\end{figure*}

\textit{Rindler time evolution}.\textemdash
The expression~\eqref{eq:app_mk_edensity_from_chain} gives the expectation value of the right-moving and left-moving Minkowski energy density for a state defined on the spacetime hyperplane $\tau=t=0$.
Under the Rinder time evolution $\exp(-\ii T\hat H)$ which we apply in our numerical simulation, the observable on the left-hand side of the expression transforms non-trivially.
Hence, after applying $\exp(-\ii T\hat H)$, which corresponds to the Lorentz boost~\eqref{eq:Lorentz_boost_in_Mk_coord}, we need to reinterpret the expectation value given by the right-hand side of~\eqref{eq:app_mk_edensity_from_chain}.
For this, there are two alternatives.

The first alternative is to interpret the results as being measured on the hyperplane $t=0$, as it stands on the left-hand side of~\eqref{eq:app_mk_edensity_from_chain}, but in a scenario where the interaction between detector and field started on the hyperplane $t=-\tanh(aT)x$ (corresponding to $\tau=-T$).
This alternative makes use of the  Minkowski vacuum being invariant under Lorentz boosts. Hence, the initial state, which is taken to be a product state between the detector initial state and the Minkowski vacuum of the field, can be boosted back in Rindler time.

The second alternative interprets the result as arising in a scenario where the interaction between detector and field starts on the hyperplane $\tau=t=0$ where the overall state is still given by the initial product state.
In this scenario, the expectation value given by the right-hand side of~\eqref{eq:app_mk_edensity_from_chain} corresponds to the observable
\begin{equation}\label{eq:edensity_lorentzboost}
    \ee{\pm2aT}  \exptval{\normord{\hat\pi^\pm(t',x')^2}},
\end{equation}
with $(t',x')=(x\sinh(aT),x\cosh(aT))$, as we derive in the following.

To derive this transformation, we first use~\eqref{eq:Mk_to_rindler_coordinates} to obtain
\begin{align}
    &\partial_t=\partialfrac\tau{t}\partial_\tau+\partialfrac\xi{t}\partial_\xi 
    = \cosh(a\tau) \ee{-a\xi}\partial_\tau-\sinh(a\tau) \ee{-a\xi}\partial_\xi, 
    \nn
    &\partial_x=\partialfrac\tau{x}\partial_\tau+\partialfrac\xi{x}\partial_\xi 
    = \cosh(a\tau) \ee{-a\xi}\partial_\xi-\sinh(a\tau) \ee{-a\xi}\partial_\tau,
    \nn &\partial_\tau = a \left(x\partial_t+t\partial_x\right),
    \nn&
    \partial_\xi=a\left(x\partial_x+t\partial_t\right).
\end{align}
Hence, with respect to Rindler coordinates the (left-moving and right-moving) field momentum on the hyperplane $t=\tau=0$ is
\begin{equation}
\begin{split}
    \left.\hat\pi^\pm(x)\right|_{t=0}
    &=\left.\frac12 \left(\partial_t\pm\partial_x\right)\hat\phi(x)\right|_{t=0}
    \\&
    =\left.\frac{\ee{-a\xi}}2 \left(\partial_\tau\pm\partial_\xi\right)\hat\phi\left(\xi= \ln(ax)/a\right)\right|_{\tau=0}.
\end{split}\end{equation}
Under the Rindler time evolution $(\tau,\xi)\to(\tau+T,\xi)$, which is nothing but the Lorentz boost~\eqref{eq:Lorentz_boost_in_Mk_coord}, this observable transforms into
\begin{equation}
\begin{split}
    &\left.\frac{\ee{-a\xi}}2 \left(\partial_\tau\pm\partial_\xi\right)\hat\phi(\xi)\right|_{\tau=T}
     \\ &= \frac{\ee{-a\xi}}2 a(x\pm t)\left(\partial_t\pm\partial_x\right) \hat\phi( t(\xi,T),x(\xi,T) )
     \\ &
    = \left.\ee{\pm aT}  \hat\pi^\pm\left(x\cosh(aT)\right)\right|_{t=x\sinh(aT)}\,.
\end{split}\end{equation}

\textit{Interpretation of Unruh radiation}.\textemdash
Both as a benchmark and for an interpretation of the energy density emitted from an accelerated detector, we can boost the energy density emitted from a detector at rest, as calculated in Sec.~\ref{ssec:numerical-example}, and pretend as if it emanates from the accelerated worldline~\eqref{eq:unif_a_observer_wordline}.
Then we compare the thereby obtained, \textit{boosted} energy density profile to the results from Sec.~\ref{sec:unruh}.

Fig.~\ref{fig:synth_vac} compares this boosted radiation to our results from Sec.~\ref{sec:unruh} for the left-moving radiation.
In detail, we perform the following transformation of the data from the resting detector:
Fig.~\ref{fig:edensities_vacuum} shows the right-moving energy density expectation value, $\exptval{\normord{\hat\pi_-^2(x,t)}}$, emitted from detectors that are at rest.
We here use the data for an interaction time of $t=7L$.
By neglecting the width of the detector, we can interpret the energy density $\exptval{\normord{\hat\pi_-^2(x,t=7L)}}$ in the interval $0\leq x\leq 7L$ %
as emanating from the detector (at $x=0$) at times $-7L \leq t \leq 0$.
Since the detector is at rest, its proper time $\tau=t$ equals the coordinate time.

It follows from the derivation of~\eqref{eq:edensity_lorentzboost} above, that the energy density $\exptval{\normord{\hat\pi^\pm}}$, which is emitted from the worldline~\eqref{eq:unif_a_observer_wordline} at proper time $-\tau$, is Doppler shifted by a factor of $\ee{\mp a\tau}$ with respect to the resting observer.
Radiation emanating from the wordlline at $\tau=-T$ has light cone coordinates $z^\pm=t\pm x= \pm \ee{\mp a T}/a$, and thus arrives on the hyperplane $t=0$ at the spatial coordinate $x=\ee{\mp a T}/a$.
In conclusion, the boosted  data $\exptval{\normord{\hat\pi^2_+}}_{\mathrm{bst}}$ plotted in Figs.~\ref{fig:synth_vac} and~\ref{fig:synth_exc} is obtained from the left-moving energy density $\exptval{\normord{\hat\pi^2_+(x,t=7L)}}_{\mathrm{res}}$ (which is related to the right-moving energy density plotted in Fig.~\ref{fig:edensities_vacuum} by $\exptval{\normord{\hat\pi^2_+(x,t)}}=\exptval{\normord{\hat\pi^2_-(-x,t)}}$)
as
\begin{equation}
    \exptval{\normord{\hat\pi^2_+(x)}}_{\mathrm{bst}} = \frac1{a^2 x^2}\exptval{\normord{\hat\pi^2_+(-\ln(ax)/a)}}_{\mathrm{res}}.
\end{equation}
The figures show that the boosted data resemble the results calculated for the accelerated detectors to a very good degree.

\section{Bogoliubov transformations between  Minkowski, Rindler and Unruh modes}\label{app:bogo_rindler}
This section reviews the calculation of the Bogoliubov transformations between Minkowski, Rindler and Unruh modes.
In order to express the Rindler modes as linear superpositions of Minkowski modes, we need to calculate the Klein-Gordon inner prodcuts of $ u_\omega^\pm$ and $v_\Omega^{S\pm}$, with $S=R,L$, and their conjugates.
To see that any inner product between a left-moving and a right-moving solution vanishes,  note that for left- and right-moving wave functions we have
\begin{align}
    \partial_t u_\omega^\pm=\pm\partial_x u_\omega^\pm,\quad \partial_t v_\Omega^{S\pm} =\pm\partial_x v_\Omega^{S\pm}.
\end{align}
Applying integration by parts and making use of the fact the boundary terms may be discarded we obtain:
\begin{equation}
\begin{split}
    \left( v_\Omega^{S\pm},u_\omega^{\mp}\right) &= -\ii\integral{x}{}{} v_\Omega^{S\pm}\partial_t u_\omega^\mp{}^*-(\partial_t v_\Omega^{S\pm})  u_\omega^\mp{}^*
    \\& 
    = -\ii\integral{x}{}{} (\mp1) v_\Omega^{S\pm}\partial_x u_\omega^\mp{}^* \mp (\partial_x v_\Omega^{S\pm})  u_\omega^\mp{}^*
    \\& 
    = \pm \ii\integral{x}{}{} v_\Omega^{S\pm }\partial_x u_\omega^\mp{}^*-v_\Omega^{S\pm}\partial_x u_\omega^\mp{}^* =0
    \end{split}
\end{equation}
For the non-vanishing inner products, we find:
\begin{equation}
    \begin{split}
    & \alpha^{R\pm}_{\Omega\omega}=\left(v_\Omega^{R\pm},u_\omega^\pm\right)
=  -\ii\integral{x}{}{} v_\Omega^{R\pm}\partial_t u_\omega^\pm{}^*-(\partial_t v_\Omega^{S\pm})  u_\omega^\pm{}^*
\\&
=  \mp2 \ii\integral{x}{}{} v_\Omega^{R\pm}\partial_x u_\omega^\pm{}^*
\\& \stackrel{t=0}=\frac{\sqrt \omega}{2\pi \sqrt \Omega} \integral{x}{0}{\infty}  \ee{\mp\ii \Omega/a\ln(ax)}     \ee{\pm \ii \omega x}
\end{split}
\end{equation}
The integral is not convergent because the Bogoliubov transformation is relating to set of improper, continuous modes. We can obtain a regularized expression by introducing a regularizing factor $\ee{-\epsilon a x}$, where we have to take the limit $\epsilon\to 0^+$ after integrating against properly normalized expressions.
With this we obtain (see 3.381.4 in~\cite{gradshteyn_table_2014}),
\begin{align}
    & \alpha^{R\pm}_{\Omega\omega}= \frac{\sqrt \omega}{2\pi \sqrt \Omega} a^{\mp\ii \Omega/a} \integral{x}{0}{\infty} x^{\mp\ii \Omega/a}     \ee{\pm \ii \omega x- \epsilon a x}
       \nn&
    = \frac{\mp\ii \sqrt{\Omega \omega}}{2\pi a} \frac{ a^{\mp\ii \Omega/a}}{\left(\epsilon a\mp\ii \omega\right)^{1\mp\ii \Omega/a}} \Gamma(\mp\ii \Omega/a)
\nn&\stackrel{\epsilon\to 0^+}\to %
 \frac{\sqrt{\Omega}}{2\pi a\sqrt \omega}  \ee{ \pi \Omega/(2a)} \left(\frac{\omega}a\right)^{\pm\ii \Omega/a } \Gamma(\mp\ii \Omega/a).
\end{align}
Similarily,
\begin{align}
&\beta^{R\pm}_{\Omega \omega}=- \left(v_\Omega^{R\pm},u_\omega^\pm{}^*\right)
=\pm 2\ii\integral{x}{}{} v_\Omega^{R\pm} \partial_x u_\omega^\pm{}
\nn& =\pm2\ii\integral{x}0\infty \frac1{4\pi\sqrt{\Omega \omega}} \ee{\mp\ii \Omega /a\ln(a x)} (\mp\ii \omega)\ee{\mp\ii \omega x}
\nn& =\frac{\sqrt \omega}{2\pi\sqrt{\Omega}} \integral{x}0\infty  \ee{\mp\ii \Omega /a\ln(a x)}  \ee{\mp\ii \omega x},
\end{align}
which is regularized to 
\begin{align}
&\beta^{R\pm}_{\Omega \omega}=\frac{\sqrt \omega}{2\pi\sqrt{\Omega}} \integral{x}0\infty  \ee{\mp\ii \Omega /a\ln(a x)}  \ee{\mp\ii \omega x-\epsilon a x}
      \nn&
    = \frac{\mp\ii \sqrt{\Omega \omega}}{2\pi a} \frac{ a^{\mp\ii \Omega/a}}{\left(\epsilon a\pm\ii \omega\right)^{1\mp\ii \Omega/a}} \Gamma(\mp\ii \Omega/a)
\nn&\stackrel{\epsilon\to 0^+}\to -\frac{\sqrt{\Omega}}{2\pi a\sqrt \omega}  \ee{ -\pi \Omega/(2a)} \left(\frac{\omega}a\right)^{\pm\ii \Omega/a } \Gamma(\mp\ii \Omega/a).
\end{align}
The coefficients for the left Rindler wedge modes are related to the right ones by complex conjugation,
\begin{align}
     \alpha^{L\pm}_{\Omega\omega}&=\left(v_\Omega^{L\pm},u_\omega^\pm\right)
=  \pm2 \ii\integral{x}{}{} v_\Omega^{L\pm}\partial_x u_\omega^\pm{}^*
\nn&
\stackrel{t=0}= -\frac{\sqrt \omega}{2\pi \sqrt \Omega} \integral{x}{-\infty}0  \ee{\pm\ii \Omega/a\ln|ax|}     \ee{\pm \ii \omega x}
=  (\alpha^{R\pm}_{\Omega\omega}{})^*
\end{align}
and similarily
\begin{align}
\beta_{\Omega \omega}^{L\pm}&= (\beta^{R\pm}_{\Omega\omega}{})^*.
\end{align}

Using the regularized expression we calculate the Minkowski vacuum expectation value for the Rindler modes to be
\begin{align}\label{eq:app_beta_thermal_expt}
&\exptval{\hat b_\Omega^{R\pm}{}^\dagger \hat b_{\Omega'}^{R\pm} }=\integral{\omega}0\infty \beta^{R\pm}_{\Omega \omega} \beta^{R\pm}_{\Omega' \omega}{}^*
\nn&=\integral{\nu}{-\infty}\infty \frac{\sqrt{\Omega \Omega'} }{4\pi^2 a^2 }\ee{-\frac{(\Omega+\Omega')\pi}{2a}} \ee{\frac{\pm\ii (\Omega-\Omega')\nu}a } \Gamma\left(\frac{\mp\ii \Omega}a\right) \Gamma\left(\frac{\pm\ii \Omega'}a\right)
\nn&
=  \frac{\delta(\Omega-\Omega') }{\ee{\frac{2\Omega \pi}a} -1}.
\end{align}

Both for left-movers and right-movers we have the relation
\begin{align}
&\ee{-\pi \Omega/(2a)} \alpha^{R\pm}_{\Omega \omega}+\ee{\pi \Omega/(2a)} \beta^{L\pm}_{\Omega \omega}{}^*
\nn&=\ee{-\pi \Omega/(2a)} \alpha^{R\pm}_{\Omega \omega}+\ee{\pi \Omega/(2a)} \beta^{R\pm}_{\Omega \omega}{}=0 
\end{align}
which is easy to read off from the regularized expressions, but also holds without regularisation but considering contour integrals.
These relations warrant that the Unruh modes are (linear combinations of only) positive Minkowski frequency modes.
With $\Omega>0$ we have
\begin{align}
    & w_{\Omega}^\pm = \frac{\ee{\Omega\pi/(2a)} v_{\Omega}^{R\pm}+\ee{-\Omega\pi/(2a)} v_\Omega^{L\pm}{}^* }{\sqrt{2\sinh(\pi\Omega/a)}}
    \nn&=\frac1{\sqrt{2\sinh(\pi\Omega/a)}}\integral\omega0\infty    \left( \ee{\frac{\Omega\pi}{2a}} \alpha^{R\pm}_{\Omega\omega}+\ee{-\frac{\Omega\pi}{2a}} \beta^{R\pm}_{\Omega\omega}\right) u^\pm_\omega
    \nn&\qquad + \underbrace{\left(\ee{\frac{\Omega\pi}{2a}} \beta^{R\pm}_{\Omega\omega}+\ee{-\frac{\Omega\pi}{2a}} \alpha^{R\pm}_{\Omega\omega}\right)}_{=0} u^\pm_\omega{}^*
    \nn&=\integral\omega0\infty  \frac{\sqrt{\Omega \sinh\left(\frac{\pi\Omega}a\right)}}{\pi a\sqrt{2 \omega} }   \left(\frac{\omega}a\right)^{\pm\ii \Omega/a } \Gamma(\mp\ii \Omega/a) u^\pm_\omega.
\end{align}
Similarly, for the Unruh modes with negative Rindler frequency, $\Omega<0$, we have
\begin{align}
   w_{\Omega}^{\pm} & =\frac{\ee{|\Omega|\pi/(2a)} v_{|\Omega|}^{L\pm}+\ee{-|\Omega|\pi/(2a)} v_{|\Omega|}^{R\pm}{}^* }{\sqrt{2\sinh(\pi|\Omega|/a)}}
   \nn&=\frac{\integral\omega0\infty    \left( \ee{\frac{|\Omega|\pi}{2a}} \alpha^{R\pm}_{|\Omega|\omega}{}^*+\ee{-\frac{|\Omega|\pi}{2a}} \beta^{R\pm}_{|\Omega|\omega}{}^* \right) u^\pm_\omega}{\sqrt{2\sinh(\pi|\Omega|/a)}}
  \nn&=\integral\omega0\infty  \frac{\sqrt{\Omega \sinh\left(\frac{\pi\Omega}a\right)}}{\pi a\sqrt{2 \omega} }   \left(\frac{\omega}a\right)^{\pm\ii \Omega/a } \Gamma(\mp\ii \Omega /a)  u^\pm_\omega
\end{align}
thus we can unify  the formulae  for negative and positive $\Omega$  and write
\begin{align}
    & w_{\Omega}^\pm =\integral\omega0\infty \gamma_{\Omega\omega}^\pm u_\omega^\pm, \\
    &\gamma^\pm_{\Omega\omega}=  \frac{\sqrt{\Omega \sinh\left(\frac{\pi\Omega}a\right)}}{\pi a\sqrt{2 \omega} }   \left(\frac{\omega}a\right)^{\pm\ii \Omega/a } \Gamma(\mp\ii \Omega/a).
    \label{eq:gamma_bogo}
\end{align}

\begin{widetext}

\section{Energy density from a resting detector coupling to a thermal field state}\label{app:thermal_edensity}
Also for a detector at rest coupling to a thermal field state, the energy density which is emitted into the field can be calculated from the numerical data for the chain modes.
In this appendix we discuss the evaluation of $\exptval{\normord{\hat \pi_\pm^2}}$ in this context and derive its expansion in terms of the chain modes obtained from the thermal double construction.

In the thermal state of the field, \ie even before the detector couples to the field, the energy density expectation values is not zero.
This is because of the thermal occupation~\eqref{eq:nk_thermal_exptval} of the field modes $\hat b_k$ (all the modes $\hat b^{(e)}_k$, $\hat b^{(o)}_k$ and $\hat b_k$ are thermal).
Hence the expectation value of the left-moving and right-moving energy density~\eqref{eq:norm_ord_pi_density_squared} in this state is
\begin{equation}\label{eq:thermal_background_density}
    \exptval{\normord{\hat\pi_\mp^2(x)}} =\integral{k}0\infty\integral{k'}0\infty \frac{\sqrt{kk'}}{2\pi} \ee{\mp \ii (k-k') x} \exptval{\hat b^\dagger_{\pm k}\hat b_{\pm k'}} 
    =\integral{k}0\infty \frac{k}{2\pi \left(\ee{\beta k}-1\right)} =\frac\pi{12\beta^2}
.
\end{equation}
Even and odd modes contribute equally to this background energy density. The contribution from the odd sector remains constant in time, whereas the contribution from the even sector is modulated due to  the interaction with the atom. Hence, we need to express 
\begin{equation}\label{eq:thermal_density_even}
   \exptval{\normord{(\hat \pi^{(e)}_\mp{})^2}} =
\integral\omega0\infty\integral{\omega'}0\infty 
    \frac{\sqrt{\omega\omega'}}{4\pi}  \ee{\mp \ii (\omega-\omega') x} \exptval{\hat b^{(e)}_{\omega}{}^\dagger \hat b^{(e)}_{\omega'} }
 \quad -\Re\integral\omega0\infty\integral{\omega'}0\infty 
    \frac{\sqrt{\omega\omega'}}{4\pi}  \ee{\pm\ii (\omega+\omega')x}\exptval{\hat b^{(e)}_{\omega} \hat b^{(e)}_{\omega'} }  
\end{equation}
in terms of the chain mode operators.

Starting from~\eqref{eq:dmodes_thermal}, and using
the inverse transformation of~\eqref{eq:thermal_chain_modes_from_poly} which is
$
    \hat d_\omega=\sum_i \ii \frac{\sgn(\omega) \sqrt{|\omega|} \ee{\frac{\beta\omega}{4}-L|\omega|} }{\sqrt{4\pi |\sinh(\beta\omega/2)|}} p_i(\omega)\hat c_i,
$
we have
\begin{equation}
    \hat b^{(e)}_\omega =\frac1{\sqrt{2\sinh(\beta\omega/2)}} \left(\ee{\frac{\beta\omega}4}\hat d_\omega+\ee{\frac{-\beta\omega}4}\hat{d}^\dagger_{-\omega}\right)
    =\sum_{i=0}^\infty\sum_{k=0}^i P_{i,k} \frac{\ii \sqrt{\omega} \ee{-L\omega} }{\sqrt{8\pi} \sinh(\omega/\theta)}\omega^k \left(    \ee{\frac\omega{\theta}}   \hat c_i + (-1)^k     \ee{-\frac\omega{\theta} }  \hat c_i^\dagger \right).    
\end{equation}
With this at hand
\begin{align}
    &\exptval{\hat b^{(e)}_{\omega}{}^\dagger \hat b^{(e)}_{\omega'} }
    = \sum_{i,j=0}^\infty\sum_{k=0}^i\sum_{l=0}^j \frac{\sqrt{\omega\omega'} \ee{-L(\omega+\omega')} P_{i,k}P_{j,l}\omega^k\omega'^l}{8\pi\sinh(\beta\omega/2)\sinh(\beta\omega'/2)} \exptval{\left(    \ee{\frac\omega{\theta}}   \hat c_i^\dagger + (-1)^k     \ee{-\frac\omega{\theta} }  \hat c_i  \right)\left(    \ee{\frac{\beta\omega'}{2}}   \hat c_j + (-1)^l     \ee{\frac{-\beta\omega'}{2} }  \hat c_j^\dagger \right)}
\end{align}
and
\begin{align}
    &\exptval{\hat b^{(e)}_{\omega}  \hat b^{(e)}_{\omega'} }
    = - \sum_{i,j=0}^\infty\sum_{k=0}^i\sum_{l=0}^j \frac{\sqrt{\omega\omega'} \ee{-L(\omega+\omega')} P_{i,k}P_{j,l}}{8\pi\sinh(\beta\omega/2)\sinh(\beta\omega'/2)}\omega^k\omega'^l \exptval{\left(    \ee{\frac{\beta\omega}2}   \hat c_i + (-1)^k     \ee{-\frac{\beta\omega}2 }  \hat c_i^\dagger  \right)\left(    \ee{\frac{\beta\omega'}2}   \hat c_j + (-1)^l     \ee{\frac{-\beta\omega'}2 }  \hat c_j^\dagger \right)}.
\end{align}
Hence, in the expression for the energy density, the following integrals appear,
\begin{align}
    I_{\mp,+}^k&:%
    = \frac1{4\pi}\integral\omega0\infty  \frac{ \ee{\mp\ii\omega x}\ee{-L\omega}\ee{\omega/\theta} }{\sinh(\omega/\theta) }\omega^{k+1} 
    = \frac{\theta^{k+2} (k+1)!}{2^{k+3}\pi}   \zeta\left[k+2,\frac{(L\pm\ii x)\theta}2\right],
\\
    I_{\mp,-}^k&:%
    =\frac1{4\pi} \integral\omega0\infty  \frac{ \ee{\mp\ii\omega x}\ee{-L\omega} \ee{-\omega/\theta} }{\sinh(\omega/\theta) }\omega^{k+1}
    = \frac{\theta^{k+2} (k+1)!}{2^{k+3}\pi}  \zeta\left[k+2,\frac{(L\pm\ii x)\theta}2+1\right],
\end{align}
where $\theta=2/\beta$, $\zeta[a,b]$ is the generalized Riemann zeta function, and we used $$\integral{x}0\infty\frac{x^{\mu-1}\ee{-\beta x}}{\sinh(x)}=2^{1-\mu} \Gamma(\mu) \zeta\left[\mu,\frac{\beta+1}2\right],\,\Re\mu>1,\,\Re\beta>-1,$$
with $\mu=k+2$ and $ \beta=\mp 1+L\theta\pm\ii\theta x$ (see 3.552.1 of~\cite{gradshteyn_table_2014}).
Observe that $\zeta[s,a]^*=\zeta[s^*,a^*]$ and $\zeta[s,a]=\zeta[s,a+1]+\frac1{(a^2)^{s/2}}$, hence,
\begin{align}
    I^k_{\mp,\circ}=\left(I^k_{\pm,\circ}\right)^*
    ,\qquad
    I^k_{\mp,+}= I^k_{\mp,-}+ \frac{\theta^{k+2}(k+1)!}{2^{k+3}\pi} \left( \left( \frac{(L\pm\ii x)\theta}2\right)^2 \right)^{-(k+2)/2}
    = I^k_{\mp,-}+ \frac{(k+1)!}{2\pi\left( L\pm\ii x  \right)^{k+2}} 
    \nn
	I^k_{\mp,+}+(-1)^k I^k_{\pm,-}= \frac{(k+1)!}{2\pi(L\pm\ii x)^{k+2}} +\left(I^k_{\mp,-}+(-1)^k I^k_{\pm,-}\right)= \frac{(k+1)!}{2\pi(L\pm\ii x)^{k+2}} + \begin{cases} 2\Re I^k_{\mp,-},\quad\text{if }k=0,2,4,... \\ 2\ii\Im I^k_{\mp,-},\quad\!\!\text{if }k=1,3,5,...\end{cases}\,.
\end{align}

Making use of this relation and inserting everything into~\eqref{eq:thermal_density_even} we obtain
\begin{equation}\label{eq:emergy_density_from_thermal_chain}
\begin{split}
    \exptval{\normord{(\pir^{(e)})^2}(x)} 
	&=\frac12 \sum_{i,j=0}^\infty\sum_{k=0}^i\sum_{l=0}^j P_{i,k}P_{j,l} \Re
    \left[ 
    \left(I_{-,+}^k + (-1)^k I_{+,-}^k\right) \left(I_{+,+}^l+(-1)^l I^l_{-,-}\right) \exptval{\hat c_i^\dagger \hat c_j} 
    \right.\\&\left.\quad 
    +\left(I_{+,+}^k+(-1)^k I_{-,-}^k\right) \left(I_{+,+}^l +(-1)^l I^l_{-,-}\right) \exptval{\hat c_i\hat c_j}+(-1)^l\left( I_{-,+}^k+(-1)^{k} I_{+,-}^k \right)I_{-,-}^l \delta_{ij}
    \right]
    \,.
 \end{split}
\end{equation}
The very last term in this expression is a constant that is independent of the field state. Hence this term corresponds to the contribution from the even sector to the thermal background energy density~\eqref{eq:thermal_background_density}.
This contribution is only exact in the limit of an infinite number of chain modes, but it is impacted by the truncation error in any numerical calculation with only finitely many modes.
Hence, in numerical calculations, it is more practical to replace this constant term by its known exact value and to only evaluate the modulations of the energy density's expectation value due to the interaction with the detector over time from the numerical data.

\section{Field energy density in terms of chain modes for detector at rest}\label{app:energy_density_coefffs_atrest}
Inserting $\hat b_{\pm\omega}=\frac1{\sqrt2}\left(\hat b_{\omega}^{(e)}\pm\hat b_{\omega}^{(o)}\right)$ into  the right-moving, normal-ordered energy density expectation value, and using that the odd field modes remain in their vacuum state,
yields
\begin{align}
&\exptval{\normord{\pir^2(x)}}=\exptval{\normord{\pil^2(-x)}}
=\integral\omega0\infty\integral{\omega'}0\infty \frac{\sqrt{\omega\omega'}}{4\pi} \left( \ee{- \ii (\omega-\omega') x} \exptval{{\hat{b}_\omega}^{(e)}{}^\dagger\hat b^{(e)}_{\omega'} } 
    - \Re\left[\ee{\ii (\omega+\omega')x} \exptval{\hat b^{(e)}_{\omega} \hat b^{(e)}_{\omega'} }\right]
    \right).
\end{align}
Using~\eqref{eq:be_from_chain}, 
we obtain
\begin{align}\label{eq:right_edensity_mkvacuum_chainladder}
\exptval{\normord{\pir^2(x)}}
   & =\integral\omega0\infty\integral{\omega'}0\infty \frac{\sqrt{\omega\omega'}}{4\pi} \left( \ee{- \ii (\omega-\omega') x} \exptval{{\hat{b}_\omega}^{(e)}{}^\dagger \hat {b^{(e)}}_{\omega'} } 
    - \Re\left[\ee{\ii (\omega+\omega')x} \exptval{\hat {b^{(e)}}_{\omega} \hat {b^{(e)}}_{\omega'} }\right]    \right)
    \nn&=\Re\sum_{i,j} \frac{L^2\sqrt{(i+1)(j+1)}}\pi \frac{(-L-\ii x)^j}{(L-\ii x)^{j+2}} \left( \frac{(-L+\ii x)^i}{ (L+\ii x)^{i+2}}  \exptval{\hat c_i^\dagger \hat c_j} +\frac{(-L-\ii x)^i}{ (L-\ii x)^{i+2} } \exptval{\hat c_i  \hat c_j}
    \right)
    \nn&=\Re\sum_{i,j} \frac{(-1)^{i+j}\sqrt{(i+1)(j+1)}}{\pi L^2 (1+x^2/L^2)^2}     \left(\ee{\ii 2(j-i)\arctan\tfrac{x}L}  \exptval{\hat c_i^\dagger \hat c_j} +\ee{\ii(4+2i+2j)\arctan\tfrac{x}L}  \exptval{\hat c_i  \hat c_j}
    \right)
\end{align}
where we expanded~\eqref{eq:pn_from_laguerre},
\begin{equation}\label{eq:pn_fron_laguerre_long}
    p_n(\omega)=\frac{L\sqrt{8\pi}}{ \sqrt{n+1}}L^1_n(2L \omega)
= \sqrt{\frac{2\pi}{n+1}} \sum_{k=0}^n \binom{n+1}{n-k}\frac{(-1)^k}{k!} 2^{k+1} L^{k+1} \omega^k,
\end{equation} used
$
    \integral{\omega}0\infty \frac{\omega}{\pi\sqrt8} \ee{-L\omega\mp\ii\omega x} p_n(\omega) 
    = \frac{L \sqrt{n+1} (-L\pm\ii x)^n (L\pm\ii x)^{-n-2}}{\sqrt{\pi }},
$
 and rewrote
$
L+\ii x= \sqrt{L^2+x^2} \ee{\ii \arctan\tfrac{x}L}
$.
The total energy density expectation value is given by $\normord{\hat T_{00}(x)}=\normord{\pir^2(x)}+\normord{\pil^2(x)}=\normord{\pir^2(x)}+\normord{\pir^2(-x)}$ yielding
\begin{equation}
\exptval{\normord{\hat T_{00}}}   = \frac{2}{\pi L^2}
\sum_{k,l} 
\frac{(-1)^{k+l} \sqrt{(k+1)(l+1)}}{(1+x^2/L^2)^2}
\Re \left(   \cos\left(2(k-l)\arctan\tfrac{x}L\right) \exptval{\hat c_k^\dagger \hat c_l} +\cos\left(2(k+l+2)\arctan\tfrac{x}L\right) \exptval{\hat c_k\hat c_l }
\right).
\end{equation}

\section{Perturbative calculation of emitted energy density}\label{app:pert_edensity}
Fig.~\ref{fig:edensities_vacuum_compare_perturbative} compares our numerical results for the emitted energy density to the results obtained within in leading order perturbation theory. This section derives the latter.

\subsection{Perturbative state expansion}
For time-dependent perturbation theory we employ the interaction picture, in which the field momentum operator reads
\begin{align}
\hat\pi(x,t)&=\integral{k}{-\infty}\infty \frac{-\ii\sqrt{|k|}}{\sqrt{4\pi}}\left(\ee{-\ii |k| t+\ii kx}\hat b_k-h.c.\right).
\end{align}
For the HO detector the interaction Hamiltonian reads
\begin{align}
\Hi(t)&= \lambda \left(\alad(t) +\alad^\dagger\right) \otimes \integral{x}{}{} f(x) \hat \pi(x,t)
= \lambda \left( \alad \ee{-\ii \Omega_{\mathrm{d}} t}+\alad^\dagger \ee{\ii \Omega_{\mathrm{d}} t} \right)\otimes\hat\Pi_f(t), \\
\hat\Pi_f(t)&=\integral{k}{-\infty}{\infty} \left(\frac{-\ii\sqrt{|k|}}{\sqrt{4\pi}} \ee{-\ii |k| t} \left(\integral{x}{-\infty}{\infty} f(x) \ee{\ii kx} \right) \hat b_k+h.c.\right)
=\integral{k}{-\infty}{\infty} \left( \ee{-\ii |k| t} f_k  \hat b_k+h.c.\right)
.
\end{align}
We assume that the initial state is a product state between an emitter number state $\ket{n}$, and  the field vacuum state $\ket{0_f}$.
The time evolved state is then expanded as $\ket{\psi_t}\sim\ket{n}\otimes\ket{0_f}+\ket{\psi_t^{(1)}}+\ket{\psi_t^{(2)}}+\mathcal{O}\left(\lambda^3\right) $ with
\begin{align}
\ket{\psi_t^{(1)}}&=-\ii\integral{t'}0t H_i(t')\ket{\psi_0}
=-\ii \lambda\integral{t'}0t  \left( \ee{-\ii \Omega_{\mathrm{d}} t'}\sqrt{n}\ket{n-1}+ \ee{\ii \Omega_{\mathrm{d}} t'}\sqrt{n+1}\ket{n+1} \right)\otimes \hat\Pi_f(t')\ket0
\\
\ket{\psi_t^{(2)}}&=-\integral{t'}0t\integral{t''}0{t'} H_i(t') H_i(t'')\ket{\psi_0}
\nn &
=- \lambda^2\integral{t'}0t\integral{t''}0{t'}\left( \ee{-\ii\Omega_{\mathrm{d}} (t'+t'')}\sqrt{n(n-1)}\ket{n-2} + \ee{\ii\Omega_{\mathrm{d}} (t'+t'')}\sqrt{(n+2)(n+1)}\ket{n+2} 
\right.\nn&\qquad\left.
+\left(2n\cos(\Omega_{\mathrm{d}} (t'-t'')) +\ee{\ii\Omega_{\mathrm{d}} (t''-t')}\right)\ket{n}   \right)
\otimes \hat\Pi_f(t')\hat\Pi_f(t'')\ket0.
\end{align}

For a TLS detector the interaction Hamiltonian reads
\begin{align}
\Hi(t)&= \lambda \sigma(t)\otimes\integral{x}{}{} f(x) \hat \pi(x,t)= \lambda \left( \ketbra{g}e \ee{-\ii \Omega_{\mathrm{d}} t}+\ketbra{e}g\ee{\ii \Omega_{\mathrm{d}} t} \right)\otimes\hat\Pi_f(t)\,.
\end{align}
We assume that the initial state is a product state between an atom eigenstate, either $\ket{g}$ or $\ket{e}$, and  the field vacuum state $\ket{0_f}$. 
The leading order correction to the joint atom and field state is
\begin{align}\label{eq:_state_first_order}
&\ket{\psi_t^{(1)}}=-\ii\integral{t'}0t H_i(t')\ket{\psi_0}
=  -\ii \lambda\integral{t'}0t \ee{\mp\ii \Omega_{\mathrm{d}} t'} \begin{cases}\ket g\\ \ket e \end{cases} \otimes \Pi_f(t')\ket0.
\end{align}
Here the upper sign/line applies to the initital state $\ket{e}\otimes \ket0$, and the lower to $\ket{g}\otimes\ket0$.
And the second order correction to the state is
\begin{align}
    \ket{\psi_t^{(2)}} &= -\ii\integral{t'}0tH_i(t')\ket{\psi_{t'}^{(1)}} 
    = -\lambda^2   \integral{t'}0t \integral{t''}0{t'} \ee{\mp\ii \Omega_{\mathrm{d}} (t''-t')} \begin{cases}\ket e\\ \ket g \end{cases} \otimes\hat\Pi_f(t')\Pi_f(t'')\ket0
\end{align}

\subsection{Perturbative calculation of energy density}
To leading order, the expectation value of the right-moving energy density is
\begin{align}
    \bra{\psi_t}\normord{\pir^2(x)}\ket{\psi_t}\sim\bra{\psi_t^{(1)}}\normord{\pir^2(x)}\ket{\psi_t^{(1)}}+2\Re \bra{\psi_0 \vphantom{\psi_t^{(2)}} }\normord{\pir^2(x)}\ket{\psi_t^{(2)}}.
\end{align}
In~\eqref{eq:norm_ord_pi_density_squared} we have for the right-moving energy density
\begin{align}
    \normord{\pir^2(x)} &=\integral\omega0\infty\integral{\omega'}0\infty \frac{\sqrt{\omega\omega'}}{4\pi} \left(2 \ee{- \ii (\omega-\omega') x} \hat b^\dagger_{\omega}\hat b_{\omega'} 
    - \ee{\ii (\omega+\omega')x} \hat b_{\omega} \hat b_{\omega'} - \ee{- \ii (\omega+\omega')x} \hat b_{\omega}^\dagger \hat b_{\omega'}^\dagger \right).
\end{align}
Note that in the present calculation we need to interpret $x$ as a null coordinate because we are working in the interaction picture.
Since the field starts out in the vacuum, the first order correction to the state is in the one-particle sector of the field. Hence the first term simplifies to
\begin{align}
    \bra{\psi_t^{(1)}}\normord{\pir^2(x)}\ket{\psi_t^{(1)}}= \integral\omega0\infty\integral{\omega'}0\infty \frac{\sqrt{\omega\omega'}}{2\pi} \ee{- \ii (\omega-\omega') x}  \bra{\psi_t^{(1)}} \hat b^\dagger_{\omega}\hat b_{\omega'} \ket{\psi_t^{(1)}}.
\end{align}
Similarily, the second term simplifies due to the vacuum in the initial state:
\begin{align}
    \bra{\psi_0 \vphantom{\psi_t^{(2)}} }\normord{\pir^2(x)}\ket{\psi_t^{(2)}} &=
    -\integral\omega0\infty\integral{\omega'}0\infty \frac{\sqrt{\omega\omega'}}{4\pi} \ee{\ii(\omega+\omega')x} \bra{\psi_0 \vphantom{\psi_t^{(2)}} } \hat b_{\omega} \hat b_{\omega'} \ket{\psi_t^{(2)}}.
\end{align}
Note that
\begin{align}
    \hat b_{\omega'} \hat\Pi_f(t')\ket0 &= \hat b_{\omega'} \integral{k}{-\infty}\infty \ee{\ii|k|t'} f_k^*\hat b_k^\dagger \ket{0}=\ee{\ii\omega' t'} f^*_{\omega'}\ket0
\end{align}
hence
\begin{align}
    \bra0\hat \Pi_f(t'')\hat b_\omega^\dagger\hat b_{\omega'}\hat\Pi(t')\ket0 =\ee{\ii (\omega't'-\omega t'')} f^*_{\omega'} f_\omega \,.
\end{align}
Similarily,
\begin{align}
    \bra0\hat b_{\omega'}\hat b_\omega\hat \Pi_f(t')\hat\Pi(t'')\ket0
    =\integral{k'}{-\infty}\infty \integral{k''}{-\infty}\infty \ee{\ii(|k'|t'+|k''|t'')} f_{k'}^* f_{k''}^* \bra0\hat b_{\omega'}\hat b_\omega \hat b_{k'}^\dagger\hat b_{k''}^\dagger\ket0
    =f_{\omega}^* f_{\omega'}^*( \ee{\ii(\omega t'+\omega 't'')}+\ee{\ii(\omega' t'+\omega t'')})
\end{align}
And from~\eqref{eq:def-fk}, we have
$
f_k={-\ii\sqrt{|k|}}\ee{-L|k|}/{\sqrt{4\pi}} .
$

For the TLS detector, using the upper sign  for the excited initial state $\ket{e}$, we obtain
\begin{align}
    &\bra{\psi_t^{(1)}} \hat b^\dagger_{\omega}\hat b_{\omega'} \ket{\psi_t^{(1)}}=
     \lambda^2 \integral{t'}0t\integral{t''}0t \ee{\pm\ii\Omega_{\mathrm{d}}(t'-t'')} \ee{\ii (\omega't''-\omega t')}\frac{\sqrt{\omega'\omega}}{4\pi}\ee{-L(\omega+\omega')}
\\
    &\bra{\psi_t^{(1)}}\normord{\pir^2(x)}\ket{\psi_t^{(1)}}
    = \frac{\lambda^2}{8\pi^2}   \left|\integral{t'}0t  \frac{\ee{\pm\ii\Omega_{\mathrm{d}} t'}}{(L+\ii (x+ t'))^2}\right|^2
\\
    &\bra{\psi_0 \vphantom{\psi_t^{(2)}} } \hat b_{\omega} \hat b_{\omega'} \ket{\psi_t^{(2)}} 
    =\lambda^2\frac{\sqrt{\omega'\omega}}{4\pi}\ee{-L (\omega+\omega')} \integral{t'}0t\integral{t''}0{t'} \ee{\mp\ii\Omega_{\mathrm{d}}(t''-t')}
    ( \ee{\ii(\omega t'+\omega 't'')}+\ee{\ii(\omega' t'+\omega t'')})
\\
    &2\Re\bra{\psi_0 \vphantom{\psi_t^{(2)}} }\normord{\pir^2(x)}\ket{\psi_t^{(2)}} 
    =
    -\frac{\lambda^2}{4\pi^2} \Re  \integral{t'}0t\integral{t''}0{t'} \frac{\ee{\mp\ii\Omega_{\mathrm{d}}(t''-t')}}{(L-\ii(x+t'))^2(L-\ii(x+t''))^2}
\end{align}
Thus, for the TLS emitter,
\begin{align}
    \bra{\psi_t}\normord{\pir^2(x)}\ket{\psi_t}\sim
     \frac{\lambda^2}{8\pi^2}   \left|\integral{t'}0t  \frac{\ee{\pm\ii\Omega_{\mathrm{d}} t'}}{(L+\ii (x+ t'))^2}\right|^2
     -\frac{\lambda^2}{4\pi^2} \Re  \integral{t'}0t\integral{t''}0{t'} \frac{\ee{\mp\ii\Omega_{\mathrm{d}}(t''-t')}}{(L-\ii(x+t'))^2(L-\ii(x+t''))^2}+\mathcal{O}(\lambda^3)
\end{align}

For the HO detector we obtain
\begin{align}
    &\bra{\psi_t^{(1)}} \hat b^\dagger_{\omega}\hat b_{\omega'} \ket{\psi_t^{(1)}}
    =\lambda^2 \frac{\sqrt{\omega\omega'}}{4\pi}\ee{-L(\omega+\omega')} \integral{t'}0t\integral{t''}0t \left(n\ee{\ii(\Omega_{\mathrm{d}}-\omega) t'} \ee{-\ii(\Omega_{\mathrm{d}}-\omega') t''} +(n+1)\ee{-\ii(\Omega_{\mathrm{d}}+\omega)t'}\ee{\ii (\Omega_{\mathrm{d}}+\omega')t''}\right) 
    ,\\
    &\bra{\psi_0 \vphantom{\psi_t^{(2)}} } \hat b_{\omega} \hat b_{\omega'} \ket{\psi_t^{(2)}}
     =\lambda^2 \frac{\sqrt{\omega\omega'}}{4\pi}\ee{-L(\omega+\omega')}  \integral{t'}0t\integral{t''}0{t'} \left(2n\cos(\Omega_{\mathrm{d}} (t'-t'')) +\ee{\ii\Omega_{\mathrm{d}} (t''-t')}\right) ( \ee{\ii(\omega t'+\omega 't'')}+\ee{\ii(\omega' t'+\omega t'')})
\end{align}

We then have, using $\integral\omega0\infty \omega \ee{\omega (\ii X-L)}=(L-\ii X)^{-2}$,
\begin{align}
    &\bra{\psi_t^{(1)}}\normord{\pir^2(x)}\ket{\psi_t^{(1)}}
    = \frac{\lambda^2 }{8\pi^2} \left( n \left| \integral{t'}0t \frac{\ee{\ii \Omega_{\mathrm{d}} t'}}{(L+\ii(t'+x))^2} \right|^2+ (n+1) \left| \integral{t'}0t \frac{\ee{-\ii \Omega_{\mathrm{d}} t'}}{(L+\ii(t'+x))^2} \right|^2\right)
,\\
   &2\Re \bra{\psi_0 \vphantom{\psi_t^{(2)}} }\normord{\pir^2(x)}\ket{\psi_t^{(2)}} 
   =
    -\lambda^2\Re \frac{1}{4\pi^2}   \integral{t'}0t\integral{t''}0{t'}   \frac{ 2n\cos(\Omega_{\mathrm{d}} (t'-t'')) +\ee{\ii\Omega_{\mathrm{d}} (t''-t')} }{(L-\ii(x+t'))^2 (L-\ii (x+t''))^2} 
    .
\end{align}

\section{Derivation of state error bound}\label{app:state_error_bound}
To bound the norm of the error $\|\ket\epsilon\|$, we first consider
\begin{equation}
\difffrac{}t \braket\epsilon\epsilon = 2\Re\bra\epsilon \difffrac{}t\ket{\epsilon}
    =2\Im\bra{\epsilon}\Delta H\ket{\psi^\epsilon}
\leq 2 \left|\braket{\epsilon}{\Delta H \psi^\epsilon} \right| \leq 2  \sqrt{\braket{\epsilon}\epsilon} \sqrt{\braket{\Delta H\psi^\epsilon}{\Delta H\psi^\epsilon}},
\end{equation}
and since
$
    \difffrac{}t\braket\epsilon\epsilon = \difffrac{}t \left\|\ket\epsilon\right\|^2=2\left\|\ket\epsilon\right\| \difffrac{}t\left\|\ket\epsilon\right\|
$,
we have
\begin{equation}
    \difffrac{}t\left\|\ket\epsilon\right\|\leq \sqrt{\braket{\Delta \hat H\psi^\epsilon}{\Delta \hat  H\psi^\epsilon}}.
\end{equation}
This expression can be evaluated in numerical calculations, because it only involves   $\ket{\psi^\epsilon}$ which we obtain from the numerical calculations.
The state $\ket{\psi^\epsilon}$ always remains a product state between the first $N$ sites and the rest of the chain, which remains in its vacuum state, 
hence
\begin{equation}
     \braket{\Delta \hat H\psi^\epsilon}{\Delta \hat  H\psi^\epsilon}=\gamma_{N-1}^2  \bra{\psi^\epsilon}\hat c_{N-1}^\dagger\hat c_{N-1}\ket{\psi^\epsilon}.
\end{equation}
At $t=0$  the error vanishes, $\ket\epsilon=0$, and therefore its norm  at time $t$ is lower or equal to the integral
\begin{equation}
    \left\|\ket\epsilon\right\|\leq 
    \epsilon_t:= %
        \left|\gamma_{N-1}\right| \integral{t'}0t \sqrt{\bra{\psi^\epsilon}\hat c_{N-1}^\dagger\hat c_{N-1}\ket{\psi^\epsilon}}.
\end{equation}

\section{Error bound for quadratic observables and harmonic emitters}\label{app:quadratic_error_bound}
When the emitter is a harmonic oscillator and the initial state is Gaussian, the system remains in a Gaussian state both under the exact and the truncated Hamiltonian, because both are quadratic.
In this scenario, we can use Gaussian state methods to derive an error bound on quadratic observables similar to the error bound~\eqref{eq:errorbound}.
We employ the Kähler structure formalism for Gaussian states (for a review, see~\cite{hackl_bosonic_2021}).

Assume that we are interested in the expectation value of a quadratic observable. Then, 
working with respect to a real symplectic basis of quadrature operators (\ie $\hat{\mat\xi}^\tra =(\hat q_1,\hat q_2,...,\hat p_1,\hat p_2,...)$ with $\comm{\hat q_i}{\hat p_j}=\ii\delta_{ij}$), we can express the observable as $\hat O=\frac12\sum_{i,j}\mat{O}_{ij} \hat\xi^i\hat\xi^j$. We may assume that the matrix $\mat{O}$ is symmetric, since any anti-symmetric part would only add an operator proportional to the identity operator to $\mat O$. Thus, the expectation value of $\hat O$ is given by
\begin{equation}
    \exptval{\hat O}=\frac14\sum_{i,j} \mat{O}_{ij}\exptval{\hat\xi^i\hat\xi^j+\hat\xi^j\hat\xi^i}
   = \frac14\sum_{i,j}\mat{O}_{ij}\mat{G}_{ij}
    =\frac14\Tr \mat{O}^\tra  \mat{G}=\Tr \mat{O^\tra \Omega J^\tra} 
    =\frac14\Tr \mat{\Omega^\tra O J}
    =\frac14\langle \mat{O^\tra \Omega},\mat J\rangle
\end{equation}
where the matrix $\mat\Omega_{ij}=\ii\comm{\hat\xi^i}{\hat\xi^j}$ represents the symplectic form, $\mat G_{ij}=\exptval{\hat\xi^i\hat\xi^j+\hat\xi^j\hat\xi^i}$ represents the covariance matrix of the state and $\mat{J}=-\mat{G \Omega^{-1}}$ represents the linear complex structure of the state (represented by a real square matrix), and we used the Frobenius scalar product $\langle \mat{A},\mat B\rangle=\Tr\mat{A^\tra B}$ for real-valued square matrices.

The linear complex structure evolves in time as
\begin{equation}
    \mat{J}(t) =\ee{t \mat K}\mat{J}(t=0) \ee{-t \mat K}\quad \Rightarrow \dot{\mat J}=\mat K \mat J(t)-\mat J(t)\mat K=\comm{\mat J(t)}{\mat K}.
\end{equation}
Here $\mat K=\mat{\Omega h}$ represents the Hamiltonian generator of the full system Hamiltonian which is $\hat H=\frac12\sum_{i,j} \mat{h}_{i,j}\hat\xi^i\hat\xi^j$.
However, due to the truncation of the chain we are not calculating the state evolution under the full Hamiltonian, but only with the truncated Hamiltonian generator $\mat{K^\epsilon}=\mat{K-\Delta K}$.
Accordingly, we only calculate the linear complex structure $\mat{J^\epsilon}=\mat{J}- \mat{\Delta J}$ with $\dot{\mat{J^\epsilon}}=\comm{\mat{K^\epsilon}}{\mat{J^\epsilon}}$.

The error in the expectation value, which we seek to bound, is
\begin{equation}
    \left|\exptval{O}-\exptval{O}^\epsilon \right|
    =\frac14\left| \left<\mat{O^\tra\Omega},\mat{\Delta J}\right>\right|
    \leq \frac14 \sqrt{\langle\mat{O^\tra\Omega},\mat{O^\tra\Omega}\rangle} \sqrt{\langle \mat{\Delta J},\mat{\Delta J}\rangle} 
    =\frac14 \left\|\mat{O^\tra \Omega}\right\| \left\|\mat{\Delta J}\right\|.
\end{equation}
The time derivative of the error in the linear complex structure is
\begin{equation}
    \dot{\mat{\Delta J}}= \comm{\mat K}{\mat J}-\comm{\mat{K^\epsilon}}{\mat{J^\epsilon}}
    = \comm{\mat{K^\epsilon+\Delta K}}{\mat{J^\epsilon+\Delta J}} -\comm{\mat{K^\epsilon}}{\mat{J^\epsilon}}
    = \comm{\mat K}{\mat{\Delta J}}+\comm{\mat{\Delta K}}{\mat J^\epsilon}.
\end{equation}
This we can use to bound
\begin{equation}
    \difffrac{}t \sqrt{\langle \mat{\Delta J},\mat{\Delta J}\rangle}
    =\frac{\difffrac{}t \langle \mat{\Delta J},\mat{\Delta J}\rangle}{2 \sqrt{\langle \mat{\Delta J},\mat{\Delta J}\rangle}} 
    =\frac{\difffrac{}t \Tr\mat{\Delta J^\tra \Delta J} }{2 \sqrt{\langle \mat{\Delta J},\mat{\Delta J}\rangle}} .
\end{equation}
Next, since $\Tr A B=\Tr B^\tra A^\tra$, we have
\begin{equation}
    \difffrac{}t \Tr\mat{\Delta J^\tra \Delta J}
    =2\Tr\dot{\mat {\Delta J^\tra}} \mat{\Delta J}
    =2 \Tr\dot{\mat {\Delta J}} \mat{\Delta J^\tra}
    =2\Tr \left( \comm{\mat K}{\mat{\Delta J}} \mat{\Delta J^\tra} +\comm{\mat{\Delta K}}{\mat J^\epsilon}\mat{\Delta J^\tra}\right).
\end{equation}
The first term $\Tr  \comm{\mat K}{\mat{\Delta J}} \mat{\Delta J^\tra}$ vanishes, because both
$    \Tr \mat{K\Delta J\Delta J^\tra}=0$ and $\Tr \mat{\Delta J K\Delta J^\tra}=0$. This is seen using the cyclicity of the trace and  $\Tr A =\Tr A^\tra$ and using that $\mat K^\tra=-\mat K$, for example,
$
    \Tr \mat{K\Delta J\Delta J^\tra}=\Tr \mat{\Delta J\Delta J^\tra K^\tra}=-\Tr \mat{\Delta J\Delta J^\tra K}=0.
$
Thus,
\begin{equation}
    \difffrac{}t \Tr\mat{\Delta J^\tra \Delta J}
    =2\Tr \comm{\mat{\Delta K}}{\mat J^\epsilon}\mat{\Delta J^\tra}
    \leq 2 \sqrt{\langle \comm{\mat{\Delta K}}{\mat J^\epsilon},\comm{\mat{\Delta K}}{\mat J^\epsilon}\rangle }\sqrt{\langle \mat{\Delta J^\tra}, \mat{\Delta J^\tra}\rangle}
\end{equation}
such that
\begin{equation}\label{eq:quadratic_bound_dt_app}
    \difffrac{}t \left\|\mat{\Delta J}\right\|=\difffrac{}t \sqrt{\langle \mat{\Delta J},\mat{\Delta J}\rangle} \leq \sqrt{\langle \comm{\mat{\Delta K}}{\mat J^\epsilon},\comm{\mat{\Delta K}}{\mat J^\epsilon}\rangle }=\left\| \comm{\mat{\Delta K}}{\mat J^\epsilon}\right\|.
\end{equation}
Note that since $\mat{G}=-\mat{J\Omega}$ we have $\|\mat G\|=\|\mat J\|$, thus, the bound directly gives a bound on the error in the covariance matrix of the calculated state.

In order to evaluate the right hand side, we express the truncation part of the Hamiltonian~\eqref{eq:DeltaH} in terms of chain mode quadrature operators.
\begin{equation}
    \Delta \hat H= \gamma_{N-1}\left(\hat c^\dagger_{N-1}\hat c_N+\hat c_N^\dagger \hat c_{N-1}\right)
    = \gamma_{N-1}\left(\hat q_{N-1}\hat q_N+\hat p_{N-1} \hat p_N \right).
\end{equation}
Hence we have
\begin{equation}
    \mat{\Delta K}=\frac12 \mat{\Omega} \left(\begin{array}{c|c}
         \begin{array}{cc}
             0 & \gamma_{N-1} \\
             \gamma_{N-1} & 0 
         \end{array} & 0 \\ \hline
          0& \begin{array}{cc}
             0 & \gamma_{N-1} \\
             \gamma_{N-1} & 0 
         \end{array} 
    \end{array}\right)
    =\frac12  \left(\begin{array}{c|c}
         0& \begin{array}{cc}
             0 & \gamma_{N-1} \\
             \gamma_{N-1} & 0 
         \end{array}  \\ \hline
           \begin{array}{cc}
             0 & -\gamma_{N-1} \\
             -\gamma_{N-1} & 0 
         \end{array} & 0
    \end{array}\right)
\end{equation}
Since we are restricting our calculation to $N$ chain modes (and one mode given by the harmonic oscillator emitter),  the matrix $\mat{J^\epsilon}$ takes the form
\begin{equation}
    \mat{J^\epsilon}=\left( \begin{array}{cc|cc}
        \mat A &  & \mat B & \\ & 0 & &\id \\ \hline
         \mat C & &\mat D & \\  0&-\id & & 0
    \end{array}\right),
\end{equation}
with $(N+1)\times (N+1)$-matrices $\mat A,\mat B,\mat C,\mat D$.
Using indices $\mat{A}_{i,j}=a_{i,j}$  with $i,1=-1,0,1,...,N-1$, and analogously for the other matrices, we can calculate the right hand side of~\eqref{eq:quadratic_bound_dt_app} from
\begin{equation}
\begin{split}
    \frac1{\gamma_{N-1}^2}\left\| \comm{\mat{\Delta K}}{\mat J^\epsilon}\right\|^2
    &=  2  (b_{N-1,N-1}-1)^2+ 2(c_{N-1,N-1}+1)^2 +\sum_{k=-1}^{N-2}\left( (b_{k,N-1})^2 +(b_{N-1,k})^2+ (c_{k,N-1})^2 +(c_{N-1,k})^2 \right)
    \\& \qquad
     +\sum_{k=-1}^{N-1}\left( (a_{k,N-1})^2 +(a_{N-1,k})^2+ (d_{k,N-1})^2 +(d_{N-1,k})^2 \right).
\end{split}
\end{equation}

In order to apply the above bound to the expectation value of the observable $\hat O$ the Frobenius norm of $\mat O$ needs to be finite.
One important example of such an operator is the number operator of a properly normalized positive frequency mode, \ie a mode that shares the vacuum state with the chain modes.

\section{Minkowski energy density from chain modes in the Unruh effect}\label{app:mk_density_from_unruh_chain}
Using the Bogoliubov transformations derived above, the Minkowski mode operators $\hat a_\omega^\pm$ can be expressed as a linear combination
\begin{align}
    &\hat a^\pm_\omega=\integral\Omega{-\infty}\infty \gamma_{\Omega\omega}^\pm \hat d_{\Omega}^\pm 
    =\frac1{\sqrt2} \integral\Omega{-\infty}\infty \gamma_{\Omega\omega}^\pm \left( \hat d_{\Omega}^e\pm d_{\Omega}^o\right)
    =   \integral\Omega{-\infty}\infty \gamma_{\Omega\omega}^\pm 
    \left( \sum_i  \frac{\sgn(\Omega) f_{|\Omega|}^* \ee{\frac{\Omega\pi}{2a}}   }{\sqrt{2|\sinh(\Omega\pi/a)|}} p_i(\Omega) \hat c_i
    \pm \frac{d_{\Omega}^o}{\sqrt2} \right)
    \nn& =\sum_i A_{\omega,i} \hat c_i +\hat O^{(o)}_\omega
\end{align}
of chain mode operators $\hat c_i$, which is a linear combination of even Unruh modes, and some operator $\hat O^{(o)}_\omega$, which is a linear combination of odd Unruh modes. The precise form of $\hat O^{(o)}_\omega$ is irrelevant to our purpose, because the odd sector remains in the vacuum state.
Formally, for the coefficients $A_{\omega,i}$ we use the regularized expression for $\gamma^\pm_{\Omega\omega}$ and, with~\eqref{eq:gamma_bogo} and writing $p_i(\Omega)=\sum_{k=0}^i P_{i,k}\Omega^k$, obtain
\begin{equation}
    \label{eq:Mk_modes_from_chain}
     A_{\omega,i}%
     =\integral\Omega{-\infty}\infty \gamma_{\Omega\omega}^\pm  \frac{\sgn(\Omega) f_{|\Omega|}^* \ee{\frac{\Omega\pi}{2a}}   }{\sqrt{2|\sinh(\Omega\pi/a)|}} p_i(\Omega)
     = \sum_{k=0}^i    \frac{\ii  P_{i,k}}{4\pi a\sqrt{ \pi \omega} }  \integral\Omega{-\infty}\infty \left(\frac{\omega}a\right)^{\pm\ii \Omega/a } \Gamma(\mp\ii \Omega/a) \ee{-L|\Omega|}  \ee{\frac{\Omega\pi}{2a}}  \Omega^{k+1}.
\end{equation}

Based on this expression, a closed form expressions for  the energy density of the field in terms of the chain modes can be obtained.
Using the notation as introduced in App.~\ref{app:unruh_review}, the expectation value of the  normal ordered, Minkowski energy density of the field~\eqref{eq:norm_ord_pi_density_squared} takes the following form, into which we insert~\eqref{eq:Mk_modes_from_chain}:
\begin{align}
    \exptval{\normord{\hat\pi_\pm^2(x)}} &=\integral\omega0\infty\integral{\omega'}0\infty \frac{\sqrt{\omega\omega'}}{4\pi} \left(2 \ee{\pm \ii (\omega-\omega') x} \exptval{\hat a_\omega^\pm{}^\dagger  \hat a_{\omega'}^\pm} - \ee{\mp\ii (\omega+\omega')x} \exptval{\hat a_{\omega}^\pm \hat a_{\omega'}^\pm }- \ee{\pm \ii (\omega+\omega')x} \exptval{ \hat a_{\omega}^\pm{}^\dagger \hat a_{\omega'}^\pm{}^\dagger } \right)
    \nn&=\Re \sum_{i,j} \integral\omega0\infty\integral{\omega'}0\infty \frac{\sqrt{\omega\omega'}}{2\pi} \left( \ee{\pm \ii (\omega-\omega') x} \exptval{A_{\omega,i}^*\hat c_i^\dagger  A_{\omega',j}\hat c_j} - \ee{\mp\ii (\omega+\omega')x} \exptval{A_{\omega,i} \hat c_i   A_{\omega',j}\hat c_j } \right)
    \nn&=\Re \sum_{i,j} J_j(x) \left( J_i^*(x) \exptval{ \hat c_i^\dagger   \hat c_j} - J_i(x) \exptval{\hat c_i  \hat c_j } \right),
\end{align}
where
\begin{align}
    J_{j}(x) &= \integral{\omega}0\infty \ee{\mp\ii\omega x} \sqrt{\frac{\omega}{2\pi}} A_{\omega,j}
    \nn&
    = \integral{\omega}0\infty \ee{\mp\ii\omega x} \sqrt{\frac{\omega}{2\pi}} 
    \left( \ii \sum_{k=0}^j   P_{j,k}  \frac{1 }{2\pi a\sqrt{ 4\pi \omega} }  \integral\Omega{-\infty}\infty \left(\frac{\omega}a\right)^{\pm\ii \Omega/a } \Gamma(\mp\ii \Omega/a) \ee{-L|\Omega|}  \ee{\frac{\Omega\pi}{2a}}  \Omega^{k+1}\right)
    \nn&= \frac{\ii}{4\pi^2 a\sqrt{2} }\sum_{k=0}^j  P_{j,k} \integral\Omega{-\infty}\infty  a^{\mp\ii \Omega/a } \Gamma(\mp\ii \Omega/a) \ee{-L|\Omega|}  \ee{\frac{\Omega\pi}{2a}}  \Omega^{k+1}   \integral{\omega}0\infty \ee{\mp\ii\omega x}   \omega^{\pm\ii \Omega/a }.
\end{align}
Now introduce a regularisation in the $\omega$-integration, 
\begin{align}
      &\integral{\omega}0\infty \ee{\mp\ii\omega x}   \omega^{\pm\ii \Omega/a }\ee{-\epsilon \omega}
        = \frac{\pm \ii \Omega }{a} \Gamma \left(\frac{\pm \ii \Omega }{a}\right) \ee{\mp\frac{\ii \Omega }{a} \ln(\epsilon\pm\ii x) } \frac1{\epsilon\pm\ii x} 
    \stackrel{\epsilon\to0}\to 
    \frac{\Omega }{a x} \Gamma \left(\frac{\pm \ii \Omega }{a}\right) \ee{\mp\frac{\ii \Omega }{a} \ln|x| } \ee{\sgn(x) \Omega \pi/(2a)},
\end{align}
then
\begin{align}
    J_{j}(x) &= 
    \frac{\ii}{4\pi^2 a^2 x \sqrt{2} }\sum_{k=0}^j  P_{j,k} \integral\Omega{-\infty}\infty   \left|\Gamma(\mp\ii \Omega/a)\right|^2 \ee{-L|\Omega|}  \Omega^{k+2} \ee{\mp\frac{\ii \Omega }{a} \ln|x a| } \ee{(1+\sgn(x))\frac{\Omega\pi}{2a} }
    \nn&= \frac{\ii}{4\pi a x \sqrt{2} }\sum_{k=0}^j  P_{j,k} \integral\Omega{-\infty}\infty   \frac{ \ee{-L|\Omega|+(1+\sgn(x))\frac{\Omega\pi}{2a}\mp\frac{\ii \Omega }{a} \ln|x a| }  \Omega^{k+1} }{ \sinh(\pi \Omega/a)} 
    \nn&=\sum_{k=0}^j  P_{j,k} \underbrace{\frac{\ii a^{k+1} }{4\pi^{k+3}  x \sqrt{2} } \integral\nu{-\infty}\infty   \frac{ \ee{-La |\nu|/\pi +\left( \frac{(1+\sgn(x))}{2}\mp\frac{\ii}{\pi} \ln|x a|\right)\nu }  \nu^{k+1} }{ \sinh(\nu)}}_{=:I^\pm_k(x)}
    =\sum_{k=0}^j P_{j,k} I_k(x)
\end{align}
with $\nu=\Omega\pi/a$. 
We can split the $\nu$-integration into two,
\begin{align}
   &\integral\nu0\infty   \frac{ \ee{  \left(-\frac{L a}\pi+ \frac{(1+\sgn(x))}{2}\mp\frac{\ii}{\pi} \ln|x a|\right)\nu }  \nu^{k+1} }{ \sinh(\nu)} 
   =2^{-k-1} \Gamma(k+2) \zeta\left[k+2, \frac12\left(\frac{La}\pi-\frac{1+\sgn(x)}2 \pm\frac{\ii\ln|xa|}{\pi} +1\right)\right]
   \nn&
   =2^{-k-1} \Gamma(k+2) \zeta\left[k+2, \frac{aL\pm\ii\ln|xa|}{2\pi}+\frac{1-\sgn(x)}{4} \right],
\\
    &\integral\nu{-\infty}0   \frac{ \ee{-La |\nu|/\pi +\left( \frac{(1+\sgn(x))}{2}\mp\frac{\ii}{\pi} \ln|x a|\right)\nu }  \nu^{k+1} }{ \sinh(\nu)} 
    =(-1)^k\integral\nu0\infty   \frac{ \ee{-La \nu/\pi -\left( \frac{(1+\sgn(x))}{2}\mp\frac{\ii}{\pi} \ln|x a|\right)\nu }  \nu^{k+1} }{ \sinh(\nu)} 
    \nn&= (-1)^k2^{-k-1}\Gamma(k+2) \zeta\left[k+2, \frac12\left( \frac{La}\pi +\frac{1+\sgn(x)}2\mp\frac\ii\pi\ln|ax| +1\right)\right]
    \nn&=(-1)^k 2^{-k-1}\Gamma(k+2) \zeta\left[k+2,   \frac{aL\mp\ii\ln|xa| }{2\pi} +\frac{3+\sgn(x)}4  \right]
\end{align}
where we used $
\integral{x}0\infty\frac{x^{\mu-1}\ee{-\beta x}}{\sinh(x)}=2^{1-\mu}\Gamma(\mu)\zeta[\mu,\frac12(\beta+1)],\,\Re\mu>1,\,\Re\beta>-1$
(see 3.552.1 of~\cite{gradshteyn_table_2014}).
Inserting above we obtain
\begin{align}
    &I^\pm_{k}(x) = \frac{\ii a^{k+1} (k+1)! }{(2\pi)^{k+3}  x \sqrt{2} } \left( \zeta\left[k+2, \frac{aL\pm\ii\ln|xa|}{2\pi}+\frac{1-\sgn(x)}{4} \right]+(-1)^k  \zeta\left[k+2,   \frac{aL\mp\ii\ln|xa| }{2\pi} +\frac{3+\sgn(x)}4  \right]\right).
\end{align}
For negative $x<0$, we obtain
\begin{align}
    &I^\pm_{k}(x) = \frac{\ii a^{k+1} (k+1)! }{(2\pi)^{k+3}  x \sqrt{2} } \begin{cases} 2\Re \zeta\left[k+2, \frac{aL\pm\ii\ln|xa|}{2\pi}+\frac{1}2 \right],& k  \text{ even}, \\
    2\ii\Im \zeta\left[k+2, \frac{aL\pm\ii\ln|xa|}{2\pi}+\frac{1}2 \right],& k  \text{ odd},
    \end{cases}
\end{align}
which, however, is not relevant for our considerations here since in our setup we only consider the energy density in the right Rindler wedge.
There, for positive $x>0$, we obtain (using $\zeta(s,a)=\zeta(s,a+1)+\frac1{\left(a^2\right)^{s/2}}$)
\begin{align}\label{eq:I_terms_unruh_edensity}
    I^\pm_k(x) &= \frac{\ii a^{k+1} (k+1)! }{(2\pi)^{k+3}  x \sqrt{2} } \left( \zeta\left[k+2, \frac{aL\pm\ii\ln|xa|}{2\pi}  \right]+(-1)^k  \zeta\left[k+2,   \frac{aL\mp\ii\ln|xa| }{2\pi} +1  \right]\right)
    \nn& = \frac{\ii a^{k+1} (k+1)! }{(2\pi)^{k+3}  x \sqrt{2} } \left( \zeta\left[k+2, \frac{aL\pm\ii\ln|xa|}{2\pi}  \right] \right.
    \nn&\qquad\left. +(-1)^k  \zeta\left[k+2,   \frac{aL\mp\ii\ln|xa| }{2\pi}   \right]+(-1)^{k+1}\left( \frac{aL\mp\ii\ln|ax| }{2\pi}  \right)^{-k-2} \right)
     \nn& =  \left( \frac{\ii(-1)^{k+1} a^{k+1} (k+1)! }{2\pi  x \sqrt{2} } \left(  aL\mp\ii\ln|xa| \right)^{-k-2} + \frac{\ii a^{k+1} (k+1)! }{(2\pi)^{k+3}  x \sqrt{2} } \begin{cases} 2\Re \zeta\left[k+2, \frac{aL\pm\ii\ln|xa|}{2\pi}  \right], & k\text{ even} \\
     2\ii\Im \zeta\left[k+2, \frac{aL\pm\ii\ln|xa|}{2\pi}  \right], & k\text{ odd}\end{cases}\right).
\end{align}
These enter the final expression for the energy density as
\begin{align}\label{eq:app_mk_edensity_from_chain}
    \exptval{\normord{\hat\pi_\pm^2(x)}} &=\Re \sum_{i,j}\sum_{k=0}^i\sum_{l=0}^j  \left( I_k^{\pm *}(x) P_{i,k} \exptval{ \hat c_i^\dagger   \hat c_j}P_{j,l}I^\pm_l(x) - I^\pm_k(x)  P_{i,k} \exptval{\hat c_i  \hat c_j }P_{j,l}I^\pm_l(x) \right).
\end{align}
Note that this expression is valid only on the hyperplane $\tau=t=0$. Under the Rindler time evolution the right-handside of this equation evolves into a transformed observable expectation value as detailed in App.~\ref{app:mk_observables_vs_rindler_evol}.
\end{widetext}

 \bibliographystyle{quantum}
\bibliography{Unruh-article,mathematica,man_refs}
\end{document}